\documentclass[fleqn,usenatbib]{mnras}
\usepackage{newtxtext,newtxmath}
\usepackage[T1]{fontenc}

\DeclareRobustCommand{\VAN}[3]{#2}
\let\VANthebibliography\thebibliography
\def\thebibliography{\DeclareRobustCommand{\VAN}[3]{##3}\VANthebibliography}

\usepackage{orcidlink}   
\usepackage{graphicx}	 
\usepackage{siunitx}     
\usepackage{booktabs}    
\usepackage{anyfontsize} 

\newcommand{\HI}{\text{\textsc{H\,i}}}

\title
[Simulations of 21\,cm emission from \HI{} galaxies]
{Simulations of the 21\,cm emission line for upcoming large-scale \HI{} galaxy surveys}

\author[J. Mayor et al.]{%
    Jo\"{e}l Mayor$^{1}$\thanks{E-mail: jmayor@ethz.ch}\orcidlink{0009-0008-6499-8013},
    Marta Spinelli$^{2,3}$\orcidlink{0000-0003-0148-3254},
    Gabriella De Lucia$^{4,5}$\orcidlink{0000-0002-6220-9104},
    Robert Yates$^{6}$\orcidlink{0000-0001-9320-4958},
    Alexandre Refregier$^{1}$\orcidlink{0000-0003-3416-9317},
    \newauthor
    Fabio Fontanot$^{4,5}$\orcidlink{0000-0003-4744-0188},
    Lizhi Xie$^{7}$\orcidlink{0000-0003-3864-068X},
    and Michaela Hirschmann$^{8,4}$\orcidlink{0000-0002-3301-3321},
\\
$^{1}$Institute for Particle Physics and Astrophysics, ETH Z\"urich, Wolfgang-Pauli-Strasse 27, 8093 Z\"urich, Switzerland\\
$^{2}$Observatoire de la C\^ote d’Azur, Laboratoire J-L Lagrange, Boulevard de l'Observatoire, Nice, France\\
$^{3}$Department of Physics and Astronomy, University of the Western Cape, Cape Town 7535, South Africa\\
$^{4}$INAF - Astronomical Observatory of Trieste, via G.B. Tiepolo 11, I-34143 Trieste, Italy\\
$^{5}$IFPU - Institute for Fundamental Physics of the Universe, via Beirut 2, 34151, Trieste, Italy\\
$^{6}$Centre for Astrophysics Research, University of Hertfordshire, Hatfield, AL10 9AB, UK\\
$^{7}$Astrophysics Center, Tianjin Normal University, Binshuixidao 393, 300387, Tianjin, China\\
$^{8}$EPFL - Institute of Physics, Laboratory for Galaxy Evolution, Observatoire de Sauverny, Chemin Pegasi 51, 1290 Versoix, Switzerland
}

\date{Accepted XXX. Received YYY; in original form ZZZ}
\pubyear{\the\year{}}

\begin{document}

\label{firstpage}
\pagerange{\pageref{firstpage}--\pageref{lastpage}}
\maketitle

\begin{abstract}
    Upcoming galaxy surveys with the SKA Observatory will detect neutral hydrogen (\HI{}) across unprecedented volumes, and their scientific return will crucially depend on predictive models for \HI{} observables.
    In this work, we present a framework to simulate the neutral hydrogen 21\,cm emission line in such large-scale \HI{}-selected galaxy surveys.
    This framework is developed as a modular layer that builds on semi-analytic models.
    In particular we use as bases the Galaxy Evolution and Assembly (GAEA) and L\nobreakdash-Galaxies semi-analytic models, coupled to merger trees from the Millennium Simulation suite.
    We compare predictions from our framework against local Universe observations, in particular statistics of the 21\,cm line width such as generalised Tully-Fisher relations.
    Comparing predictions based on GAEA and L\nobreakdash-Galaxies allows us to quantify and characterise the modelling uncertainties associated with the distinct underlying physical prescriptions.
    We construct forward-modelled \HI{}-selected galaxy catalogues that incorporate selection, velocity dispersion, inclination, and redshift broadening effects, to further investigate to what extent our approach can reproduce the statistical distributions from recent observational data.
    Finally, we present forecasts for the redshift distribution of future \HI{}-selected galaxy surveys with the SKA Observatory.
    Our framework provides a flexible and controlled basis on which to build the essential tools that will be needed for interpreting upcoming large-scale radio surveys and studying \HI{} line observables as cosmological probes.
\end{abstract}



\section{Introduction}

    Observations of atomic neutral hydrogen (\HI{}) through its characteristic 21\,cm hyperfine transition emission line are key for extragalactic astrophysics \citep[e.g.][]{kerr_1954,giovanelli_1988}, providing unique insights that complement optical and infrared observations, especially in tracing diffuse, dust-obscured gas \citep[e.g.][]{wilson_2013}.

    Within galaxies, \HI{} is the dominant atomic phase of the cold interstellar medium and fuels the baryon cycle between the circumgalactic medium and the star-forming disc via accretion, cooling, and feedback-driven recycling \citep[e.g.][]{keres_2005,dekel_2006,fraternali_2008,sancisi_2008,tumlinson_2017}.
    In sufficiently dense, shielded and metal-enriched regions, \HI{} converts to molecular hydrogen ($\mathrm{H_2}$), providing the immediate fuel for star formation \citep[e.g.][]{hollenbach_1971,blitz_2006,krumholz_2009,gnedin_2011,cortese_2017}.
    Mapping the \HI{} content and distribution in individual galaxies therefore provides a crucial tool for understanding galactic assembly and evolution, the regulation of star formation, and phase transitions within the interstellar medium \citep[e.g.][]{gunn_1972,boselli_2006,bigiel_2008,kennicutt_2012}.
    
    Moreover, the 21\,cm emission line enables kinematic studies.
    For spatially resolved interferometric observations, synthesis imaging delivers rotation curves and gas‑velocity dispersion fields that constrain dynamical mass profiles and the impact of pressure support \citep[e.g.][]{begeman_1989,sofue_2001,deblok_2008,lelli_2016a}.
    For unresolved detections, integrated \HI{} profiles capture global dynamics \citep[e.g.][]{tully_1985} and yield rotational velocity estimates through measurements of their width.

    A central example is the Tully-Fisher (TF) relation, a correlation between \HI{} profile widths and optical magnitudes of spiral galaxies which was originally defined by \citet{tully_1977}.
    Its generalisations include any empirical correlation between proxies of the rotational motion and proxies of the intrinsic luminosity of a galaxy, such as the baryonic Tully-Fisher (bTF) relation \citep[e.g.][]{mcgaugh_2000,bell_2001,mcgaugh_2005,lelli_2019}.
    In galaxy evolution studies, the TF/bTF relations are used to constrain the coupling between haloes, discs, and feedback, the role of angular momentum, and the balance of rotation and pressure support.
    In the frame of cosmological studies, the TF/bTF relations provide luminosity distance estimates \citep[e.g.][]{sorce_2014}, enabling the construction of peculiar‑velocity catalogues \citep[e.g.][]{springob_2007, tully_2023}.
    These catalogues can be employed to probe the cosmic growth rate via velocity–density comparisons, test gravity at cosmological scales, and contribute to $H_0$ determinations \citep[e.g.][]{kourkchi_2022, watkins_2023, boubel_2024b, boubel_2024c, boubel_2025a}.
    The precision of such distance estimates depends critically on the accuracy with which \HI{} line profiles can be modelled, making a robust predictive framework a prerequisite for cosmology studies with peculiar velocities.

    \HI{}-selected galaxy surveys can thus provide independent and synergistic constraints on cosmological models alongside optical/near‑infrared spectroscopic programs such as
    SDSS/BOSS/eBOSS \citep{york_2000,dawson_2013,dawson_2016,almeida_2023},
    DESI \citep{desi_2016,desi_2025},
    4HS \citep{taylor_2023},
    and Euclid \citep{laureijs_2011,euclid_2025} through `classical' galaxy clustering probes like baryon acoustic oscillations and redshift‑space distortions, and via peculiar velocity measurements \citep[e.g.][]{koda_2014, howlett_2017b, howlett_2019, qin_2019, qin_2025}.

    The observational landscape for \HI{} galaxy studies, built on surveys such as the \HI{} Parkes All-Sky Survey \citep[HIPASS,][]{barnes_2001,meyer_2004},
    and the Arecibo Legacy Fast ALFA Survey \citep[ALFALFA,][]{giovanelli_2005,haynes_2011,haynes_2018}, is undergoing rapid transformation with the advent of SKA Observatory (SKAO) pathfinder instruments, such as the Australian SKA Pathfinder (ASKAP) and the Meer-Karoo Array Telescope (MeerKAT).
    Ongoing and planned programs, including WALLABY and DINGO \citep[ASKAP,][]{johnston_2008, duffy_2012a,duffy_2012b,koribalski_2020}, LADUMA and MIGHTEE-HI \citep[MeerKAT,][]{blyth_2016, jarvis_2016,maddox_2021}, CHILES \citep[VLA,][]{fernandez_2013,fernandez_2016}, and CRAFTS/FASHI \citep[FAST,][]{zhang_2019,zhang_2024}, are delivering unprecedented combinations of sensitivity, area, and depth.
    Looking ahead, future infrastructures, specifically the SKAO, promise routine detections to higher redshifts and statistical samples suitable for studies of galaxy evolution and the cosmic large‑scale structure \citep[e.g.][]{yaha_2015, murphy_2018, braun_2019, weltman_2020}.
    
    As ongoing and next-generation \HI{}-selected galaxy surveys enter this new era of statistical samples, fully exploiting the growing datasets requires theoretical frameworks that include the physical processes underlying the emergence of scaling relations such as the TF/bTF, and that are capable of predicting their full shape, i.e. the median relations, their scatter, outliers, and evolution with redshift and environment.
    In particular, such frameworks are essential for understanding and constraining systematics in the TF/bTF relations as distance indicators and ultimately in peculiar velocity cosmology.
    This motivates modern, predictive models for \HI{} observables, capable of both reproducing currently available data and forecasting future surveys.
    
    Simulation-based modelling of the \HI{} distribution in galaxies presents unique challenges.
    Hydrodynamical simulations, e.g. EAGLE \citep{schaye_2015}, IllustrisTNG \citep{springel_2018}, BAHAMAS \citep{mccarthy_2017}, SIMBA \citep{dave_2019}, NEWHORIZON \citep{dubois_2021}, FLAMINGO \citep{schaye_2023}, CROCODILE \citep{oku_2024}, COLIBRE \citep{schaye_2026}, incorporate detailed baryonic physics and now reach particle numbers, volumes and resolutions comparable to those of the N-body simulations on which semi-analytic models are commonly run.
    In most of them, the cold interstellar medium (ISM) is not modelled explicitly and \HI{} remains a post-processing output \citep[e.g.][]{diemer_2018}.
    A notable exception is COLIBRE, which follows the cold gas phase and the non-equilibrium chemistry of hydrogen on the fly in volumes of up to $(400~\mathrm{cMpc})^3$ \citep{schaye_2026}.
    The computational cost of such runs nevertheless limits the number of model variations that can be explored at full volume.
    The calibration and exploration of their sub-grid parameter space typically rely on emulators trained on much smaller volumes \citep[e.g.][]{chaikin_2026}.

    Semi-analytic models (SAMs) offer a complementary approach, enabling efficient simulations of the large-scale assembly and evolution of galaxies across cosmic time.
    At their core, contemporary SAMs, such as GAEA \citep{delucia_2024b,fontanot_2025}, L\nobreakdash-Galaxies \citep{yates_2024}, GALFORM \citep{lacey_2016}, SHARK \citep{lagos_2018}, SAG \citep{cora_2018}, GABE \citep{jiang_2019}, the Santa Cruz SAM \citep{yung_2022,yung_2023}, and DarkSAGE \citep{stevens_2024}, are built upon merger trees derived from large and high-resolution dark-matter-only N-body simulations.
    Within this dark matter structure, baryonic physics is incorporated in a post-processing stage using a set of prescriptions, based on phenomenologically and/or theoretically motivated arguments, for a wide array of astrophysical processes.
    SAMs are designed to rapidly explore the parameter space associated with these baryonic processes, tracing the evolution of key galaxy properties and observables for large statistical samples at a small fraction of the computational cost of a hydrodynamical simulation.
    This makes it possible to run many model variations on identical merger trees, to extend predictions to still larger volumes, and to generate suites of realisations with which to quantify cosmic variance.

    In their seminal work, \citet{obreschkow_2009a} used empirical relations to partition, in post-processing, cold hydrogen as predicted by a then state-of-the-art SAM \citep[][a precursor of GAEA]{delucia_2007}.
    The approach allowed the construction of large mock catalogues including \HI{} masses and an explicit modelling of the 21\,cm line profiles.
    Later independent efforts led to an explicit and self-consistent inclusion of the partition of the cold gas in its molecular and atomic components in several SAMs \citep[e.g.][]{fu_2010,lagos_2011b,sommerville_2015b,xie_2017}, which also yielded advancements in prescriptions for star formation and associated processes.
    In particular, some of these models now implement star formation prescriptions based on the $\mathrm{H}_2$ surface density, which is a better tracer of star forming regions \citep{bigiel_2008}.

    A few studies have developed frameworks inspired by \citet{obreschkow_2009a, obreschkow_2009b} to simulate \HI{}-selected galaxy observations.
    For example, \citet{trujillo_2011} and \citet{desmond_2012} assigned modelled 21\,cm lines to galaxy catalogues built using a Halo Abundance Matching (HAM) technique.
    More recently, predictions based on the SHARK SAM, in combination with the SURFS simulations suite \citep{elahi_2018} and a modelling of the \HI{} line profiles, have been used to carry out comparisons with ALFALFA data \citep{chauhan_2019}, and to make forecast studies for the WALLABY survey \citep{koribalski_2020}.
    In \citet{brooks_2023}, the Sibelius-DARK N-body simulation \citep{mcalpine_2022} and GALFORM have been used to create a mock ALFALFA survey, study key observational effects on the \HI{} mass and velocity function estimates, and provide updated ALFALFA measurements correcting for these effects.
    
    In this work, we develop and test a framework to associate \HI{} emission lines with galaxies from two state-of-the-art semi-analytic models, GAEA \citep{delucia_2024b} and L\nobreakdash-Galaxies \citep{yates_2024}, evaluated in post-processing of the Millennium Simulation suite \citep{springel_2005,boylan-kolchin_2009}.
    With it, we aim to assemble the first essential building blocks towards a simulation pipeline that would bridge modern SAM outputs with predictions of cosmological observables, with a focus on peculiar velocities that can be inferred using the Tully-Fisher relation.
    
    Relative to the existing literature, our contribution thus distinguishes itself in its scope, focusing on testing whether the approach is sufficient to produce realistic simulated samples for these specific observables.
    We explicitly develop our framework such that it natively integrates with multiple distinct SAMs.
    We test it using two state-of-the-art ones, GAEA and L\nobreakdash-Galaxies, that are based on the \emph{same} underlying $N$-body merger trees.
    While predictions based on these models are novel, differences between the two also provide a characterisation of the uncertainties associated with specific physical modelling assumptions.
    We compare model results using the latest observational estimates, namely from the Cosmicflows-4 catalogue, the ALFALFA survey and the first data release of the ongoing FASHI survey.

    The paper is structured as follows:
    In section~\ref{sec:sims}, we summarise the semi-analytic models and implementations relevant for this study.
    In section~\ref{sec:HELM}, we describe our model for the \HI{} 21\,cm emission line.
    In section~\ref{sec:MockSky} we build mock \HI{}-selected galaxy catalogues and compare our predictions with empirical data, first at the SAM output level, and then also by forward-modelling observational effects.
    We discuss assumptions and the limitations they induce in section~\ref{sec:discussion}.
    Finally, we leverage our predictive framework to forecast the redshift distribution for potential SKAO \HI{}-selected galaxy surveys in section~\ref{sec:forecasts}.

    \section{Simulations}\label{sec:sims}
    \subsection{Dark Matter N-body Simulations}\label{sec:Nbody}
    
        In this work, SAMs are coupled with the reconstructed merger history of dark matter haloes, i.e. merger trees, extracted from two different underlying dark matter $N$-body simulations, namely those of the Millennium Simulation I \citep[MSI,][]{springel_2005}, and of the Millennium Simulation II \citep[MSII,][]{boylan-kolchin_2009}.
        These merger trees are structured as a series of `snapshots' from the simulation volume at discrete redshifts, where each snapshot is a catalogue of dark-matter haloes.
        The cosmological model adopted in the MSI and MSII is based on a combined analysis of the 2dFGRS \citep{colless_2001} and first-year WMAP data \citep{spergel_2003} with parameters $\Omega_{\rm b} = 0.0455$, $\Omega_{\rm m} = 0.25$, $\Omega_\Lambda = 0.75$, $h=0.73$, $\sigma_8 =0.9$, and $n_{\rm s}=1$.
        The MSI traces $2160^3 \simeq 10^{10}$ DM particles of mass $8.61 \cdot 10^8 \mathrm{M}_\odot~h^{-1}$ over redshifts $z \in [0, 127]$ in a cubic volume of $(500\, \mathrm{Mpc}~h^{-1})^3$ with periodic boundary conditions.
        For the MSII, the number of DM particles is identical but the box size is $(100\, \mathrm{Mpc}~h^{-1})^3$, which brings the particle mass resolution to $6.88 \cdot 10^6 \mathrm{M}_\odot~h^{-1}$.

    \subsection{Semi-analytic Models}\label{sec:SAM}

        In this study, we develop predictions based on two distinct semi-analytic models, GAEA and L\nobreakdash-Galaxies.
        The fiducial cosmological models used in both SAMs are also distinct.
        However, this has limited impact on the observables this study focuses on since the parameters of both SAMs are tuned based on similar empirical data, i.e. the differences in cosmology are absorbed in the tuning of the baryonic processes.
        SAM standard outputs are built on top of merger trees in the form of `snapshot' galaxy catalogues, i.e. galaxy properties are saved at the same discrete redshift intervals.
        
        \subsubsection{GAEA}\label{sec:GAEA}
        
        For the GAlaxy Evolution and Assembly (GAEA) semi-analytic model \citep{delucia_2014,hirschmann_2016,xie_2017,fontanot_2020,xie_2020,delucia_2024b,fontanot_2025}, we use the version presented in \cite{delucia_2024b} as reference.
        It reproduces several important observational constraints:
        (i) the evolution of the galaxy stellar mass function up to $z \sim 7$ and of the cosmic star formation rate density up to $z \sim 10$ \citep{fontanot_2017b};
        (ii) the measured correlation between stellar mass/luminosity and metal content of galaxies in the local Universe, down to the scale of Milky Way satellites \citep{delucia_2014,hirschmann_2016};
        (iii) the evolution of the galaxy mass--gas metallicity relation up to redshift $z \sim 3.5$ \citep{fontanot_2021};
        (iv) the measured \HI{} mass function and clustering of \HI{}-rich galaxies \citep{zoldan_2017,spinelli_2020,fontanot_2025};
        (v) the size–mass relation and the specific angular momentum–mass relation of late type and early type galaxies in the local Universe \citep{zoldan_2018}, as well as the evolution of the size–mass relation up to $z \sim 2$ \citep{zoldan_2019};
        (vi) the \HI{} mass -- stellar mass scaling relation for both central and satellite galaxies at $z=0$ \citep{xie_2020};
        (vii) the declining trend of \HI{} fractions of satellite galaxies in cluster haloes \citep{chen_2024};
        (viii) the emission line properties of both star-forming and AGN galaxies \citep{scharre_2024}.
        In this study, we use model outputs based on the MSI and MSII that, as discussed above, assume a cosmological model consistent with WMAP1. \footnote{In the recent work by \citet[][see also \citealt{cantarella_2025}]{fontanot_2025}, the GAEA SAM has also been run on merger trees from the P-Millennium simulation \citep{baugh_2019} that adopts cosmological parameters consistent with 2013 Planck results.}

        \subsubsection{L-Galaxies}\label{sec:L-Galaxies}
        
        Like GAEA, L\nobreakdash-Galaxies \citep{guo_2011,henriques_2015,henriques_2020,yates_2024} is built to run on the merger trees generated from the MSI and MSII N-body simulations.
        However, the corresponding merger trees have been re-scaled to match a \textit{Planck 2013} cosmology: $\Omega_{\Lambda}=0.685$, $\Omega_{\rm m}=0.315$, $\Omega_{\rm b}=0.0487$, $\sigma_8=0.826$, $n_{\rm s}=0.96$, $h=0.673$, using the technique described in \citet{angulo_2010} and \citet{angulo_2015}.

        The L\nobreakdash-Galaxies version adopted in this work \citep{yates_2024} includes the following advanced prescriptions: (i) an \HI{} -- H$_2$ partitioning model following the \citet{mckee_2010} metallicity- and density-dependent formalism \citep{fu_2010}; (ii) radially resolved gas and stellar discs which are self-consistently evolved across all time steps \citep{fu_2013}; (iii) a galaxy chemical enrichment (GCE) model accounting for binary stellar evolution and 118 individual chemical elements \citep{yates_2013,yates_2024}; and (iv) an explicit dust production and destruction model \citep{vijayan_2019,yates_2024}.

        L\nobreakdash-Galaxies' free parameters are constrained on the observed stellar mass function and quenched fraction at $z=0$ and $2$, as well as the \HI{} mass function at $z=0$, using a Markov Chain Monte Carlo formalism \citep{henriques_2020}. The model reproduces several other important observations, including (i) the \HI{} mass -- stellar mass scaling relation at $z=0$ \citep{henriques_2020}; (ii) the chemical composition of the stars, interstellar medium (ISM), and circumgalactic medium (CGM) in local galaxies, including radial profiles \citep{yates_2021a}; (iii) the cosmic metal and SFR density back to $z \sim 5$ \citep{yates_2021b}; and (iv) the evolution of the mass -- metallicity relation and dust-to-gas ratios back to $z \sim 4$ \citep{yates_2024}.

\section{Emission Line Model}\label{sec:HELM}

    Accurate modelling of \HI{} line profiles underpins Tully-Fisher distance estimates and their precision, and is therefore central to cosmological applications.
    In the following, we present the method we adopt, inspired by \citet{obreschkow_2009a} and \citet{chauhan_2019}, to associate a 21\,cm emission line to the galaxies in SAM output catalogues.
    We detail all specific assumptions in our implementation.

    \subsection{\texorpdfstring{Model 21\,cm flux}{Model 21cm flux}} \label{sec:model_flux}

        The rest-frame flux density profile of an \HI{} galaxy's 21\,cm emission line is constructed as the product of a distribution, the normalised line profile $\Psi_\HI{}(V)$, and a scale, the velocity-integrated 21\,cm flux $S_\mathrm{21\,cm}^V$.
        \begin{equation} \label{eq:fluxdens_model}
                s(V) = S_\mathrm{21\,cm}^V \cdot \Psi_\HI{}(V)
        \end{equation}
        
        The velocity-integrated flux, which fundamentally carries an observational dependence, is modelled as
        \begin{equation}
                S_\mathrm{21\,cm}^V = \frac{(1+z) L_\mathrm{21\,cm}^V}{4\pi D_{_L}^2(z)}
        \end{equation}
        where $D_{_L}(z)$ is the luminosity distance between the source and the observer, and $L_\mathrm{21\,cm}^V$ is the velocity-integrated rest-frame luminosity of the 21\,cm emission line.
        The latter, neglecting self-absorption or rather assuming the \HI{} distribution to be optically thin, can be derived from fundamental atomic physics considerations as being proportional to the number of neutral hydrogen atoms, hence to the \HI{} mass \citep[e.g.][]{meyer_2004,catinella_2010}.
        In the following we use
        \begin{equation}
            L_\mathrm{21\,cm}^V \ [\mathrm{Jy~km\,s^{-1}~Mpc^2}]= 5.319 \cdot 10^{-5} \left( M_\HI{}/\mathrm{M}_\odot\right).
        \end{equation}

        The normalised line profile $\Psi_\HI{}(V)$ encodes, simultaneously, information about both internal dynamics and the \HI{} content. Therefore, its modelling depends both on galaxy properties and structure, as we describe next.

    \subsection{\texorpdfstring{Model \HI{} galaxies}{Model HI galaxies}}\label{sec:model_gal}

        Dark matter N-body simulations and semi-analytic models provide a number of properties for the simulated host DM haloes and the associated galaxies.
        We use a set of these properties in our 21\,cm line modelling.
        
        Specifically, from N-body simulations we use:
        \begin{itemize}
            \item $R_\mathrm{vir}$ and $M_\mathrm{vir}$, the dark matter halo virial radius and mass.
            \item $V_\mathrm{max}$, the dark matter halo maximum circular velocity.
        \end{itemize}
        In the Millennium-based merger trees used by GAEA and L\nobreakdash-Galaxies, these are defined as the radius $R_\mathrm{vir} \equiv R_{200}$ of the largest sphere centred on the gravitational potential minimum which contains an over-density larger than 200 times the critical density, and the total mass $M_\mathrm{vir} \equiv M_{200}$ within this sphere.

        From the semi-analytic models, we use:
        \begin{itemize}
            \item $M_{\star,\mathrm{b}}$ and $r_{\star,\mathrm{b}}$, the stellar `bulge' mass and its radial scale length
            \item $M_{\star,\mathrm{d}}$ and $r_{\star,\mathrm{d}}$, the stellar disc mass and its radial scale length
            \item $M_g$ and $r_g$, the total cold gas disc mass and its radial scale length
            
            \item $M_\mathrm{H}$ and $R_\mathrm{mol}$ (or $f_\mathrm{mol}$), the total cold hydrogen mass and the ratio of molecular to atomic cold hydrogen (or the fraction of cold hydrogen in the molecular form).
        \end{itemize}
        Detailed definitions of these galaxy properties, specific to each SAM, can be found in \citet{delucia_2024b} for GAEA, and \citet{yates_2024} for L\nobreakdash-Galaxies.
        Coupling them with simple structural assumptions allows us to build models for the distribution of \HI{} within galaxies and associated dynamics, namely surface density profiles and circular velocity profiles, as developed next in sections~\ref{sec:surfacedensityprof} and \ref{sec:circularvelocityprof}.

    \subsubsection{Neutral hydrogen surface density profile}\label{sec:surfacedensityprof}
    
        GAEA assumes axially symmetric thin discs with exponential surface density profiles, both for the cold gas and the stellar disc components.
        \begin{equation} \label{eq:Sigma_exp}
            \Sigma_{\mathrm{d}}(r) = \frac{M_\mathrm{d}}{2\pi r_\mathrm{d}^2} \exp{(-\frac{r}{r_\mathrm{d}})}
        \end{equation}
        where $M_\mathrm{d}$ is the total mass contained in the disc, i.e. $M_g$ or $M_{\star,\mathrm{d}}$, and $r_\mathrm{d}$ is the corresponding scale length, i.e. $r_g$ or $r_{\star,\mathrm{d}}$.
        In L\nobreakdash-Galaxies, an exponential profile is also assumed for the infalling gas onto the disc, while the density profiles of the stellar and cold gas discs are not constrained by any functional form.
        They are divided up into concentric rings, and the profile evolves self-consistently due to radial gas flows, star formation, etc.

        The cold hydrogen mass $M_\mathrm{H}$, i.e. the dominant component of cold gas mass $M_\mathrm{g}$, is also estimated using different prescriptions in the two SAMs considered.
        In GAEA, it is computed assuming, at all cosmic epochs, the relative cosmic abundance of each element as predicted by big bang nucleosynthesis.
        Thus, hydrogen is assumed to account for a fixed 74~per~cent of cold gas ($M_\mathrm{H} \equiv 0.74\, M_\mathrm{g}$), with the remaining fraction assumed to be in the form of helium (He), dust and ionised gas.
        Cold hydrogen is assumed to be homogeneously distributed within the gaseous disc, hence also yielding an exponential surface density profile $\Sigma_\mathrm{H}(r)$ (Eq.~(\ref{eq:Sigma_exp})) with scale length $r_\mathrm{H} = r_g$.
        
        In L\nobreakdash-Galaxies, pristine gas in the intergalactic medium is also assumed to be 74~per~cent hydrogen, but the chemical composition of cold gas inside haloes and galaxies is then explicitly modelled and traced via the chemical evolution model \citep{yates_2013}, thus directly yielding $M_\mathrm{H}$ and $\Sigma_\mathrm{H}(r)$ for discrete ring bins.

        Finally, the cold hydrogen disc is partitioned in its atomic (\HI{}) and molecular (H$_2$) components.
        Recently published SAMs include different prescriptions for the cold gas partitioning and the evolution of the galaxies' radial profiles \citep[e.g.][]{fu_2010,fu_2013,lagos_2011b,sommerville_2015b,xie_2017}.
        The partition is typically expressed in terms of the ratio of molecular to atomic hydrogen surface densities $ R_\mathrm{mol}(r) = \Sigma_{\mathrm{H}_2}(r) / \Sigma_\HI{}(r)$, or equivalently in terms of the fraction of the hydrogen surface density in molecular form.
        \begin{equation}
            f_\mathrm{mol}(r) = \frac{R_\mathrm{mol}(r)}{\left( 1 + R_\mathrm{mol}(r) \right)} = \frac{\Sigma_{\mathrm{H_2}}(r)}{\Sigma_{\mathrm{H}}(r)}
        \end{equation}
        Thus yielding
        \begin{align}
            \Sigma_{\mathrm{H_2}}(r) &= f_\mathrm{mol}(r) \cdot \Sigma_{\mathrm{H}}(r) \\
            \Sigma_{\HI{}}(r) &= \left[ 1 - f_\mathrm{mol}(r) \right] \cdot \Sigma_{\mathrm{H}}(r) \label{eq:SigmaHI}
        \end{align}

        Multiple hydrogen partitioning prescriptions have been explored \citep[e.g.][]{blitz_2006,krumholz_2009,krumholz_2013,mckee_2010} and shown to produce equivalent predictions for the global properties and distributions of model galaxies, both within the GAEA and L\nobreakdash-Galaxies frameworks \citep{xie_2017,fu_2013}.
        In the following, we focus on the methods implemented in the reference SAMs used in this work.
        
        GAEA relies on the observed relation between $R_\mathrm{mol}(r)$ and the mid-plane pressure acting on the galactic disc $P_\mathrm{ext}(r)$, assuming it applies at all redshifts \citep{blitz_2006}.
        \begin{equation}
            R_\mathrm{mol}(r) = \left( \frac{P_\mathrm{ext}(r)}{P_0} \right)^\alpha
        \end{equation}
        where $P_0$ is the external pressure of molecular clumps, and
        \begin{equation}\label{eq:BRpressure}
            P_\mathrm{ext}(r) = \frac{\pi}{2} G \Sigma_\mathrm{g}(r) \left[ \Sigma_\mathrm{g}(r) + f_\sigma(r) \Sigma_{\star,\mathrm{d}}(r) \right]
        \end{equation}
        In the previous equation, $f_\sigma(r) = \sigma_\mathrm{g}(r) / \sigma_{\star,\mathrm{d}}(r)$ is the ratio of the velocity dispersion profiles in the cold gas and stellar discs.
        The cold gas velocity dispersion in the disc is assumed to be constant, such that $\sigma_\mathrm{g}(r) \equiv \sigma_\mathrm{g} = 10 \,\mathrm{km\, s^{-1}}$, while that of the stellar disc is modelled as $\sigma_{\star,\mathrm{d}}(r) = \sqrt{\pi G h_\star \Sigma_{\star,\mathrm{d}}(r)}$ with $h_\star = r_{\star,\mathrm{d}}/ 7.3$ \citep{xie_2017}.
        
        L\nobreakdash-Galaxies uses the modelling framework described in \citet{mckee_2010}. In this case, the molecular fraction depends on cold gas surface density and metallicity.
        \begin{equation}
            f_\mathrm{mol}(r) = \left\{ \begin{matrix} \frac{2[2-s]}{4+s} & s < 2 \\ 0 & s \geq 2 \end{matrix} \right.
            \ ; \quad
            s(r) = \frac{\ln\left(1+0.6\chi + 0.01\chi^2 \right)}{0.6\tau_c(r)}
        \end{equation}
        \begin{equation}
            \chi = \frac{3.1}{4.1}\left(1+3.1 Z'^{0.365} \right)
            \ ; \quad
            \tau_c(r) = 0.066 Z' \cdot (c_f\Sigma_g(r))
        \end{equation}
        where $Z'$ is the gas-phase metallicity in solar units and $c_f$ is the clumping factor, for which it is assumed \citep{henriques_2020}:
        \begin{equation}
            c_f = \left\{ \begin{matrix} 0.01^{-0.7} & Z' < 0.01 \\ Z'^{-0.7} & 0.01 \leq Z' < 1 \\ 1 & Z' \geq 1 \end{matrix} \right.
        \end{equation}
        
        Both GAEA and L\nobreakdash-Galaxies compute partitions and save the relevant information at each snapshot, for a predefined number of radial bins.
        The resulting total fraction $f_\mathrm{mol}$ (respectively ratio $R_\mathrm{mol}$), i.e. over the total disc mass, is also saved.

        \begin{figure}
            \centering
            \includegraphics[width=\linewidth]{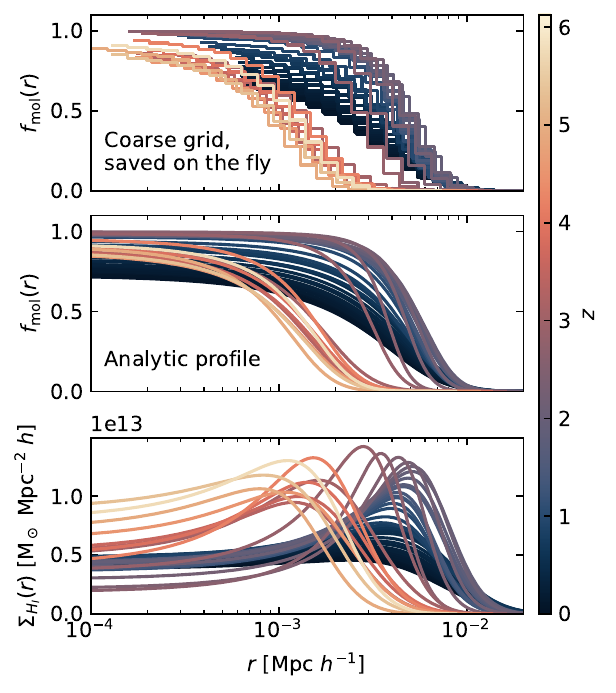}
            \caption{Example of the molecular hydrogen fraction profile (upper and central panels) and atomic hydrogen surface density profile (lower panel) for a GAEA model galaxy as a function of redshift. The graduated colour scale on the right indicates the redshift for each curve.
            \textit{Upper panel:} $f_\mathrm{mol}(r)$ as output by GAEA, where the hydrogen partition is computed at every time step and saved on the fly for a predefined number of radial bins.
            \textit{Central panel:} $f_\mathrm{mol}(r)$ re-evaluated on an arbitrarily smoother scale in post-processing with the \citet{blitz_2006} prescription.
            \textit{Lower panel:} $\Sigma_{\HI{}}(r)$, resulting from Eq.~(\ref{eq:SigmaHI}) with the analytic $f_\mathrm{mol}(r)$ from the central panel.}
            
            \label{fig:fmol}
        \end{figure}
    
        Crucially, to have control over the radial resolution of surface density profiles, hence ultimately over that of our emission line model, we explicitly implement both \citet{blitz_2006} and \citet{mckee_2010} prescriptions in post-processing of our reference SAMs outputs.
        Indeed, while the internal hydrogen partition schemes in both GAEA and L\nobreakdash-Galaxies are computed on coarse radial grids of the order of $\sim20$ bins, which is sufficient to derive star formation dynamics and associated processes, our specific objective, modelling the emission line, requires a finer resolution. 
        By mirroring the internal hydrogen partition schemes of both SAMs, our approach thus allows to re-evaluate the profile on an arbitrarily smoother scale while maintaining consistency with respect to the stellar disc and gas disc properties we use as inputs.
        In other words, by reproducing the partition schemes analytically, we are able to evaluate them with high resolution in post-processing.

        An example is presented in Figure~\ref{fig:fmol}, where the partition's redshift evolution for a simulated galaxy from GAEA's output is compared between the coarsely binned partition used internally and saved on the fly, and the corresponding analytic profile \citep{blitz_2006}, re-evaluated a posteriori on each snapshot.

    \subsubsection{Neutral hydrogen circular velocity profile}\label{sec:circularvelocityprof}

        To model circular velocity profiles traced by the \HI{} content, we assume the disc of model galaxies to be rotationally supported.
        Thus, the kinematics are closely tied to the gravitational potential in the disc plane.
        The potential is readily modelled from combining the mass distributions of the DM halo, stellar `bulge', stellar disc and gas disc.

        Here, we make the additional assumption that dark matter haloes and stellar bulges have spherically symmetric profiles.
        Since our primary aim is ultimately to predict statistics over observable properties of rotationally supported \HI{}-rich systems, this assumption implies that we are modelling the average radial distributions across galaxy populations.
        In this context, asymmetries introduce noise around the average (see section~\ref{sec:discussion} for a detailed discussion).
    
        In the Newtonian limit, the gravitational potential yields circular velocities
        \begin{equation} \label{eq:v_c}
            v_\mathrm{c}(r) = \sqrt{\frac{G M(<r)}{r}}
        \end{equation}
        where $G$ is the gravitational constant and $M(<r)$ the total mass inside a radius $r$ in galactocentric coordinates.
    
        For dark matter haloes, we follow the modelling of \cite{obreschkow_2009a} that we summarise below.
        We assume a Navarro-Frenk-White \citep[][NFW]{navarro_1996} spherically symmetric profile.
        \begin{equation}
            \rho_\mathrm{h}(r) = \rho_0 \left[ \frac{r}{r_s} (1+\frac{r}{r_s})^2 \right]^{-1}
        \end{equation}
        where $\rho_0$ is a normalisation factor and $r_s$ is the characteristic scale radius of the halo.
        We then obtain the total mass profile by integrating over the density inside a sphere of radius $r$.
        \begin{align}
            M_\mathrm{h}(<r) &= 4 \pi \int_0^r \mathrm{d}\tilde{r}\ \tilde{r}^2 \rho_\mathrm{h}(\tilde{r})  \\
            &= 4 \pi r_s^3\, \rho_0 \left[ \ln\left(1 + \frac{r}{r_s}\right) - \frac{r/r_s}{1 + r/r_s} \right]
        \end{align}
        The normalisation factor $\rho_0$ is expressed in terms of the virial mass $M_\mathrm{vir}$ and virial radius $R_\mathrm{vir}$ by using the fact that $M_\mathrm{vir} = M_\mathrm{h}(<R_\mathrm{vir})$.
        \begin{equation}
            \rho_0 = \frac{M_\mathrm{vir}}{4\pi r_s^3} \cdot \left[ \ln\left(1 + \frac{R_\mathrm{vir}}{r_s}\right) - \frac{R_\mathrm{vir}/r_s}{1 + R_\mathrm{vir}/r_s} \right]^{-1}
        \end{equation}

        The fraction $R_\mathrm{vir} / r_s$ parametrises the halo concentration $c_h$ that can be inferred from the maximum circular velocity of the halo $V_\mathrm{max}$ and the circular velocity at the virial radius $V_\mathrm{vir} = \left[ GM_\mathrm{vir}/R_\mathrm{vir} \right]^{1/2}$ \citep{prada_2012}. Indeed, for a NFW density profile, the following relation holds:
        \begin{equation}
            \frac{V_\mathrm{max}}{V_\mathrm{vir}} =\left[ \frac{0.216 c_\mathrm{h}}{\ln(1+c_\mathrm{h}) - c_\mathrm{h}/(1+c_\mathrm{h})} \right]^{\frac{1}{2}}
        \end{equation}
        which we can solve numerically for $c_\mathrm{h}$.

        We model the stellar `bulge' with a spherically symmetric Hernquist \citep{hernquist_1990} profile.
        \begin{equation}
            \rho_{\star,\mathrm{b}}(r) = \frac{M_{\star,\mathrm{b}}}{2\pi} \frac{r_{\star,\mathrm{b}} / r}{\left( r_{\star,\mathrm{b}} + r \right)^3}
        \end{equation}
        The total mass profile is again obtained by integrating over the density inside a sphere of radius $r$.
        \begin{align}
            M_{\star,\mathrm{b}}(<r) &= 4 \pi \int_0^r \mathrm{d}\tilde{r}\ \tilde{r}^2 \rho_{\star,\mathrm{b}}(\tilde{r})  \\
            &=  M_{\star,\mathrm{b}} \left[ \frac{r}{r_{\star,\mathrm{b}} + r} \right]^2 \\
            &=  M_{\star,\mathrm{b}} \left[ \frac{r}{r_{\star,\mathrm{b}}} \cdot \frac{1}{1 + r/r_{\star,\mathrm{b}}} \right]^2
        \end{align}
        
        Finally, for the stellar and cold gas discs, the gravitational potential in the equatorial plane of a disc with axially symmetric exponential surface density profile can be integrated analytically \citep{freeman_1970,binney_2008}, yielding circular velocity contributions of the following form.
        \begin{equation}
            v_{c, \mathrm{d}}^2(r) = \frac{2 G M_\mathrm{d}}{r_\mathrm{d}} y^2 \left[ I_0(y) K_0(y) - I_1(y) K_1(y) \right] ; \quad y = \frac{r}{2 r_\mathrm{d}}
        \end{equation}
        where $I_n$ and $K_n$ are the modified Bessel functions of the first and second kind, respectively.

        The total circular velocity profile is obtained by generalising Eq.~(\ref{eq:v_c}) as a sum in quadrature, i.e. summing the squared contribution of each component.
        \begin{equation}\label{eq:quadrule}
            v_\mathrm{c}^2(r) = \sum_k v_{\mathrm{c}, k}^2(r).
        \end{equation}

    \subsection{Model line profile}

        The normalised line profile $\Psi_\HI{}(V)$, as defined in section~\ref{sec:model_flux}, is modelled through integration of the \HI{} disc's projected velocity field, weighted by its \HI{} surface density, onto the line-of-sight of an observer.
        Thanks to symmetry assumptions, this reduces to a radial integration
        \begin{align} \label{eq:normalised_flux}
            \Psi_\HI{}(V) = \frac{\int_0^\infty \mathrm{d}r \ r \ \Sigma_\HI{}(r) \cdot \psi (V, r)}{\int_0^\infty \mathrm{d}r \ r \ \Sigma_\HI{}(r)}
        \end{align}
        where $\psi(V,r)$ is the line-of-sight velocity distribution (LOSVD) yielded by a thin ring portion of the disc at radius $r$, normalised such that $\int_V \psi(V,r) dV = 1\ \forall r$, and with the denominator ensuring the normalisation condition, $\int_V \mathrm{d}V\ \Psi_\HI{}(V) = 1$.
        
        The LOSVD, in the rest frame of the system, can be derived from geometrical considerations \citep[e.g.][]{obreschkow_2009a}.
        \begin{equation} \label{eq:psitilde}
            \tilde{\psi} (V,r) = \left\{ \begin{matrix}
                \frac{1}{\pi} \left[ (v_c(r)\sin{i})^2 - V^2 \right]^{-\frac{1}{2}} & \mathrm{if} |V| < v_c(r)\sin{i} \\
                0 & \mathrm{if} |V| \geq v_c(r)\sin{i}
            \end{matrix} \right.
        \end{equation}
        where $i$ is the inclination angle between the line-of-sight and the galactic plane’s rotational axis, i.e. $i=0$ when the galaxy is viewed `face-on' and $i=\pi/2$ when it is `edge-on'. In the following, for brevity, we do not write the inclination dependency explicitly.

        This construction generates singular artefacts in the limit $V \rightarrow v_c(r)\sin{i}$, which are resolved by taking into account the effect of velocity dispersion.
        This is done by convolving the LOSVD of each thin ring with a Gaussian kernel, preserving the normalisation.
        In line with the assumptions made for the hydrogen partition, i.e. in Eq.~(\ref{eq:BRpressure}), we assume a fixed dispersion $\sigma_g \equiv 10\,\mathrm{km\,s^{-1}}$ in the gas disc, leading to:
        \begin{equation} \label{eq:LOSVD}
            \psi (V,r) = \frac{1}{\sqrt{2\pi \sigma_g^2}} \int \mathrm{d}V' \exp \left[ \frac{(V-V')^2}{-2 \sigma_g^2} \right] \cdot \tilde{\psi} (V,r)
        \end{equation}

        Finally, the LOSVD of the disc can be combined by integrating over the infinitesimal radial contributions. The normalised \HI{} flux density is then obtained by weighting each thin ring by the surface density $\Sigma_\HI{}(r)$.
        
    \subsubsection{Emission line parametrisation}

        \begin{figure}
                \centering
                \includegraphics[width=\linewidth]{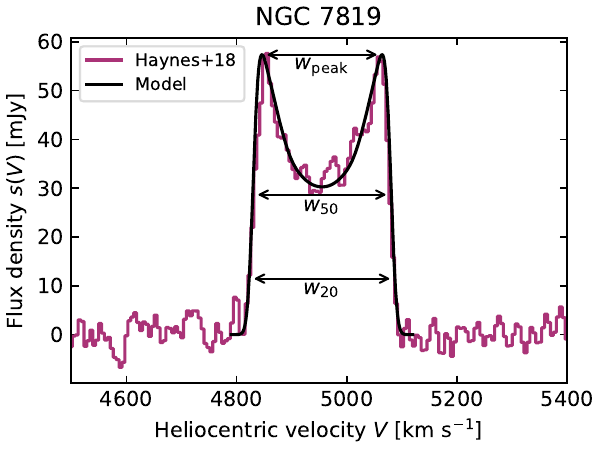}
                \caption{\HI{} 21\,cm line profile of galaxy NGC 7819 measured in the ALFALFA survey \citep[][magenta curve]{haynes_2018}, and a manual fit of our model as an example (black curve), displaying the model line width parameters, namely  $w_\mathrm{peak}$, $w_{50}$ and $w_{20}$.}
                \label{fig:alfalfa_line}
        \end{figure}

        In observational studies, the information contained in the shape of the normalised line profile $\Psi_\HI{}(V)$, and equivalently in that of the rest-frame flux density $s(V)$, is commonly parametrised by measurements of their widths, typically at the peak ($w_\mathrm{peak}$), at 50~per~cent of the peak ($w_{50}$), and at 20~per~cent of the peak ($w_{20}$).
        While observationally these measurements pose definition and robustness challenges, the rest-frame symmetry in our model simplifies the derivation by ensuring that these quantities are always well-defined.
        As an example, Figure~\ref{fig:alfalfa_line} shows the 21\,cm line profile of galaxy NGC 7819 as measured in the ALFALFA survey \citep{haynes_2018}, to which we fit our emission line model, Eq.~(\ref{eq:fluxdens_model}), by manually tuning the parameters, i.e. those from the set $\{M_\mathrm{vir}, M_\mathrm{\star,b}, M_\mathrm{\star,d}, M_\mathrm{g}, M_\mathrm{H}, R_\mathrm{vir}, r_\mathrm{\star,b}, r_\mathrm{\star,d}, r_\mathrm{g}, f_\mathrm{mol} \}$ described at the start of section~\ref{sec:model_gal}. We also show the line width parameters, namely $w_\mathrm{peak}$, $w_{50}$ and $w_{20}$.

    \subsection{Simulation and observational effects} \label{sec:systematics}

        To compare the predictions of our emission line model with observational data, we need to take into account various effects that arise from either modelling assumptions or observational conditions.
        We examine and implement models for these effects in a modular fashion.
        This allows to take a forward-modelling approach on our prediction framework, i.e. to predict data at low levels of processing, essentially including controlled effects in the simulation.
        In the broader picture, this kind of control is crucial for simulation pipelines used in precision cosmology.

        \subsubsection{Integration limits}

            As per Eq.~(\ref{eq:normalised_flux}), the model features a radial integration that, in practice, is computed over a finite range $[0, r_\mathrm{max}]$.
            We find that choosing $r_\mathrm{max} = 9\cdot r_g$, where $r_g$ is the gas disc scale length, is an appropriate limit beyond which the marginal contribution to the emission line profile becomes negligible.
            In the following, we adopt this setting as default.

        \subsubsection{Inclination}\label{sec:incl}

            \begin{figure}
                \centering
                \includegraphics[width=\linewidth]{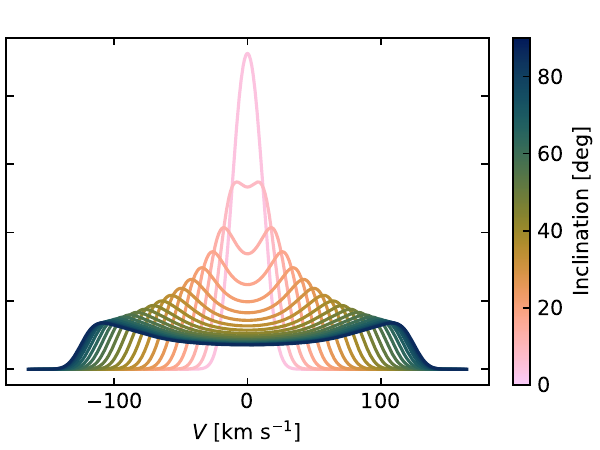}
                \caption{Example line-of-sight velocity distributions (LOSVD) of a modelled galaxy observed at different degrees of inclination. The graduated colour scale on the right indicates the inclination angle in degrees, with $0^\circ$ corresponding to the `face-on' case, and $90^\circ$ to the `edge-on' case.}
                \label{fig:incl_effect}
            \end{figure}

            In observations, galactic discs are viewed at an inclination angle $i$, which we take into account in our geometrical construction of the LOSVD.
            For galactic discs observed `edge-on', i.e. when $i=\pi/2$, the dominant contribution to the profile shape is the rotational motion, typically inducing a double-peak such as in Figure~\ref{fig:alfalfa_line}.
            In the case of a galaxy observed `face-on', i.e. when $i=0$, the contribution from the rotational motion is completely suppressed and the remaining emission line spread is purely due to velocity dispersion in the cold gas disc.
            Visually, the effect of inclination is a shrinking of the distance between the double peaks, hence a shrinking of the line width, and a boost to the line's peak(s).
            Figure~\ref{fig:incl_effect} shows this effect on the LOSVD of an example galaxy when varying only the inclination.
            Since the line-of-sight projection of the rotation velocity scales with $\sin(i)$ (Eq.~(\ref{eq:psitilde})), observational studies sometimes apply a $1/\sin{i}$ correction to their line width measurements.
            In our framework, we account for the presence of this correction in data that we compare our predictions to by simply setting $i=\pi/2$, i.e. comparing the `edge-on' profile.
            Conversely, when modelling the effect, we retrieve for each galaxy the spin orientation of the cold gas disc, a quantity provided by both SAMs, and compute its corresponding inclination angle with respect to the observer.

        \subsubsection{Redshift}\label{sec:red}

            The effect of redshift on emission line profiles can conceptually be divided into two parts.
            First, it offsets the rest-frame emission line centre according to the systemic relative motion with respect to the observer.
            Second, it also has a broadening effect on line profiles since it is, by construction, linearly dependent on frequency, hence the lower frequency tail of the profile is shifted more than the higher frequency tail.
            This broadening effect is sometimes corrected for in observational surveys \citep[e.g. in][]{kourkchi_2020,kourkchi_2022} by dividing observed line widths by a factor $(1+z)$.
            As our model directly predicts line profiles in the observed galaxy's rest frame, we can readily include the redshift broadening effect by applying the inverse correction, namely multiplying rest-frame line widths by a factor $(1+z)$, when simulating uncorrected data.
        
\section{\texorpdfstring{Simulated \HI{} galaxy catalogues}{Simulated HI galaxy catalogues}}\label{sec:MockSky}

    Before the framework can serve for cosmological forecasts, its predictions must be validated against existing data.
    To this end, we construct mock observational catalogues of \HI{} galaxies and compare them with available survey measurements.

    \subsection{Snapshot catalogues}\label{sec:box}

        Snapshot galaxy catalogues output by SAMs provide idealised representations of the simulated system, in the sense that they are observer-independent.
        To compare predictions from these catalogues with observational estimates, the latter are typically corrected for observational effects.
        In the following, we define distinct catalogues, each based on $z=0$ snapshots for combinations of underlying $N$-body simulation, i.e. MSI or MSII, and SAM, i.e. GAEA or L\nobreakdash-Galaxies.

        Any simulation becomes incomplete below a mass scale set by its finite resolution.
        A first limit is set at the halo level by the particle mass: haloes and sub-haloes are identified with at least 20 particles \citep{springel_2005, boylan-kolchin_2009}, corresponding to minimum masses of $1.7 \cdot10^{10}\ \mathrm{M}_\odot~h^{-1}$ in the MSI and $1.4 \cdot10^8\ \mathrm{M}_\odot~h^{-1}$ in the MSII.
        How this limit propagates to galaxy properties depends on the physical model, and is most directly assessed by running the same model on a higher-resolution simulation, a role that the MSII plays for the MSI.
        For our purpose, we want to study simulated samples that are close to complete.
        We limit the low-mass range of our snapshot catalogues with cuts in stellar and \HI{} mass.
        These cuts are catalogue thresholds rather than resolution limits.
        They are deliberately placed below the latter, so that resolution effects remain identifiable in the following figures.
        For catalogues based on the MSI, we set $M_\star ,\, M_\HI{} > 8.61 \cdot10^7\ \mathrm{M}_\odot~h^{-1}$.
        For those based on the MSII, we set $M_\star ,\, M_\HI{} > 6.88 \cdot10^5\ \mathrm{M}_\odot~h^{-1}$.
    
        \subsubsection{Mass functions}\label{sec:boxMF}

            \begin{figure}
                \centering
                \includegraphics[width=\linewidth]{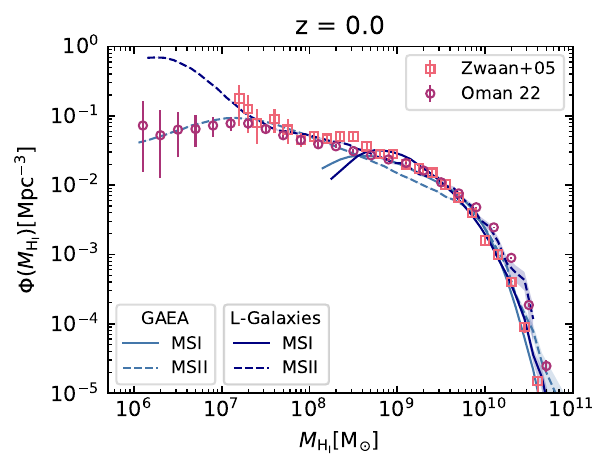}
                \caption{\HI{} mass functions on $z=0$ snapshot outputs. Teal blue lines correspond to GAEA outputs, and dark blue lines to L-Galaxies outputs. Results are displayed for outputs from the MSI (solid lines), and MSII (dashed lines). These are compared to observational data derived from the HIPASS \citep{zwaan_2005} and ALFALFA \citep[][reanalysis of the original \citealt{jones_2018} measurements]{oman_2022} surveys.}
                \label{fig:BOX_MF}
            \end{figure}

            The local ($z \approx 0$) \HI{} mass function (\HI{}~MF) is used to calibrate both\footnote{Observational data from the HIPASS survey \citep{zwaan_2005} are used as a reference in both cases.} GAEA and L\nobreakdash-Galaxies.
            Figure~\ref{fig:BOX_MF} shows how the snapshot catalogues, by construction, reproduce the \citet{zwaan_2005} data to within $\sim$0.2 dex over $M_\HI{} \sim 10^{7.5}$\nobreakdash-$10^{10.5}~\mathrm{M}_\odot$.
            It also compares them with the more recent reanalysis of ALFALFA by \citet{oman_2022}, where a similar level of agreement is observed over $M_\HI{} \sim 10^{7.5}$\nobreakdash-$10^{10}~\mathrm{M}_\odot$.
            In the high-\HI{}-mass tail ($M_\HI{}\gtrsim10^{10}~\mathrm{M}_\odot$), divergences of $\sim$0.5 dex between the ALFALFA and HIPASS \HI{} MF estimates are observed.
            Additionally, these observational estimates are obtained through volume/density estimation techniques that factor-in survey (in)completeness effects, whose associated uncertainties become larger at the lower mass end \citep[e.g.][]{oman_2022}.
            
            The resolution limit for the MSI samples ($M_\HI{} \sim 10^{8.5}~\mathrm{M}_\odot$) can be inferred by comparing model predictions based on MSI with those based on MSII.
            To evaluate a similar limit for the MSII simulation, one would need to run the same models on another simulation with higher resolution.
            We note that at $M_\HI{}\lesssim10^{7.5}~\mathrm{M}_\odot$, the two MSII predictions diverge, with L\nobreakdash-Galaxies predicting a substantially higher number density of low-$M_\HI{}$ galaxies than GAEA.
            We emphasize that this range lies below the calibration target of both SAMs, where observational estimates themselves start diverging and are likely significantly affected by incompleteness.
            Model predictions in this regime are therefore largely unconstrained.

        \subsubsection{Velocity/Width functions}\label{sec:boxWF}

            \begin{figure}
                \centering
                \includegraphics[width=\linewidth]{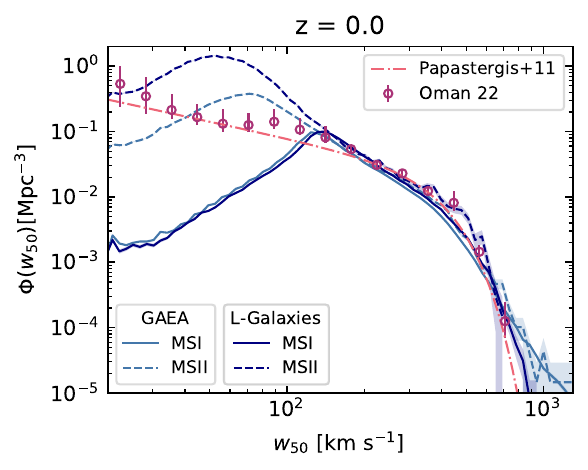}
                \caption{\HI{} velocity width ($w_{50}$) functions on $z=0$ snapshot outputs. Teal blue lines correspond to GAEA outputs, and dark blue lines to L-Galaxies outputs. Results are displayed for outputs from the MSI (solid lines), and MSII (dashed lines). These are compared to observational data derived from the 40 per cent and 100 per cent releases of the ALFALFA survey \citep{papastergis_2011,oman_2022}.}
                \label{fig:BOX_WF}
            \end{figure}

            Using our emission line model, we investigate the `width functions' (WF) obtained for each snapshot catalogue, i.e. the volume-normalised distribution of emission line widths.
            This quantity essentially probes an observable equivalent to the velocity function (VF), using 21\,cm line widths as tracers for circular velocities instead of e.g. $v_\mathrm{max}$.
            We note that while the \HI{} mass function considered in the previous section was used to fine-tune SAM parameters, the width function considered here is a true prediction of our simulations.
            Figure~\ref{fig:BOX_WF} shows the $w_{50}$-functions derived from GAEA and L\nobreakdash-Galaxies snapshots at $z=0$ compared to those derived from ALFALFA 40 per cent \citep{papastergis_2011} and ALFALFA 100 per cent data \citep[reanalysis by][]{oman_2022}.
            The observational estimates shown in this figure are corrected neither for inclination nor for redshift effects.
            For proper comparison, we thus model these effects on our predictions, as described in sections~\ref{sec:incl} and \ref{sec:red}.
            
            In the $w_{50} \sim 125$\nobreakdash-$350~\mathrm{km\,s^{-1}}$ regime, the four SAM curves overlap with the ALFALFA-based estimates of \citet{papastergis_2011} and \citet{oman_2022} to within $\sim$0.2 dex, and the GAEA and L\nobreakdash-Galaxies predictions are very close to each other.
            Outside of this regime, discrepancies between SAMs and larger uncertainties in the data are visible.
            The MSI-based width functions peak at $w_{50} \sim 10^{2.1}~\mathrm{km\,s^{-1}}\approx125~\mathrm{km\,s^{-1}}$ and then decline towards lower velocity widths, detaching clearly from the MSII curves, which extend the `plateau' power-law portion of the WF.
            This occurs at the same width for GAEA and L\nobreakdash-Galaxies and is driven by the halo mass resolution limit of the MSI.
            Below this limit, the MSII predictions from GAEA and L\nobreakdash-Galaxies snapshot catalogues differ significantly.
            The MSII curves both exhibit a `turnover' at $\sim50~\mathrm{km\,s^{-1}}$ (L\nobreakdash-Galaxies) and $\sim 70~\mathrm{km\,s^{-1}}$ (GAEA) which are unlikely to be due to halo mass resolution effects.
            Indeed, the MSII resolves haloes down to $V_\mathrm{max}\sim10~\mathrm{km\,s^{-1}}$, so that particle resolution is expected to affect the width function only around $w_{50} \sim20~\mathrm{km\,s^{-1}}$.
            Instead, these `turnovers' reflect which haloes host \HI{} content above our catalogue threshold ($M_\HI{} > 6.88 \cdot10^5\ \mathrm{M}_\odot~h^{-1}$).
            Their location and scale thus depends on the physical model and its effective `resolution' in \HI{}-mass.
            In this regime, the L\nobreakdash-Galaxies WF remains above the GAEA one, which is consistent with the \HI{}-mass function, and the difference between the two reflects the modelling uncertainty.
            
            The predicted slopes of the width function's power law `plateau' are seemingly steeper than those of the observational estimates.
            This mismatch with observations is a long-known issue, already reported in \citet{papastergis_2011}, which may have several origins.
            First, the density estimators used in observational studies cannot correct for regions of the $M_\HI{}$--$w_{50}$ plane in which a survey has no detections, an effect to which the WF is more sensitive than the \HI{} MF.
            \citet{oman_2022} estimates that this may make the measured low-velocity slope too shallow by $\sim 0.2$.
            This is however small compared with the difference between the MSII-based predictions and the data below $\sim100~\mathrm{km\,s^{-1}}$, and cannot account for it on its own.
            In their study based on the SHARK SAM, \citet{chauhan_2019} showed that this effect can be addressed in part by comparing models and data at the catalogue level.
            Second, in the low-width regime, \HI{} 21\,cm line profiles gradually become poor tracers of the velocity profile.
            Comparing predictions based on EAGLE hydrodynamical simulations to a simplified symmetric assumption, \citet{oman_2024} notably showed that, below $\sim 100~\mathrm{km\,s^{-1}}$, both measurements and detections of \HI{} 21\,cm line widths are biased by effects that suppress the correlation between line profiles and the underlying velocity profile, particularly near detection limits.
            In this regime, in particular at the snapshot level, predictions of our framework are thus limited by construction.
            We also note that these snapshot-level comparisons concern the whole galaxy population at a given width, whereas the galaxies entering an \HI{}-selected survey are biased towards \HI{}-rich systems. We return to this point in section \ref{sec:LC_MF_WF}.
            
        \subsubsection{Scaling relations}\label{sec:boxTF}

            \begin{figure}
                \centering
                \includegraphics[width=\linewidth]{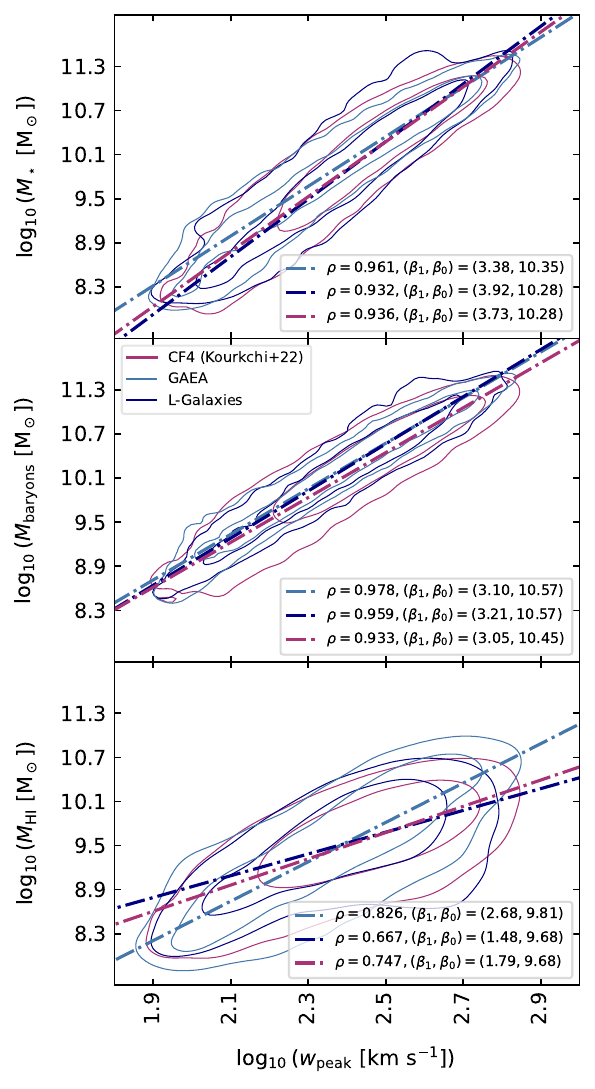}
                \caption{Generalised Tully-Fisher relations between `edge-on' peak line widths ($w_\mathrm{peak}$) and stellar, baryonic, and \HI{} masses on $z=0$ snapshot outputs.
                Contours show the 68~per~cent and 95~per~cent confidence levels of sample distributions for each simulation, namely the GAEA MSI (teal blue), and L-Galaxies MSI (dark blue).
                Magenta contours present distributions derived from the Cosmicflows-4 baryonic Tully-Fisher catalogue \citep{kourkchi_2022}.
                For each sample, the Pearson correlation coefficient $\rho$ is measured and a power law of the form $\log{M} = \beta_0 + \beta_1(\log{w} -2.5)$ is fitted.}
                
                \label{fig:BOX_TF}
            \end{figure}

            \begin{figure}
                \centering
                \includegraphics[width=\linewidth]{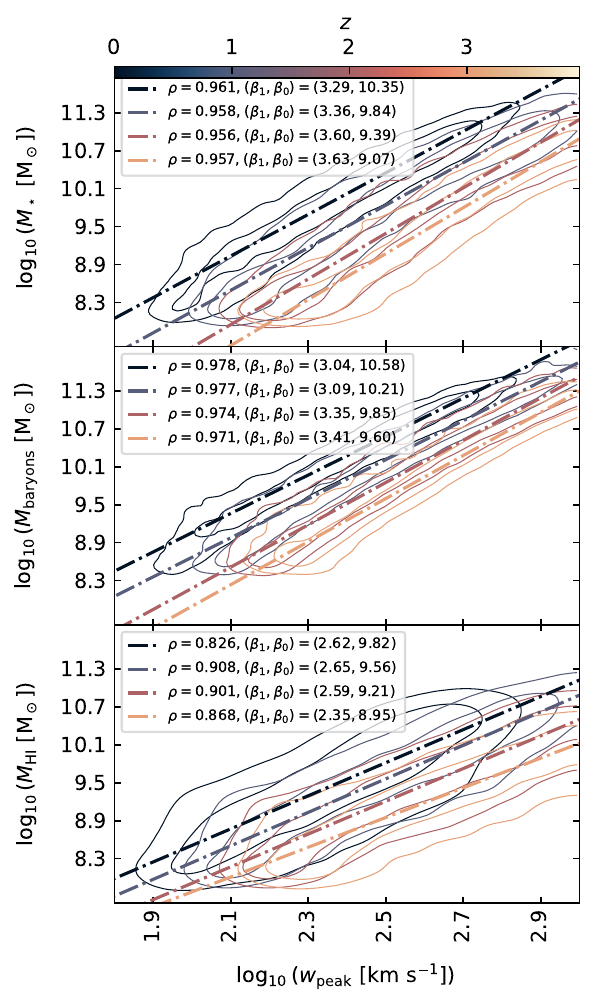}
                \caption{Generalised Tully-Fisher relations between `edge-on' peak line widths ($w_\mathrm{peak}$) and stellar, baryonic, and \HI{} masses on GAEA MSI snapshot outputs at redshifts $0 < z <  3$.
                The graduated colour gradient on top indicates the redshift for each snapshot sample.
                Contours show the 68~per~cent and 95~per~cent confidence levels of sample distributions.
                For each sample, the Pearson correlation coefficient $\rho$ is measured and a power law of the form $\log{M} = \beta_0 + \beta_1(\log{w} -2.5)$ is fitted.}
                \label{fig:BOX_TFz}
            \end{figure}

            Both the \HI{} mass functions and the width functions may be thought of as proportional to marginals of higher dimensional distributions of correlated galaxy properties.
            For example, up to a volume normalisation, the generalised Tully-Fisher relation between $M_\HI{}$ and $w_{50}$ represents a 2D distribution of which the \HI{} mass function and \HI{} 21\,cm line width function are marginal distributions.
            
            Figure~\ref{fig:BOX_TF} shows generalised Tully-Fisher scaling relations between stellar, baryonic, and \HI{} masses and `edge-on' peak line widths, obtained using `equi-representative' subsets of snapshot catalogues at $z=0$.
            The `equi-representative' subsets, representing $\sim10^4$ simulated galaxies each, are constructed by thinly binning the data in \HI{} mass, and randomly selecting the same number of samples from each bin.
            The contours represent the 68~per~cent and 95~per~cent confidence levels of the distributions.
            The MSI subset catalogues for GAEA and L\nobreakdash-Galaxies are compared to observational data from the Cosmicflows-4 (hereafter CF4) bTF catalogue \citep{kourkchi_2022}.
            The CF4 catalogue is the result of a broad combination of observations from multiple instruments at various frequencies, where the data are corrected for inclination, redshift and instrumental effects. Their line width parametrisation $W_\mathrm{mx}$ \citep[see][]{courtois_2009,courtois_2011} is defined as `statistically descriptive of the peak-to-peak maximum rotation velocity of a galaxy', i.e. is a proxy of $w_\mathrm{peak}$.

            Vertical and horizontal offsets between GAEA, L\nobreakdash-Galaxies and CF4 contours reflect differences observed in the 1D marginal distributions, the origin of which are discussed in sections~\ref{sec:boxMF} and \ref{sec:boxWF}.
            The additional information contained in the 2D distributions is the correlation between properties of the galaxies and their 21\,cm emission lines, i.e. the predicted slope and scatter of the (generalised) Tully-Fisher scaling relations.
            To quantify these, we fit power laws of the form $\log{M} = \beta_0 + \beta_1(\log{w} -2.5)$ and estimate, in log-space, the Pearson correlation coefficient $\rho$ for each sample.
            Using `equi-representative' subsets has the advantage of probing the relations uniformly over the mass ranges, i.e. removing the `window' effect of the mass function on the full distribution.
            
            Comparing the two predicted samples, the fact that slopes deviate in opposite directions in the $M_\star$\nobreakdash-$w_\mathrm{peak}$ and $M_\HI{}$\nobreakdash-$w_\mathrm{peak}$ planes, and agree more closely in the $M_\mathrm{baryons}$\nobreakdash-$w_\mathrm{peak}$ plane, is consistent with a difference in the cold hydrogen partition prescription.
            Indeed, the \citet{mckee_2010} method used in L\nobreakdash-Galaxies likely predicts more $\mathrm{H_2}$, fuelling star formation, and less \HI{} in massive galaxies than \citet{blitz_2006} used in GAEA, while the baryonic mass is not affected as it comprises both the full hydrogen and stellar masses.
            A similar argument holds for scatter predictions, which are closer to one another in the $M_\mathrm{baryons}$\nobreakdash-$w_\mathrm{peak}$ plane.
            Overall, divergences between the two simulated samples highlight the sensitivity of the predicted relations to the underlying \HI{} modelling, and thus the uncertainty associated with predictions of our framework at the snapshot level.

            With respect to CF4 data, the important modelling uncertainty limits the interpretation.
            However, in the $M_\star$\nobreakdash-$w_\mathrm{peak}$ and $M_\HI{}$\nobreakdash-$w_\mathrm{peak}$ planes, the CF4 slope and scatter values are within the interval spanned by predictions, hence the approach can in principle generate samples consistent with data.
            In the $M_\mathrm{baryons}$\nobreakdash-$w_\mathrm{peak}$ plane, the simulated samples predict the slope and correlation values observed in the CF4 sample within a $\sim5$ per cent margin.
            The residual discrepancies which constitute this error margin might originate from the nature of the comparison.
            Indeed, the CF4 catalogue is composed of heterogeneous data from instruments with different sensitivities, at different frequency ranges, i.e. combining radio with optical/infrared, needed in particular to estimate baryonic masses and correct for effects such as inclinations.
            For example, in \citet{kourkchi_2022}, estimates of the bTF slope vary by about $\sim3$ per cent between measurements in the optical and infrared bands.
            Comparisons presented in Figure~\ref{fig:BOX_TF} are thus simultaneously limited by the accuracy of these estimates/corrections in the CF4 data, and by the approximate characterisation of the slope and scatter we use in this work.
            Distinguishing the origin of the discrepancies between a genuine limitation of the model and one of the data comparison motivates a forward modelling approach to test the predictions of our framework at the catalogue level.

            Figure~\ref{fig:BOX_TFz} presents the predicted redshift evolution of the generalised Tully-Fisher scaling relations between stellar, baryonic, and \HI{} masses and `edge-on' peak line widths, obtained using our equi-representative subsets at redshifts $0 < z < 3$. The 68~per~cent and 95~per~cent confidence level contours of the  distributions are displayed for GAEA outputs based on the MSI.
            Our simulations predict that the principal effect of redshift evolution is an offset of the relations towards higher widths, while the slopes and scatters remain relatively stable.
            \citet{obreschkow_2009a} argued that this is a consequence of the `mass-radius-velocity relation' for dark matter haloes in the Millennium simulations \citep{croton_2006,mo_1998}, which assumes, for a fixed halo mass, a scaling of the circular velocity $v_c$ asymptotically proportional to $\sqrt{1+z}$.
            Whether this predicted evolution is fully physical or partly an artefact of the simulation assumptions thus remains to be tested.

    \subsection{Light-cone catalogues}\label{sec:LCcats}

            From GAEA MSI snapshot outputs we construct mock survey catalogues, projecting and redshifting model galaxies on light-cones, using the framework presented in \citet{blaizot_2005} and \citet{zoldan_2017}.
            In particular, we define two geometries: the first one with $0 < z < 0.06$, which we use hereafter to simulate existing surveys, and the second one with $0 < z < 0.4$, which we use in section~\ref{sec:forecasts} to forecast a SKAO \HI{} galaxy survey.
            The corresponding solid angles covered by the simulated light-cones are constrained by the size of the MSI box.
            We obtain $\Omega \simeq 5230~\deg^2$ for our first light-cone, and $\Omega \simeq 500~\deg^2$ for our second one.

        \subsubsection{Selection function}\label{sec:flux_cut}

            To forward-model selection effects, we define masks for sensitivity limits and apply them to the light-cone catalogues.
            To avoid the significant artificial detection bias which may affect symmetric line profile models at low inclination \citep[][]{oman_2024}, we apply the sensitivity limit before the inclination correction.
            To test our approach with respect to data from the ALFALFA survey \citep{haynes_2018}, we adopt their reported $5\sigma$ survey sensitivity in terms of velocity-integrated flux: $0.72~\mathrm{Jy~km\,s^{-1}}$ for $w_{50} = 200~\mathrm{km\,s^{-1}}$, which we translate into a $720~\mathrm{mJy\,km\,s^{-1}}/200~\mathrm{km\,s^{-1}} =3.6~\mathrm{mJy}$ threshold on peak flux density.
            This approximation is reasonable as the mode of the observed distribution of line widths in the ALFALFA sample \citep{haynes_2018} is around $w_{50}\simeq 200~\mathrm{km\,s^{-1}}$.
            Moreover, by comparing redshift distributions (see Figure~\ref{fig:dNdz} hereafter), we find that the first data release from the FASHI survey \citep{zhang_2024} provides a distinct catalogue with comparable completeness in the overlapping redshift range.
            Comparing simultaneously to both observational data samples allows us to keep under control the uncertainties associated with these measurements, as differences between them highlight how sensitive each statistic is to the details of the selection.
            However the samples therefore do not share the same effective flux limit, which should be kept in mind when comparing their distributions near the detection thresholds.

        \subsubsection{Redshift distribution}\label{sec:dNdz}

            \begin{figure*}
                \centering
                \includegraphics[width=\linewidth]{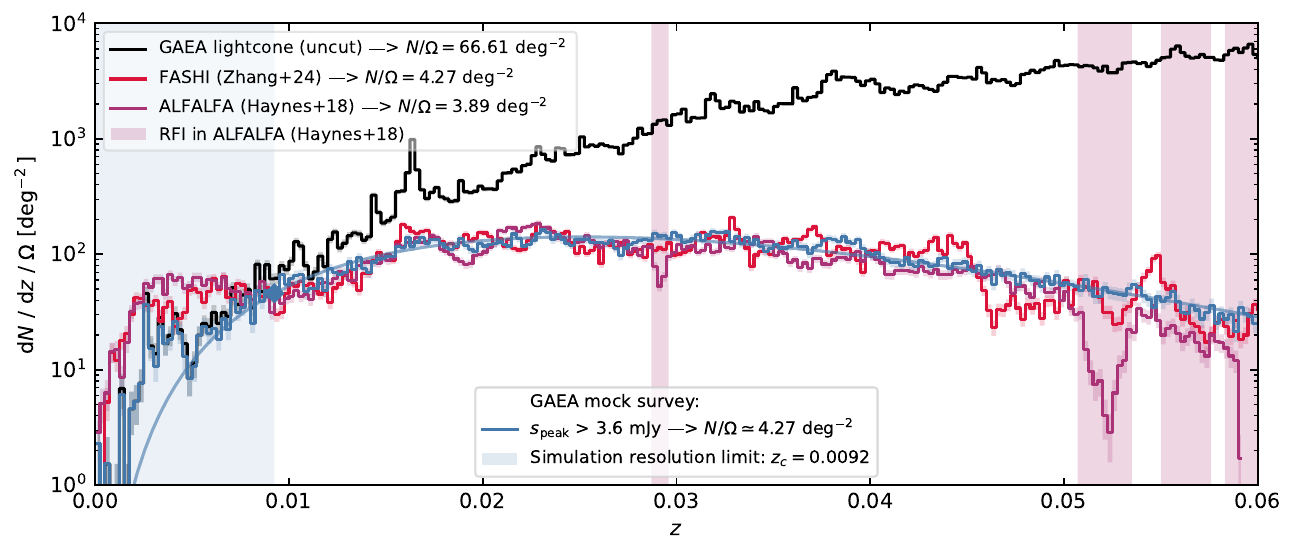}
                \caption{Redshift distribution of \HI{} galaxy samples, normalised by survey area $\Omega$.
                The black histogram shows the uncut GAEA light-cone, i.e. before applying any selection.
                The blue histogram shows the sample obtained by applying a peak flux density limit $s_\mathrm{peak} > 3.6~\mathrm{mJy}$.
                The blue curve shows the best fit estimate to the mock survey sample.
                The blue shaded region highlights the redshift range for which the simulation is expected to under-predict the number of detections due to its finite resolution. The higher limit $z_c$ is defined as the redshift of the first bin, passing through from high to low redshift, in which the fraction of sources with $M_\HI{} < 10^8~\mathrm{M_\odot}$ exceeds $1$ per cent.
                The magenta histogram shows data from the ALFALFA survey \citep{haynes_2018}.
                Magenta shaded regions highlight redshift ranges affected by RFI in the ALFALFA survey.
                The crimson histogram shows data from the first (partial) data release of the FASHI survey \citep{zhang_2024}, in which RFI bands are not identified.
                For all samples, we derive the average number of galaxies detected, in the range $0.01<z<0.05$, per square degree of survey sky.}
                \label{fig:dNdz}
            \end{figure*}

            Figure~\ref{fig:dNdz} shows the redshift distribution of \HI{} galaxies for our mock survey catalogue compared to data from the ALFALFA \citep{haynes_2018} and FASHI surveys \citep{zhang_2024}.
            It also highlights ranges affected by radio frequency interference (RFI) for ALFALFA as reported in \citet{haynes_2018}. The equivalent has not been reported for FASHI.

            The $\mathrm{d}N/\mathrm{d}z$ distribution, normalised by the solid angle $\Omega$ spanned by the survey area, is well described by an exponentially damped power law parametrisation \citep[e.g.][]{obreschkow_2009b}.
            \begin{equation} \label{eq:model_dndz}
                \frac{\mathrm{d}N}{\mathrm{d}z} / \Omega \simeq 10^{c_1} z^{c_2} \exp{(-c_3 z)}
            \end{equation}
            The best fit of this parametrisation for the GAEA mock survey catalogue is displayed in Figure~\ref{fig:dNdz}, with parameters: $c_1 = 8.37$, $c_2 = 3.07$, $c_3 = 121.77$.
            The resulting curve can be integrated over the redshift range to estimate the average number of galaxies detected, normalised by the survey sky area $\Omega$.
            \begin{align}
                N(z < z_\mathrm{max}) / \Omega &= \int_0^{z_\mathrm{max}} \left[10^{c_1} z^{c_2} \exp(- z c_3)\right] \mathrm{d}z\\
                &= \frac{10^{c_1}}{c_3^{(c_2 + 1)}} \gamma(c_2 + 1, {z_\mathrm{max}} c_3) \label{eq:N(z)}
            \end{align}
            where $\gamma(s,x)$ is the lower incomplete gamma function.
            This method is employed to derive an estimate for the GAEA mock survey sample.
            The estimate corresponds within a 10 per cent margin to the value measured for the ALFALFA sample, while coincidentally matching very closely the one for the FASHI sample.

            Results presented in Figure~\ref{fig:dNdz} are limited in several aspects.
            First, the mock survey light-cone suffers from the same simulation resolution limits discussed for the snapshot catalogues.
            The deficiency at low \HI{} masses translates into an under-prediction of the number counts below $z\lesssim0.01$, where real observational sensitivity allows the detection of galaxies with lower \HI{} mass.
            To characterise this limit, we define the limit $z_c$ as the redshift of the first bin, passing through from high to low redshift, in which the fraction of sources with $M_\HI{} < 10^8~\mathrm{M_\odot}$ exceeds $1$ per cent.
            This estimate corresponds consistently to the limit of the regime in which our mock survey under-predicts the number counts with respect to data.
            In the regime below $z_c$, the predicted $\mathrm{d}N/\mathrm{d}z$ may be interpreted as a lower bound.
            This is also a regime where the sampled volume is small and thus the number counts are more sensitive to cosmic variance.
            Second, above $z \sim 0.05$, ALFALFA RFI flags explain the large discrepancies observed between predictions and data.
            Although similar flags are not included in the FASHI first data release \citep{zhang_2024}, localised large discrepancies in the same regime are likely of similar origin.
            Third, in the intermediate redshift range $0.01 < z < 0.05$, the scatter of the simulated redshift distribution around the best-fit curve is smaller than in both ALFALFA and FASHI.
            This highlights a limitation of our selection function forward-model, which does not capture the full scatter contributions from observational effects, nor, as will be discussed in detail in section~\ref{sec:discussion}, from asymmetries in line profiles.
        
        \subsubsection{Mass and width distributions on the light-cone} \label{sec:LC_MF_WF}

            \begin{figure}
                \centering
                \includegraphics[width=\linewidth]{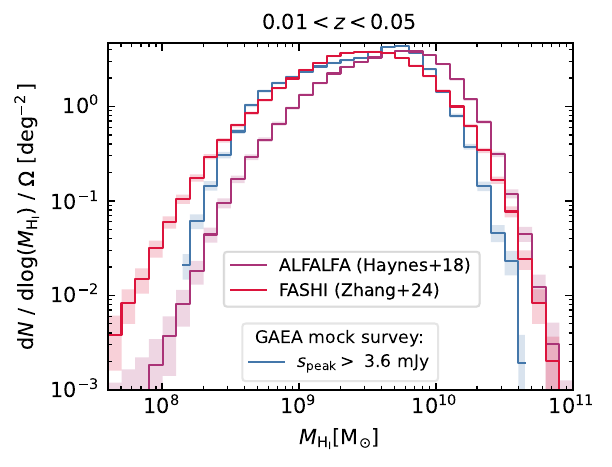}
                \caption{Observable \HI{} mass distributions on the light-cone, normalised by survey area.
                The blue histogram corresponds to the GAEA mock survey catalogue obtained with a peak flux density cut.
                The magenta histogram shows the \HI{} mass distribution of the ALFALFA catalogue \citep{haynes_2018}.
                The crimson histogram shows the \HI{} mass distribution of the FASHI first (partial) data release catalogue \citep{zhang_2024}.}
                \label{fig:LC_MF}
            \end{figure}

            \begin{figure}
                \centering
                \includegraphics[width=\linewidth]{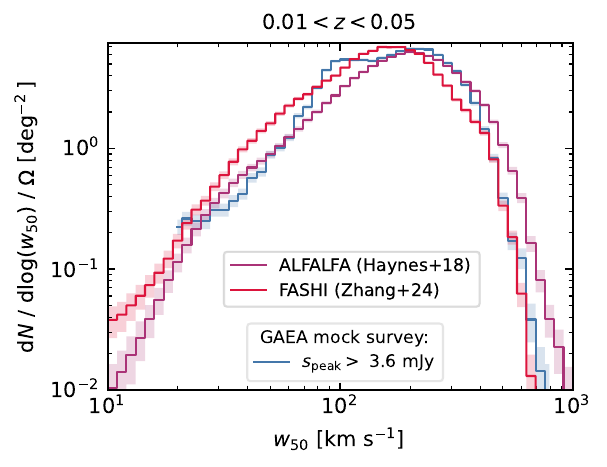}
                \caption{Observable width ($w_{50}$) distributions on the light-cone, normalised by survey area.
                The blue histogram corresponds to the GAEA mock survey catalogue obtained with a peak flux density cut.
                The magenta histogram shows the width ($w_{50}$) distribution of the ALFALFA catalogue \citep{haynes_2018}.
                The crimson histogram shows the width ($w_{50}$) distribution of the FASHI first (partial) data release catalogue \citep{zhang_2024}.}
                \label{fig:LC_WF}
            \end{figure}

            In contrast with results obtained in sections~\ref{sec:boxMF} and \ref{sec:boxWF} that rely on corrections to observational estimates, we leverage our forward-modelled selection function to analyse predictions closer to the direct observables, comparing the \HI{} mass and line width distributions at the catalogue level.
            To regularise comparisons, we restrict the redshift range to $0.01 < z < 0.05$, and normalise number counts by the survey area, i.e. the solid angle $\Omega$ described by the light-cone's sky coverage.
            For ALFALFA data, we use $\Omega = 7000~\deg^2$.
            For FASHI data, we use $\Omega = 7600~\deg^2$.
            Figures~\ref{fig:LC_MF} and \ref{fig:LC_WF} show the resulting mass and width distributions for our mock survey catalogue, compared to those from the ALFALFA and FASHI surveys.
            Since the catalogued line widths are not corrected for velocity dispersion, inclination, or redshift broadening\footnote{The ALFALFA line widths are corrected for instrumental broadening only \citep{haynes_2018}.}, we include these effects in the forward-model, as described in sections~\ref{sec:incl}, \ref{sec:red} and \ref{sec:flux_cut}.
            
            The predicted distributions display dynamic ranges and peak locations that are broadly consistent with the ALFALFA and FASHI data, within $\sim$0.2 dex in $M_\HI{}$ and $\sim$0.2 dex in $w_{50}$.
            However, important discrepancies are also evident.

            In comparison with \HI{} mass functions at the snapshot level (Figure~\ref{fig:BOX_MF}), \HI{} mass distributions at the catalogue level (Figure~\ref{fig:LC_MF}) capture similar high-mass deficiencies due to the calibration of GAEA on HIPASS \citep{zwaan_2005} data.
            Observed number counts in the low \HI{}-mass range vary by factors of order $\sim$1 dex between ALFALFA and FASHI data.
            This is likely an effect of differences in selection functions between the two samples near detection thresholds to which the \HI{} mass distribution is highly sensitive in this regime.
            Discrepancies between predictions and either data samples in the low-mass regime highlight the limitations of our simple selection function forward-model, where obtaining closer agreement with a particular sample may require characterising and tuning complex effects with survey-specific constraints.

            Width distributions at the catalogue level (Figure~\ref{fig:LC_WF}) display seemingly lower discrepancies between ALFALFA and FASHI samples, of order $\sim$0.5 dex, but across the whole width range.
            Below $w_{50} \sim 10^{2.1}~\mathrm{km\,s^{-1}} \simeq 125~\mathrm{km\,s^{-1}}$, the predicted distribution is in closer agreement with data than the width function at the snapshot level (Figure~\ref{fig:BOX_WF}).
            However, this should not be read as a resolution of the snapshot-level discrepancies.
            \HI{}-selected catalogues are biased towards \HI{}-massive galaxies: as shown in Figure~\ref{fig:LC_MF}, the bulk of the detections in the range $0.01 < z < 0.05$ has $M_\HI{} \sim 10^9$ to $10^{10}~\mathrm{M_\odot}$.
            This is the population on which both SAMs are calibrated and which is well resolved in the MSI.
            The low-\HI{}-mass galaxies that dominate the volume-weighted width function at low $w_{50}$, and that are missing from the MSI at the snapshot level, contribute little to such samples, in which narrow profiles can also arise from more massive galaxies observed at low inclination.
            The two comparisons thus probe different galaxy populations.
            This is not a limitation for the applications targeted in this work, since Tully-Fisher samples are drawn from the same \HI{}-rich population.
            Residual discrepancies between predicted and observed sample distributions in this regime can be attributed to simulation resolution limits and limitations of our selection function forward model in this regime.
            Indeed, using the SHARK SAM on the SURFS N-body simulation suite, which features smaller volumes ($40^3$ and $210^3$~$\mathrm{Mpc}^3\,h^{-3}$) but better particle mass resolutions ($4.13 \times10^7$ and $2.21\times10^8 ~\mathrm{M_\odot}\,h^{-1}$) than the MSI, coupled to an empirical model of the ALFALFA selection function, \citet{chauhan_2019} obtained qualitatively similar simulated 21\,cm line width distributions but with higher accuracy in the low widths regime, notably showing how simulation resolution plays an important role even at the catalogue level.
        
        \subsubsection{Scaling relations on the light-cone}

            \begin{figure}
                \centering
                \includegraphics[width=\linewidth]{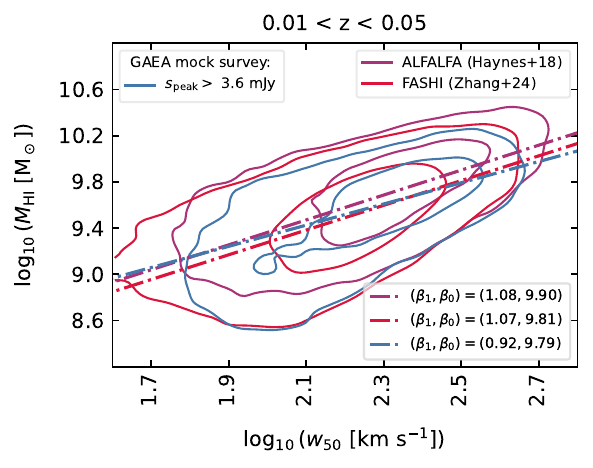}
                \caption{Scaling relation between galaxy line widths ($w_{50}$) and galaxy \HI{} masses. Contours show the 68~per~cent and 95~per~cent confidence levels of sample distributions. Blue contours correspond to the GAEA mock survey sample obtained with a peak flux density cut. Magenta and crimson contours correspond to the raw distributions from the ALFALFA \citep{haynes_2018} and FASHI \citep{zhang_2024} catalogues respectively.
                For each sample, a power law of the form $\log{M} = \beta_0 + \beta_1(\log{w} -2.5)$ is fitted.}
                \label{fig:LC_TF}
            \end{figure}
        
            Similarly to the argument developed in section~\ref{sec:boxTF}, the \HI{} mass and width distributions may also be thought of as marginals of higher dimensional distributions.
            Leveraging our forward-model of the selection function, velocity dispersion, inclination and redshift effects, we compare predictions for the generalised Tully-Fisher scaling relation between line widths ($w_{50}$) and \HI{} masses with observational data from the ALFALFA \citep{haynes_2018} and FASHI \citep{zhang_2024} surveys at the catalogue level.
            Figure~\ref{fig:LC_TF} shows the 2D contours obtained, as well as fitted power laws of the form $\log{M} = \beta_0 + \beta_1(\log{w} -2.5)$ for each sample.
            Horizontal and vertical offsets between the contours correspond to discrepancies observed in the marginal distributions, i.e. Figures~\ref{fig:LC_MF} and \ref{fig:LC_WF}, and the additional information lies in the slope and scatter of the scaling relation.
            
            With respect to data the predicted slope is off by around $15$ per cent, showing how, compared to the GAEA MSI snapshot level predictions in Figure~\ref{fig:BOX_TF}, building mock survey light-cones and forward-modelling the selection function allows to test the model predictions against data in a more controlled manner.
            The orientation of this slope discrepancy is also consistent with the fact that the predicted width distribution (Figure~\ref{fig:LC_WF}) agrees more closely with the ALFALFA sample in the low-width range, and with the FASHI sample in the high-width range.

            Scatters display a $w_{50}$-dependence which renders the use of a simple correlation estimator, like what was done in Figure~\ref{fig:BOX_TF}, uninformative.
            However, in the range above $w_{50} \sim 10^{2.1}~\mathrm{km\,s^{-1}} \simeq 125~\mathrm{km\,s^{-1}}$, discrepancies between observations are of the same order as those with respect to the predicted sample.
            
            In this regime and for these observables, the modelling approach developed in our framework may thus well be sufficient to generate mock \HI{}-selected samples comparable, at the catalogue level, to observations.
            
\section{Modelling limitations}\label{sec:discussion}

    Our framework rests on several central assumptions, out of which we need to distinguish the ones inherited from the underlying N-body simulations, SAMs and the additional ones in our line profile modelling.

    In both GAEA and L\nobreakdash-Galaxies, the cold gas and part of the stellar content of galaxies are assumed to settle in rotationally supported discs with axis-symmetric surface density profiles.
    Since we explicitly reproduce both SAMs' hydrogen partition prescriptions, these are also inherited as assumptions.
    Moreover, in the pressure-based prescription used in GAEA, the cold gas dispersion is fixed to a constant $\sigma_\mathrm{g} = 10~\mathrm{km\ s^{-1}}$ \citep{xie_2017}, consistent with typical values found in (limited) observed samples \citep[e.g.][]{leroy_2008}.
    For consistency, this inherited assumption is reused in Eq.~(\ref{eq:LOSVD}) where the cold gas dispersion has a smoothing effect on the LOSVD.
    The consequence is a fixed lower bound for the line width.

    To model circular velocity profiles, we take into account contributions from only a few dominant components, namely dark matter haloes, stellar bulges, stellar discs and gas discs, thus assuming that other potential components are negligible.
    For dark matter haloes, we assume NFW density profiles.
    While the choice of symmetric profiles pairs naturally with the disc components, NFW profiles are parametrised by halo concentrations.
    To determine the latter, we assume the $V_\mathrm{max}/V_\mathrm{vir}$ method of \citet{prada_2012}.
    Compared to an alternative approach that could consist e.g. in using a reference power law fitted to the median concentration as a function of halo mass \citep[e.g.][]{neto_2007}, this has the advantage of capturing the intrinsic scatter in the relation.
    For stellar bulges we assume Hernquist density profiles, similar to the modelling of e.g. \citet{lacey_2016} and \citet{brooks_2023}.
    While a simplification with respect to `de Vaucouleurs' profiles, which best describe observed bulges, these have very similar shapes and the advantage of being analytically integrable.
    
    As a consequence of these assumptions, our model only produces symmetric 21\,cm line profiles, which is an important simplification.
    Real observed galactic discs can feature arms, bars, warps, turbulence, non-circular motion, etc. which break the symmetry.
    Moreover, a number of observational effects can introduce asymmetries in line-profile measurements, such as instrumental noise, foregrounds, confusion and radio-frequency interference.

    Asymmetries in the shape of line profiles, irrespective of their origin, may be thought of as `noise' perturbations around a symmetric `signal'.
    The effect of this `noise' propagates on line parameters (e.g. width) as uncertainties, i.e. contributing to scatter in their distributions.
    In observational studies, asymmetries in line-profile measurements are typically handled with definitions for parameter estimators designed for statistical robustness against them \citep[e.g.][]{courtois_2009,courtois_2011}.
    The symmetric prediction from our framework may thus be interpreted as precisely the true value these aim to estimate.

    Selection effects, on the other hand, also depend on the shapes of 21\,cm line profiles.
    Asymmetry-induced scatter in survey completeness adds on top of the intrinsic one captured by symmetric profiles.
    The effect is strongest near the detection threshold, where the `noise' realisation that perturbs the profile shape may be the dominant determinant of whether a source enters the survey catalogue or not.
    To address this limitation, observational studies, in particular Tully-Fisher oriented ones \citep[e.g.][]{kourkchi_2022}, typically define conservative sample-selection cuts, for example in inclination, asymmetry/quality, or line width (lower bound).

    In our framework, the symmetry assumption inherited from SAMs implies that these effects cannot easily be included.
    On the other hand, our primary focus is on modelling (generalised) Tully-Fisher observables that are typically limited to systems where the 21\,cm line profile traces well the underlying velocity profile.
    In principle, similar (extended) frameworks can be built on hydrodynamical simulations \citep[e.g.][]{oman_2024}, in which asymmetries arise naturally.
    Recent runs reach volumes and resolutions comparable to those of the simulations used here, so that mock surveys of the kind presented in this work are within their reach.
    The features under discussion are however likely to depend strongly on the particular physical prescriptions adopted, for star formation and stellar feedback in particular.
    Quantifying this dependence requires, in principle, rerunning the galaxy formation model under several alternative assumptions, which remains far more expensive for a hydrodynamical simulation than for a SAM at comparable volume.

    Observational studies typically characterise survey completeness empirically \citep[e.g.][]{papastergis_2011}, which has the advantage of indiscriminately taking into account all observational effects.
    In this work, to build a forward-model agnostic to specific survey configurations, which we can test on existing data and extrapolate for instruments with a priori unknown sensitivities, we assume and assess a simple peak flux density threshold as a basis.
    Indeed, predicting statistical properties of the 21\,cm line within the regime where it traces well the underlying velocity profile, i.e. well above the detection threshold, only requires, in principle, an approximate model of the selection function.
    To avoid unrealistically boosting the peak(s) of line profiles above the detection threshold when including the effect of inclination in the forward-model \citep[an effect described in][]{oman_2024} we assume a counter-intuitive order, namely applying the sensitivity selection first.
    For an isotropic distribution of galactic disc orientations, the distribution of their inclinations, $i$, with respect to the line of sight has a probability density function proportional to $\sin(i)$, meaning that this approximate `correction' is small for the bulk of the distribution, and is important only for a small fraction of the sample, precisely in the limit where the symmetry in our model would induce an artificial detection bias.
    
\section{\texorpdfstring{Forecast for a SKAO \HI{} galaxy survey}{Forecast for a SKAO HI galaxy survey}} \label{sec:forecasts}

\begin{figure*}
            \centering
            \includegraphics[width=\linewidth]{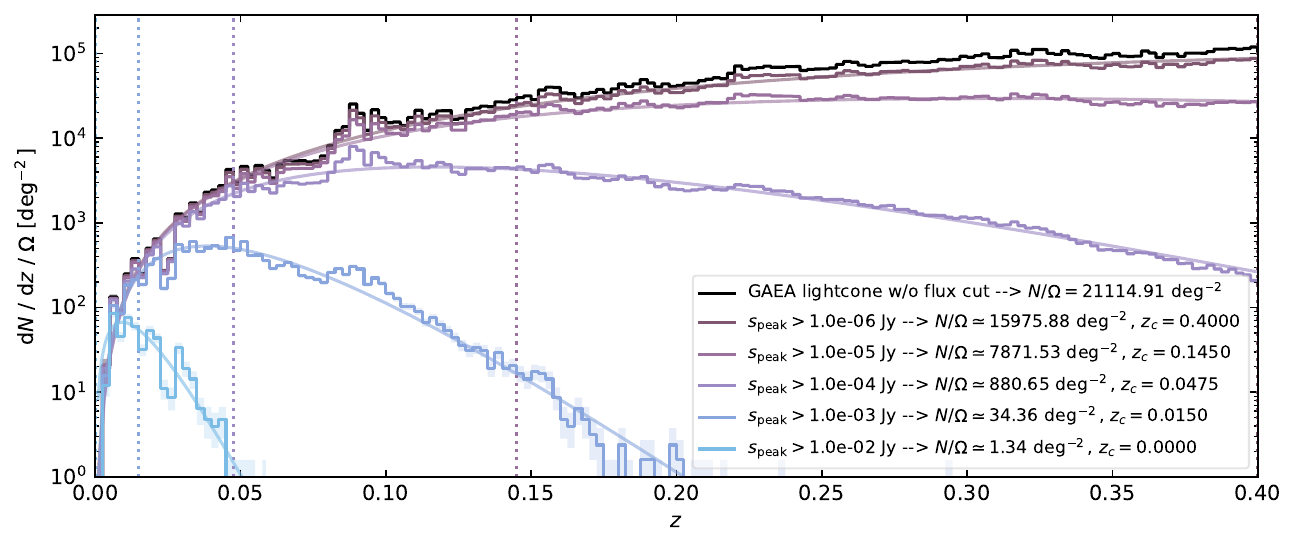}
            \caption{Redshift distribution of \HI{} galaxies for GAEA mock SKAO observations, with a range of peak flux density thresholds covering orders of magnitudes in spectral sensitivities from $10~\mathrm{mJy}$ to $1~\mu\mathrm{Jy}$. The black histogram shows the redshift distribution of the uncut GAEA projected light-cone, i.e. before applying any flux cut. Coloured histograms show the sub-samples obtained by applying the respective peak flux density detection limit to the mock catalogue.
            Dotted vertical lines of the corresponding colour mark, for each sub-sample, the redshift limit $z_c$ under which the simulation is expected to under-predict the number of detections due to its finite resolution, defined as the redshift of the first bin, passing through from high to low redshift, in which the fraction of sources with $M_\HI{} < 10^8~\mathrm{M_\odot}$ exceeds $1$ per cent.
            For all samples, we derive the average number of galaxies detected, in the range $0<z<0.4$, per square degree of survey sky.}
            \label{fig:dNdz_SKA}
        \end{figure*}

    Bearing in mind the limitations of our modelling approach, we investigate forecasts for future SKAO \HI{}-selected galaxy surveys with our second GAEA light-cone catalogue.
    To quantify the projected survey catalogue size as a function of sensitivity, we compute the $\mathrm{d}N/\mathrm{d}z$ functions for a logarithmically spaced range of peak flux density limits, between $1~\mu\mathrm{Jy}$ and $10~\mathrm{mJy}$.
    Figure~\ref{fig:dNdz_SKA} shows the results obtained, where, for all samples, we derive the corresponding average number of galaxies, in the range $0<z<0.4$, per square degree of survey sky.
    For the raw light-cone, i.e. before applying selection cuts, this is obtained by dividing the total number of galaxies in the catalogue by the survey area.
    For mock \HI{} surveys, we fit the parametric model (Eq.~(\ref{eq:model_dndz})) and find best fit parameters, which we report in Table~\ref{tab:bestfits_dNdz_SKA}.
    The fitted curves are integrated over redshift (Eq.~(\ref{eq:N(z)})) to estimate the number counts per square degree reported in the figure.
    Highlighted with dotted vertical lines, the simulation resolution limits $z_c$ associated with each flux density threshold, computed as in Figure~\ref{fig:dNdz}, give an overview of the current limitation of our approach in predicting redshift distributions for surveys of high sensitivity.
        
        \begin{table}
        \begin{center}
        \caption{Best-fit parameters for model $\mathrm{d}N/\mathrm{d}z/\Omega$ (Eq.~(\ref{eq:model_dndz})) on mock SKAO \HI{}-selected galaxy surveys ($0 < z <0.4$) for a range of spectral sensitivities.}
        \label{tab:bestfits_dNdz_SKA}
        \begin{tabular}{|llll|}
            \hline
            & $\mathbf{c_1}$ & $\mathbf{c_2}$ & $\mathbf{c_3}$ \\
            \hline
            $s_\mathrm{peak} > 10^{-6} \mathrm{Jy}$ : & $6.35$ & $2.11$ & $3.23$ \\
            $s_\mathrm{peak} > 10^{-5} \mathrm{Jy}$ : & $6.60$ & $2.25$ & $7.30$ \\
            $s_\mathrm{peak} > 10^{-4} \mathrm{Jy}$ : & $6.91$ & $2.38$ & $20.41$ \\
            $s_\mathrm{peak} > 10^{-3} \mathrm{Jy}$ : & $7.18$ & $2.41$ & $62.62$ \\
            $s_\mathrm{peak} > 10^{-2} \mathrm{Jy}$ : & $5.64$ & $1.55$ & $165.71$ \\
            \hline
        
        \end{tabular}
        \end{center}
        \end{table}

\section{Conclusions}
    We have presented a framework that associates \HI{} 21\,cm emission-line profiles with galaxies in the outputs of semi-analytic models, and forward-models the observational effects needed to compare its predictions with current \HI{}-selected galaxy surveys at the catalogue level. Our main results are the following.

    \begin{enumerate}
        \item The modelling uncertainties associated with semi-analytic hydrogen partition prescriptions can be controlled by comparing results based on distinct SAMs.
        In particular, comparing results for GAEA and L\nobreakdash-Galaxies, we find that these uncertainties become important in the low end of the \HI{} 21\,cm line width function at $z=0$, below $w_{50} \sim 125~\mathrm{km\,s^{-1}}$.
        In this regime, MSI-based predictions are limited by halo mass resolution, while MSII-based predictions depend on the effective \HI{}-mass resolution of the SAMs.
        The uncertainties on slopes and scatters of generalised Tully-Fisher relations are, consistently, important on the \HI{} component and limited on the baryonic relation.

        \item Comparisons at the snapshot level with the Cosmicflows-4 catalogue highlight that the adopted modelling approach may, in principle, produce samples with generalised Tully-Fisher relations consistent with observational estimates.
        Indeed, the two SAM predictions deviate from data in opposite directions in the stellar- and \HI-mass relations, and reproduce the baryonic relation's observed slope and correlation coefficient to within $\sim$5 per cent.

        \item Within the resolution limits of the underlying simulations, our framework can produce mock \HI-selected samples whose statistical properties such as redshift distributions, \HI{}\nobreakdash-mass distributions, and 21\,cm line width distributions, agree with current observations at a level comparable to the differences between independent surveys.
        In particular, despite limitations due to symmetry assumptions, the presented SAM-based modelling approach may be sufficient to simulate samples in the regimes where 21\,cm emission line profiles trace well the galactic discs' velocity profiles, i.e. the regimes relevant for Tully-Fisher studies.
    
        \item Light-cones extending to $z\sim0.4$ provide forecasts of the redshift distribution of \HI-selected galaxies for potential SKAO surveys.
        At high survey sensitivity, the effective \HI{}-mass resolution needed becomes an important limitation for such forecasts.
    \end{enumerate}
    
    Towards further development of our framework, results in the present work are informative on where effort could be invested.
    The limiting factor depends on the observable considered.
    Number-count forecasts at SKAO sensitivities are primarily limited by the effective \HI{}-mass resolution of the underlying simulation.
    Predictions involving \HI{} masses are limited by the modelling of the \HI{} component within the SAMs, and catalogue-level distributions near the detection threshold by the modelling of line profiles and selection effects.
    For cosmological applications, a natural extension would be to perform Tully-Fisher analyses on mock catalogues, deriving distances and peculiar velocities with the same estimators and corrections as applied to real data, and comparing them with the known input.
    Such analyses would propagate modelling uncertainties identified in the present work at the cosmological observable level, where they would be needed to characterise accurately the constraining power of these observables on cosmological models.

    Results presented in this work do not point to the abstractions inherent to our SAM-based approach as a limiting factor in the regime most relevant for Tully-Fisher analyses.
    The features that our framework does not describe mainly add scatter to line parameters and around detection thresholds, i.e. matter most near detection limits and at low widths, where line profiles become poorer tracers of the rotation curve \citep{oman_2024}.
    In this regime, hydrodynamical simulations can play a valuable role.
    In the spirit of \citet{oman_2024}, one could compare our predictions with line profiles extracted directly from their gas component.
    This would quantify what our framework misses and provide the statistics needed to calibrate a potential stochastic layer that could be added to the selection function forward-model.
    
    Hydrodynamical simulations are, however, not a straightforward substitute for SAMs as a basis for survey mocks.
    State-of-the-art runs now combine volumes and resolutions comparable to those used in this work, in some cases with an explicit treatment of the cold interstellar medium \citep[e.g.][]{schaye_2026}, and the mocks presented here could in principle be built on them.
    Their cost per realisation nevertheless limits the number of model variations that can be explored.
    The dependence of predictions on sub-grid prescriptions for star formation and feedback is also harder to isolate from resolution effects on the structure and kinematics of the cold gas than in SAMs, where alternative models can be run on identical merger trees as in this work.
    SAMs can in addition be run at low cost on still larger volumes, or on suites of realisations to quantify cosmic variance.
    We therefore regard the two approaches as complementary.
    SAM-based mocks provide an affordable and controlled exploration of physical uncertainties, and scale readily in volume and in number of realisations, while hydrodynamical simulations can inform the small-scale effects that SAMs do not describe.

    SKAO \HI{} galaxy surveys are expected to lift the limitations in sample size and redshift coverage that have, historically, restricted Tully-Fisher-based cosmology.
    Simulation pipelines of the kind motivated in this work will then be essential to exploit these surveys as cosmological probes.

\section*{Acknowledgements}

    This work was supported in part by grant CRSII5\_193826 from the Swiss National Science Foundation, by the Swiss State Secretariat for Education, Research and Innovation (SERI) as part of the SKACH consortium, and by the French government through the France 2030 investment plan managed by the National Research Agency (ANR), as part of the Initiative of Excellence Universit\'{e} C\^{o}te d’Azur under reference number ANR-15-IDEX-01.

\section*{Data and Code Availability}

    This work uses previously published data as referenced and described in the text.
    Details of the analysis and code can be made available upon reasonable request to the corresponding author.


\bibliographystyle{mnras}
\bibliography{references}

@ARTICLE{almeida_2023,
       author = {{Almeida}, Andr{\'e}s and {Anderson}, Scott F. and {Argudo-Fern{\'a}ndez}, Maria and {Badenes}, Carles and {Barger}, Kat and {Barrera-Ballesteros}, Jorge K. and {Bender}, Chad F. and {Benitez}, Erika and {Besser}, Felipe and {Bird}, Jonathan C. and {Bizyaev}, Dmitry and {Blanton}, Michael R. and {Bochanski}, John and {Bovy}, Jo and {Brandt}, William Nielsen and {Brownstein}, Joel R. and {Buchner}, Johannes and {Bulbul}, Esra and {Burchett}, Joseph N. and {Cano D{\'\i}az}, Mariana and {Carlberg}, Joleen K. and {Casey}, Andrew R. and {Chandra}, Vedant and {Cherinka}, Brian and {Chiappini}, Cristina and {Coker}, Abigail A. and {Comparat}, Johan and {Conroy}, Charlie and {Contardo}, Gabriella and {Cortes}, Arlin and {Covey}, Kevin and {Crane}, Jeffrey D. and {Cunha}, Katia and {Dabbieri}, Collin and {Davidson}, James W. and {Davis}, Megan C. and {de Andrade Queiroz}, Anna Barbara and {De Lee}, Nathan and {M{\'e}ndez Delgado}, Jos{\'e} Eduardo and {Demasi}, Sebastian and {Di Mille}, Francesco and {Donor}, John and {Dow}, Peter and {Dwelly}, Tom and {Eracleous}, Mike and {Eriksen}, Jamey and {Fan}, Xiaohui and {Farr}, Emily and {Frederick}, Sara and {Fries}, Logan and {Frinchaboy}, Peter and {G{\"a}nsicke}, Boris T. and {Ge}, Junqiang and {Gonz{\'a}lez {\'A}vila}, Consuelo and {Grabowski}, Katie and {Grier}, Catherine and {Guiglion}, Guillaume and {Gupta}, Pramod and {Hall}, Patrick and {Hawkins}, Keith and {Hayes}, Christian R. and {Hermes}, J.~J. and {Hern{\'a}ndez-Garc{\'\i}a}, Lorena and {Hogg}, David W. and {Holtzman}, Jon A. and {Ibarra-Medel}, Hector Javier and {Ji}, Alexander and {Jofre}, Paula and {Johnson}, Jennifer A. and {Jones}, Amy M. and {Kinemuchi}, Karen and {Kluge}, Matthias and {Koekemoer}, Anton and {Kollmeier}, Juna A. and {Kounkel}, Marina and {Krishnarao}, Dhanesh and {Krumpe}, Mirko and {Lacerna}, Ivan and {Lago}, Paulo Jakson Assuncao and {Laporte}, Chervin and {Liu}, Chao and {Liu}, Ang and {Liu}, Xin and {Lopes}, Alexandre Roman and {Macktoobian}, Matin and {Majewski}, Steven R. and {Malanushenko}, Viktor and {Maoz}, Dan and {Masseron}, Thomas and {Masters}, Karen L. and {Matijevic}, Gal and {McBride}, Aidan and {Medan}, Ilija and {Merloni}, Andrea and {Morrison}, Sean and {Myers}, Natalie and {M{\'e}sz{\'a}ros}, Szabolcs and {Negrete}, C. Alenka and {Nidever}, David L. and {Nitschelm}, Christian and {Oravetz}, Daniel and {Oravetz}, Audrey and {Pan}, Kaike and {Peng}, Yingjie and {Pinsonneault}, Marc H. and {Pogge}, Rick and {Qiu}, Dan and {Ramirez}, Solange V. and {Rix}, Hans-Walter and {Fern{\'a}ndez Rosso}, Daniela and {Runnoe}, Jessie and {Salvato}, Mara and {Sanchez}, Sebastian F. and {Santana}, Felipe A. and {Saydjari}, Andrew and {Sayres}, Conor and {Schlaufman}, Kevin C. and {Schneider}, Donald P. and {Schwope}, Axel and {Serna}, Javier and {Shen}, Yue and {Sobeck}, Jennifer and {Song}, Ying-Yi and {Souto}, Diogo and {Spoo}, Taylor and {Stassun}, Keivan G. and {Steinmetz}, Matthias and {Straumit}, Ilya and {Stringfellow}, Guy and {S{\'a}nchez-Gallego}, Jos{\'e} and {Taghizadeh-Popp}, Manuchehr and {Tayar}, Jamie and {Thakar}, Ani and {Tissera}, Patricia B. and {Tkachenko}, Andrew and {Hernandez Toledo}, Hector and {Trakhtenbrot}, Benny and {Fern{\'a}ndez-Trincado}, Jos{\'e} G. and {Troup}, Nicholas and {Trump}, Jonathan R. and {Tuttle}, Sarah and {Ulloa}, Natalie and {Vazquez-Mata}, Jose Antonio and {Vera Alfaro}, Pablo and {Villanova}, Sandro and {Wachter}, Stefanie and {Weijmans}, Anne-Marie and {Wheeler}, Adam and {Wilson}, John and {Wojno}, Leigh and {Wolf}, Julien and {Xue}, Xiang-Xiang and {Ybarra}, Jason E. and {Zari}, Eleonora and {Zasowski}, Gail},
        title = "{The Eighteenth Data Release of the Sloan Digital Sky Surveys: Targeting and First Spectra from SDSS-V}",
      journal = {\apjs},
         year = 2023,
        month = aug,
       volume = {267},
       number = {2},
          eid = {44},
        pages = {44},
          doi = {10.3847/1538-4365/acda98},
archivePrefix = {arXiv},
       eprint = {2301.07688},
 primaryClass = {astro-ph.GA},
       adsurl = {https://ui.adsabs.harvard.edu/abs/2023ApJS..267...44A}
}

@ARTICLE{barnes_2001,
       author = {{Barnes}, D.~G. and {Staveley-Smith}, L. and {de Blok}, W.~J.~G. and {Oosterloo}, T. and {Stewart}, I.~M. and {Wright}, A.~E. and {Banks}, G.~D. and {Bhathal}, R. and {Boyce}, P.~J. and {Calabretta}, M.~R. and {Disney}, M.~J. and {Drinkwater}, M.~J. and {Ekers}, R.~D. and {Freeman}, K.~C. and {Gibson}, B.~K. and {Green}, A.~J. and {Haynes}, R.~F. and {te Lintel Hekkert}, P. and {Henning}, P.~A. and {Jerjen}, H. and {Juraszek}, S. and {Kesteven}, M.~J. and {Kilborn}, V.~A. and {Knezek}, P.~M. and {Koribalski}, B. and {Kraan-Korteweg}, R.~C. and {Malin}, D.~F. and {Marquarding}, M. and {Minchin}, R.~F. and {Mould}, J.~R. and {Price}, R.~M. and {Putman}, M.~E. and {Ryder}, S.~D. and {Sadler}, E.~M. and {Schr{\"o}der}, A. and {Stootman}, F. and {Webster}, R.~L. and {Wilson}, W.~E. and {Ye}, T.},
        title = "{The HI Parkes All Sky Survey: southern observations, calibration and robust imaging}",
      journal = {\mnras},
         year = 2001,
        month = apr,
       volume = {322},
       number = {3},
        pages = {486-498},
          doi = {10.1046/j.1365-8711.2001.04102.x},
       adsurl = {https://ui.adsabs.harvard.edu/abs/2001MNRAS.322..486B}
}

@ARTICLE{baugh_2019,
       author = {{Baugh}, C.~M. and {Gonzalez-Perez}, Violeta and {Lagos}, Claudia D.~P. and {Lacey}, Cedric G. and {Helly}, John C. and {Jenkins}, Adrian and {Frenk}, Carlos S. and {Benson}, Andrew J. and {Bower}, Richard G. and {Cole}, Shaun},
        title = "{Galaxy formation in the Planck Millennium: the atomic hydrogen content of dark matter haloes}",
      journal = {\mnras},
         year = 2019,
        month = mar,
       volume = {483},
       number = {4},
        pages = {4922-4937},
          doi = {10.1093/mnras/sty3427},
archivePrefix = {arXiv},
       eprint = {1808.08276},
 primaryClass = {astro-ph.GA},
       adsurl = {https://ui.adsabs.harvard.edu/abs/2019MNRAS.483.4922B}
}

@ARTICLE{begeman_1989,
       author = {{Begeman}, K.~G.},
        title = "{HI rotation curves of spiral galaxies. I. NGC 3198.}",
      journal = {\aap},
         year = 1989,
        month = oct,
       volume = {223},
        pages = {47-60},
       adsurl = {https://ui.adsabs.harvard.edu/abs/1989A&A...223...47B}
}

@ARTICLE{bell_2001,
       author = {{Bell}, Eric F. and {de Jong}, Roelof S.},
        title = "{Stellar Mass-to-Light Ratios and the Tully-Fisher Relation}",
      journal = {\apj},
         year = 2001,
        month = mar,
       volume = {550},
       number = {1},
        pages = {212-229},
          doi = {10.1086/319728},
archivePrefix = {arXiv},
       eprint = {astro-ph/0011493},
 primaryClass = {astro-ph},
       adsurl = {https://ui.adsabs.harvard.edu/abs/2001ApJ...550..212B}
}

@ARTICLE{bigiel_2008,
       author = {{Bigiel}, F. and {Leroy}, A. and {Walter}, F. and {Brinks}, E. and {de Blok}, W.~J.~G. and {Madore}, B. and {Thornley}, M.~D.},
        title = "{The Star Formation Law in Nearby Galaxies on Sub-Kpc Scales}",
      journal = {\aj},
         year = 2008,
        month = dec,
       volume = {136},
       number = {6},
        pages = {2846-2871},
          doi = {10.1088/0004-6256/136/6/2846},
archivePrefix = {arXiv},
       eprint = {0810.2541},
 primaryClass = {astro-ph},
       adsurl = {https://ui.adsabs.harvard.edu/abs/2008AJ....136.2846B}
}

@BOOK{binney_2008,
       author = {{Binney}, James and {Tremaine}, Scott},
        title = "{Galactic Dynamics: Second Edition}",
         year = 2008,
       adsurl = {https://ui.adsabs.harvard.edu/abs/2008gady.book.....B},
    publisher = {Princeton University Press}
}

@ARTICLE{blaizot_2005,
       author = {{Blaizot}, J{\'e}r{\'e}my and {Wadadekar}, Yogesh and {Guiderdoni}, Bruno and {Colombi}, St{\'e}phane T. and {Bertin}, Emmanuel and {Bouchet}, Fran{\c{c}}ois R. and {Devriendt}, Julien E.~G. and {Hatton}, Steve},
        title = "{MoMaF: the Mock Map Facility}",
      journal = {\mnras},
         year = 2005,
        month = jun,
       volume = {360},
       number = {1},
        pages = {159-175},
          doi = {10.1111/j.1365-2966.2005.09019.x},
archivePrefix = {arXiv},
       eprint = {astro-ph/0309305},
 primaryClass = {astro-ph},
       adsurl = {https://ui.adsabs.harvard.edu/abs/2005MNRAS.360..159B}
}

@ARTICLE{blitz_2006,
       author = {{Blitz}, Leo and {Rosolowsky}, Erik},
        title = "{The Role of Pressure in GMC Formation II: The H$_{2}$-Pressure Relation}",
      journal = {\apj},
         year = 2006,
        month = oct,
       volume = {650},
       number = {2},
        pages = {933-944},
          doi = {10.1086/505417},
archivePrefix = {arXiv},
       eprint = {astro-ph/0605035},
 primaryClass = {astro-ph},
       adsurl = {https://ui.adsabs.harvard.edu/abs/2006ApJ...650..933B}
}

@INPROCEEDINGS{blyth_2016,
       author = {{Blyth}, S. and {Baker}, A.~J. and {Holwerda}, B. and {Bouchard}, A. and {Catinella}, B. and {Chemin}, L. and {Cunnama}, D. and {Dav{\'e}}, R. and {Faltenbacher}, A. and {February}, S. and {Fern{\'a}ndez}, X. and {Gawiser}, E. and {Heywood}, I. and {Kere{\v{s}}}, D. and {Kl{\"o}ckner}, H.~R. and {Lah}, P. and {Lochner}, M. and {Maddox}, N. and {Makhathini}, S. and {Moodley}, K. and {Morganti}, R. and {Obreschkow}, D. and {Oh}, S.~H. and {Pisano}, D.~J. and {Popping}, A. and {Popping}, G. and {Ravindranath}, S. and {Schinnerer}, E. and {Sheth}, K. and {Skelton}, R. and {Smith}, M. and {Srianand}, R. and {Staveley-Smith}, L. and {Vaccari}, M. and {Vaisanen}, P. and {Walter}, F. and {Rawlings}, S. and {Bassett}, B.~A. and {Bershady}, M.~A. and {Briggs}, F.~H. and {Crawford}, S.~M. and {Cress}, C.~M. and {Darling}, J.~K. and {Deane}, R.~P. and {de Blok}, G. and {Elson}, E.~C. and {Frank}, B.~S. and {Henning}, P.~A. and {Hess}, K.~M. and {Hughes}, J.~P. and {Jarvis}, M.~J. and {Kannappan}, S.~J. and {Katz}, N.~S. and {Kraan-Korteweg}, R.~C. and {Lehnert}, M.~D. and {Leroy}, A.~K. and {Meurer}, G.~R. and {Meyer}, M.~J. and {Pisano}, D.~J. and {Schr{\"o}der}, A.~C. and {Smirnov}, O.~M. and {Somerville}, R.~S. and {Stewart}, I.~M. and {van der Heyden}, K.~J. and {Verheijen}, M.~A.~W. and {Wilcots}, E.~M. and {Williams}, T.~B. and {Woudt}, P.~A. and {Wu}, J.~F. and {Zwaan}, M.~A. and {Zwart}, J.~T.~L. and {Oosterloo}, T.~A. and {van Drie}, W.},
        title = "{LADUMA: Looking at the Distant Universe with the MeerKAT Array}",
    booktitle = {MeerKAT Science: On the Pathway to the SKA},
         year = 2016,
        month = jan,
          eid = {4},
        pages = {4},
          doi = {10.22323/1.277.0004},
       adsurl = {https://ui.adsabs.harvard.edu/abs/2016mks..confE...4B}
}

@ARTICLE{boselli_2006,
       author = {{Boselli}, Alessandro and {Gavazzi}, Giuseppe},
        title = "{Environmental Effects on Late-Type Galaxies in Nearby Clusters}",
      journal = {\pasp},
         year = 2006,
        month = apr,
       volume = {118},
       number = {842},
        pages = {517-559},
          doi = {10.1086/500691},
archivePrefix = {arXiv},
       eprint = {astro-ph/0601108},
 primaryClass = {astro-ph},
       adsurl = {https://ui.adsabs.harvard.edu/abs/2006PASP..118..517B}
}

@ARTICLE{boubel_2025a,
       author = {{Boubel}, Paula and {Colless}, Matthew and {Said}, Khaled and {Staveley-Smith}, Lister},
        title = "{Testing anisotropic Hubble expansion}",
      journal = {\jcap},
         year = 2025,
        month = mar,
       volume = {2025},
       number = {3},
          eid = {066},
        pages = {066},
          doi = {10.1088/1475-7516/2025/03/066},
archivePrefix = {arXiv},
       eprint = {2412.14607},
 primaryClass = {astro-ph.CO},
       adsurl = {https://ui.adsabs.harvard.edu/abs/2025JCAP...03..066B}
}

@ARTICLE{boubel_2024b,
       author = {{Boubel}, Paula and {Colless}, Matthew and {Said}, Khaled and {Staveley-Smith}, Lister},
        title = "{Large-scale motions and growth rate from forward-modelling Tully-Fisher peculiar velocities}",
      journal = {\mnras},
         year = 2024,
        month = jun,
       volume = {531},
       number = {1},
        pages = {84-109},
          doi = {10.1093/mnras/stae1122},
archivePrefix = {arXiv},
       eprint = {2301.12648},
 primaryClass = {astro-ph.CO},
       adsurl = {https://ui.adsabs.harvard.edu/abs/2024MNRAS.531...84B}
}

@ARTICLE{boubel_2024c,
       author = {{Boubel}, Paula and {Colless}, Matthew and {Said}, Khaled and {Staveley-Smith}, Lister},
        title = "{An improved Tully-Fisher estimate of H$_{0}$}",
      journal = {\mnras},
         year = 2024,
        month = sep,
       volume = {533},
       number = {2},
        pages = {1550-1559},
          doi = {10.1093/mnras/stae1925},
archivePrefix = {arXiv},
       eprint = {2408.03660},
 primaryClass = {astro-ph.CO},
       adsurl = {https://ui.adsabs.harvard.edu/abs/2024MNRAS.533.1550B}
}

@ARTICLE{boylan-kolchin_2009,
       author = {{Boylan-Kolchin}, Michael and {Springel}, Volker and {White}, Simon D.~M. and {Jenkins}, Adrian and {Lemson}, Gerard},
        title = "{Resolving cosmic structure formation with the Millennium-II Simulation}",
      journal = {\mnras},
         year = 2009,
        month = sep,
       volume = {398},
       number = {3},
        pages = {1150-1164},
          doi = {10.1111/j.1365-2966.2009.15191.x},
archivePrefix = {arXiv},
       eprint = {0903.3041},
 primaryClass = {astro-ph.CO},
       adsurl = {https://ui.adsabs.harvard.edu/abs/2009MNRAS.398.1150B}
}

@ARTICLE{braun_2019,
       author = {{Braun}, Robert and {Bonaldi}, Anna and {Bourke}, Tyler and {Keane}, Evan and {Wagg}, Jeff},
        title = "{Anticipated Performance of the Square Kilometre Array -- Phase 1 (SKA1)}",
      journal = {arXiv e-prints},
         year = 2019,
        month = dec,
          eid = {arXiv:1912.12699},
        pages = {arXiv:1912.12699},
          doi = {10.48550/arXiv.1912.12699},
archivePrefix = {arXiv},
       eprint = {1912.12699},
 primaryClass = {astro-ph.IM},
       adsurl = {https://ui.adsabs.harvard.edu/abs/2019arXiv191212699B}
}

@ARTICLE{brooks_2023,
       author = {{Brooks}, Richard A.~N. and {Oman}, Kyle A. and {Frenk}, Carlos S.},
        title = "{The north-south asymmetry of the ALFALFA H I velocity width function}",
      journal = {\mnras},
         year = 2023,
        month = jul,
       volume = {522},
       number = {3},
        pages = {4043-4058},
          doi = {10.1093/mnras/stad1191},
archivePrefix = {arXiv},
       eprint = {2211.08092},
 primaryClass = {astro-ph.GA},
       adsurl = {https://ui.adsabs.harvard.edu/abs/2023MNRAS.522.4043B}
}

@ARTICLE{cantarella_2025,
       author = {{Cantarella}, Sebastiano and {De Lucia}, Gabriella and {Fontanot}, Fabio and {Hirschmann}, Michaela and {Xie}, Lizhi and {Franco}, Maximilien and {Plat}, Ad{\`e}le},
        title = "{Probing the dawn of galaxies: star formation and feedback in the JWST era through the GAEA model}",
      journal = {arXiv e-prints},
         year = 2025,
        month = nov,
          eid = {arXiv:2511.03787},
        pages = {arXiv:2511.03787},
          doi = {10.48550/arXiv.2511.03787},
archivePrefix = {arXiv},
       eprint = {2511.03787},
 primaryClass = {astro-ph.GA},
       adsurl = {https://ui.adsabs.harvard.edu/abs/2025arXiv251103787C}
}

@ARTICLE{catinella_2010,
       author = {{Catinella}, Barbara and {Schiminovich}, David and {Kauffmann}, Guinevere and {Fabello}, Silvia and {Wang}, Jing and {Hummels}, Cameron and {Lemonias}, Jenna and {Moran}, Sean M. and {Wu}, Ronin and {Giovanelli}, Riccardo and {Haynes}, Martha P. and {Heckman}, Timothy M. and {Basu-Zych}, Antara R. and {Blanton}, Michael R. and {Brinchmann}, Jarle and {Budav{\'a}ri}, Tam{\'a}s and {Gon{\c{c}}alves}, Thiago and {Johnson}, Benjamin D. and {Kennicutt}, Robert C. and {Madore}, Barry F. and {Martin}, Christopher D. and {Rich}, Michael R. and {Tacconi}, Linda J. and {Thilker}, David A. and {Wild}, Vivienne and {Wyder}, Ted K.},
        title = "{The GALEX Arecibo SDSS Survey - I. Gas fraction scaling relations of massive galaxies and first data release}",
      journal = {\mnras},
         year = 2010,
        month = apr,
       volume = {403},
       number = {2},
        pages = {683-708},
          doi = {10.1111/j.1365-2966.2009.16180.x},
archivePrefix = {arXiv},
       eprint = {0912.1610},
 primaryClass = {astro-ph.CO},
       adsurl = {https://ui.adsabs.harvard.edu/abs/2010MNRAS.403..683C}
}

@ARTICLE{chaikin_2026,
       author = {{Chaikin}, Evgenii and {Schaye}, Joop and {Schaller}, Matthieu and {Ploeckinger}, Sylvia and {Bah{\'e}}, Yannick M. and {Ben{\'\i}tez-Llambay}, Alejandro and {Correa}, Camila and {Forouhar Moreno}, Victor J. and {Frenk}, Carlos S. and {Hu{\v{s}}ko}, Filip and {Kugel}, Roi and {McGibbon}, Robert and {Richings}, Alexander J. and {Trayford}, James W. and {Borrow}, Josh and {Crain}, Robert A. and {Helly}, John C. and {Lacey}, Cedric G. and {Ludlow}, Aaron and {Nobels}, Folkert S.~J.},
        title = "{COLIBRE: calibrating subgrid feedback in cosmological simulations that include a cold gas phase}",
      journal = {\mnras},
         year = 2026,
        month = may,
       volume = {548},
       number = {1},
          eid = {stag300},
        pages = {stag300},
          doi = {10.1093/mnras/stag300},
archivePrefix = {arXiv},
       eprint = {2509.04067},
 primaryClass = {astro-ph.GA},
       adsurl = {https://ui.adsabs.harvard.edu/abs/2026MNRAS.548ag300C}
}

@ARTICLE{chauhan_2019,
       author = {{Chauhan}, Garima and {Lagos}, Claudia del P. and {Obreschkow}, Danail and {Power}, Chris and {Oman}, Kyle and {Elahi}, Pascal J.},
        title = "{The H I velocity function: a test of cosmology or baryon physics?}",
      journal = {\mnras},
         year = 2019,
        month = oct,
       volume = {488},
       number = {4},
        pages = {5898-5915},
          doi = {10.1093/mnras/stz2069},
archivePrefix = {arXiv},
       eprint = {1906.06130},
 primaryClass = {astro-ph.GA},
       adsurl = {https://ui.adsabs.harvard.edu/abs/2019MNRAS.488.5898C}
}

@ARTICLE{chen_2024,
       author = {{Chen}, Hongxing and {Xie}, Lizhi and {Wang}, Jie and {Hu}, Wenkai and {De Lucia}, Gabriella and {Fontanot}, Fabio and {Hirschmann}, Michaela},
        title = "{Environmental effects on satellite galaxies from the perspective of cold gas}",
      journal = {\mnras},
         year = 2024,
        month = feb,
       volume = {528},
       number = {2},
        pages = {2451-2463},
          doi = {10.1093/mnras/stae162},
archivePrefix = {arXiv},
       eprint = {2401.07158},
 primaryClass = {astro-ph.GA},
       adsurl = {https://ui.adsabs.harvard.edu/abs/2024MNRAS.528.2451C}
}

@ARTICLE{colless_2001,
       author = {{Colless}, Matthew and {Dalton}, Gavin and {Maddox}, Steve and {Sutherland}, Will and {Norberg}, Peder and {Cole}, Shaun and {Bland-Hawthorn}, Joss and {Bridges}, Terry and {Cannon}, Russell and {Collins}, Chris and {Couch}, Warrick and {Cross}, Nicholas and {Deeley}, Kathryn and {De Propris}, Roberto and {Driver}, Simon P. and {Efstathiou}, George and {Ellis}, Richard S. and {Frenk}, Carlos S. and {Glazebrook}, Karl and {Jackson}, Carole and {Lahav}, Ofer and {Lewis}, Ian and {Lumsden}, Stuart and {Madgwick}, Darren and {Peacock}, John A. and {Peterson}, Bruce A. and {Price}, Ian and {Seaborne}, Mark and {Taylor}, Keith},
        title = "{The 2dF Galaxy Redshift Survey: spectra and redshifts}",
      journal = {\mnras},
         year = 2001,
        month = dec,
       volume = {328},
       number = {4},
        pages = {1039-1063},
          doi = {10.1046/j.1365-8711.2001.04902.x},
archivePrefix = {arXiv},
       eprint = {astro-ph/0106498},
 primaryClass = {astro-ph},
       adsurl = {https://ui.adsabs.harvard.edu/abs/2001MNRAS.328.1039C}
}

@ARTICLE{cora_2018,
       author = {{Cora}, Sof{\'\i}a A. and {Vega-Mart{\'\i}nez}, Cristian A. and {Hough}, Tom{\'a}s and {Ruiz}, Andr{\'e}s N. and {Orsi}, {\'A}lvaro A. and {Mu{\~n}oz Arancibia}, Alejandra M. and {Gargiulo}, Ignacio D. and {Collacchioni}, Florencia and {Padilla}, Nelson D. and {Gottl{\"o}ber}, Stefan and {Yepes}, Gustavo},
        title = "{Semi-analytic galaxies - I. Synthesis of environmental and star-forming regulation mechanisms}",
      journal = {\mnras},
         year = 2018,
        month = sep,
       volume = {479},
       number = {1},
        pages = {2-24},
          doi = {10.1093/mnras/sty1131},
archivePrefix = {arXiv},
       eprint = {1801.03883},
 primaryClass = {astro-ph.GA},
       adsurl = {https://ui.adsabs.harvard.edu/abs/2018MNRAS.479....2C}
}

@ARTICLE{cortese_2017,
       author = {{Cortese}, Luca and {Catinella}, Barbara and {Janowiecki}, Steven},
        title = "{ALMA Shows that Gas Reservoirs of Star-forming Disks over the Past 3 Billion Years Are Not Predominantly Molecular}",
      journal = {\apjl},
         year = 2017,
        month = oct,
       volume = {848},
       number = {1},
          eid = {L7},
        pages = {L7},
          doi = {10.3847/2041-8213/aa8cc3},
archivePrefix = {arXiv},
       eprint = {1709.07933},
 primaryClass = {astro-ph.GA},
       adsurl = {https://ui.adsabs.harvard.edu/abs/2017ApJ...848L...7C}
}

@ARTICLE{courtois_2009,
       author = {{Courtois}, H{\'e}l{\`e}ne M. and {Tully}, R. Brent and {Fisher}, J. Richard and {Bonhomme}, Nicolas and {Zavodny}, Maximilian and {Barnes}, Austin},
        title = "{The Extragalactic Distance Database: All Digital H I Profile Catalog}",
      journal = {\aj},
         year = 2009,
        month = dec,
       volume = {138},
       number = {6},
        pages = {1938-1956},
          doi = {10.1088/0004-6256/138/6/1938},
archivePrefix = {arXiv},
       eprint = {0902.3670},
 primaryClass = {astro-ph.CO},
       adsurl = {https://ui.adsabs.harvard.edu/abs/2009AJ....138.1938C}
}

@ARTICLE{courtois_2011,
       author = {{Courtois}, H{\'e}l{\`e}ne M. and {Tully}, R. Brent and {Makarov}, D.~I. and {Mitronova}, S. and {Koribalski}, B. and {Karachentsev}, I.~D. and {Fisher}, J. Richard},
        title = "{Cosmic Flows: Green Bank Telescope and Parkes H I observations}",
      journal = {\mnras},
         year = 2011,
        month = jul,
       volume = {414},
       number = {3},
        pages = {2005-2016},
          doi = {10.1111/j.1365-2966.2011.18515.x},
archivePrefix = {arXiv},
       eprint = {1101.3802},
 primaryClass = {astro-ph.CO},
       adsurl = {https://ui.adsabs.harvard.edu/abs/2011MNRAS.414.2005C}
}

@ARTICLE{croton_2006,
       author = {{Croton}, Darren J. and {Springel}, Volker and {White}, Simon D.~M. and {De Lucia}, G. and {Frenk}, C.~S. and {Gao}, L. and {Jenkins}, A. and {Kauffmann}, G. and {Navarro}, J.~F. and {Yoshida}, N.},
        title = "{The many lives of active galactic nuclei: cooling flows, black holes and the luminosities and colours of galaxies}",
      journal = {\mnras},
         year = 2006,
        month = jan,
       volume = {365},
       number = {1},
        pages = {11-28},
          doi = {10.1111/j.1365-2966.2005.09675.x},
archivePrefix = {arXiv},
       eprint = {astro-ph/0508046},
 primaryClass = {astro-ph},
       adsurl = {https://ui.adsabs.harvard.edu/abs/2006MNRAS.365...11C}
}

@ARTICLE{dave_2019,
       author = {{Dav{\'e}}, Romeel and {Angl{\'e}s-Alc{\'a}zar}, Daniel and {Narayanan}, Desika and {Li}, Qi and {Rafieferantsoa}, Mika H. and {Appleby}, Sarah},
        title = "{SIMBA: Cosmological simulations with black hole growth and feedback}",
      journal = {\mnras},
         year = 2019,
        month = jun,
       volume = {486},
       number = {2},
        pages = {2827-2849},
          doi = {10.1093/mnras/stz937},
archivePrefix = {arXiv},
       eprint = {1901.10203},
 primaryClass = {astro-ph.GA},
       adsurl = {https://ui.adsabs.harvard.edu/abs/2019MNRAS.486.2827D}
}

@ARTICLE{dawson_2013,
       author = {{Dawson}, Kyle S. and {Schlegel}, David J. and {Ahn}, Christopher P. and {Anderson}, Scott F. and {Aubourg}, {\'E}ric and {Bailey}, Stephen and {Barkhouser}, Robert H. and {Bautista}, Julian E. and {Beifiori}, Alessandra and {Berlind}, Andreas A. and {Bhardwaj}, Vaishali and {Bizyaev}, Dmitry and {Blake}, Cullen H. and {Blanton}, Michael R. and {Blomqvist}, Michael and {Bolton}, Adam S. and {Borde}, Arnaud and {Bovy}, Jo and {Brandt}, W.~N. and {Brewington}, Howard and {Brinkmann}, Jon and {Brown}, Peter J. and {Brownstein}, Joel R. and {Bundy}, Kevin and {Busca}, N.~G. and {Carithers}, William and {Carnero}, Aurelio R. and {Carr}, Michael A. and {Chen}, Yanmei and {Comparat}, Johan and {Connolly}, Natalia and {Cope}, Frances and {Croft}, Rupert A.~C. and {Cuesta}, Antonio J. and {da Costa}, Luiz N. and {Davenport}, James R.~A. and {Delubac}, Timoth{\'e}e and {de Putter}, Roland and {Dhital}, Saurav and {Ealet}, Anne and {Ebelke}, Garrett L. and {Eisenstein}, Daniel J. and {Escoffier}, S. and {Fan}, Xiaohui and {Filiz Ak}, N. and {Finley}, Hayley and {Font-Ribera}, Andreu and {G{\'e}nova-Santos}, R. and {Gunn}, James E. and {Guo}, Hong and {Haggard}, Daryl and {Hall}, Patrick B. and {Hamilton}, Jean-Christophe and {Harris}, Ben and {Harris}, David W. and {Ho}, Shirley and {Hogg}, David W. and {Holder}, Diana and {Honscheid}, Klaus and {Huehnerhoff}, Joe and {Jordan}, Beatrice and {Jordan}, Wendell P. and {Kauffmann}, Guinevere and {Kazin}, Eyal A. and {Kirkby}, David and {Klaene}, Mark A. and {Kneib}, Jean-Paul and {Le Goff}, Jean-Marc and {Lee}, Khee-Gan and {Long}, Daniel C. and {Loomis}, Craig P. and {Lundgren}, Britt and {Lupton}, Robert H. and {Maia}, Marcio A.~G. and {Makler}, Martin and {Malanushenko}, Elena and {Malanushenko}, Viktor and {Mandelbaum}, Rachel and {Manera}, Marc and {Maraston}, Claudia and {Margala}, Daniel and {Masters}, Karen L. and {McBride}, Cameron K. and {McDonald}, Patrick and {McGreer}, Ian D. and {McMahon}, Richard G. and {Mena}, Olga and {Miralda-Escud{\'e}}, Jordi and {Montero-Dorta}, Antonio D. and {Montesano}, Francesco and {Muna}, Demitri and {Myers}, Adam D. and {Naugle}, Tracy and {Nichol}, Robert C. and {Noterdaeme}, Pasquier and {Nuza}, Sebasti{\'a}n E. and {Olmstead}, Matthew D. and {Oravetz}, Audrey and {Oravetz}, Daniel J. and {Owen}, Russell and {Padmanabhan}, Nikhil and {Palanque-Delabrouille}, Nathalie and {Pan}, Kaike and {Parejko}, John K. and {P{\^a}ris}, Isabelle and {Percival}, Will J. and {P{\'e}rez-Fournon}, Ismael and {P{\'e}rez-R{\`a}fols}, Ignasi and {Petitjean}, Patrick and {Pfaffenberger}, Robert and {Pforr}, Janine and {Pieri}, Matthew M. and {Prada}, Francisco and {Price-Whelan}, Adrian M. and {Raddick}, M. Jordan and {Rebolo}, Rafael and {Rich}, James and {Richards}, Gordon T. and {Rockosi}, Constance M. and {Roe}, Natalie A. and {Ross}, Ashley J. and {Ross}, Nicholas P. and {Rossi}, Graziano and {Rubi{\~n}o-Martin}, J.~A. and {Samushia}, Lado and {S{\'a}nchez}, Ariel G. and {Sayres}, Conor and {Schmidt}, Sarah J. and {Schneider}, Donald P. and {Sc{\'o}ccola}, C.~G. and {Seo}, Hee-Jong and {Shelden}, Alaina and {Sheldon}, Erin and {Shen}, Yue and {Shu}, Yiping and {Slosar}, An{\v{z}}e and {Smee}, Stephen A. and {Snedden}, Stephanie A. and {Stauffer}, Fritz and {Steele}, Oliver and {Strauss}, Michael A. and {Streblyanska}, Alina and {Suzuki}, Nao and {Swanson}, Molly E.~C. and {Tal}, Tomer and {Tanaka}, Masayuki and {Thomas}, Daniel and {Tinker}, Jeremy L. and {Tojeiro}, Rita and {Tremonti}, Christy A. and {Vargas Maga{\~n}a}, M. and {Verde}, Licia and {Viel}, Matteo and {Wake}, David A. and {Watson}, Mike and {Weaver}, Benjamin A. and {Weinberg}, David H. and {Weiner}, Benjamin J. and {West}, Andrew A. and {White}, Martin and {Wood-Vasey}, W.~M. and {Yeche}, Christophe and {Zehavi}, Idit and {Zhao}, Gong-Bo and {Zheng}, Zheng},
        title = "{The Baryon Oscillation Spectroscopic Survey of SDSS-III}",
      journal = {\aj},
         year = 2013,
        month = jan,
       volume = {145},
       number = {1},
          eid = {10},
        pages = {10},
          doi = {10.1088/0004-6256/145/1/10},
archivePrefix = {arXiv},
       eprint = {1208.0022},
 primaryClass = {astro-ph.CO},
       adsurl = {https://ui.adsabs.harvard.edu/abs/2013AJ....145...10D}
}

@ARTICLE{dawson_2016,
       author = {{Dawson}, Kyle S. and {Kneib}, Jean-Paul and {Percival}, Will J. and {Alam}, Shadab and {Albareti}, Franco D. and {Anderson}, Scott F. and {Armengaud}, Eric and {Aubourg}, {\'E}ric and {Bailey}, Stephen and {Bautista}, Julian E. and {Berlind}, Andreas A. and {Bershady}, Matthew A. and {Beutler}, Florian and {Bizyaev}, Dmitry and {Blanton}, Michael R. and {Blomqvist}, Michael and {Bolton}, Adam S. and {Bovy}, Jo and {Brandt}, W.~N. and {Brinkmann}, Jon and {Brownstein}, Joel R. and {Burtin}, Etienne and {Busca}, N.~G. and {Cai}, Zheng and {Chuang}, Chia-Hsun and {Clerc}, Nicolas and {Comparat}, Johan and {Cope}, Frances and {Croft}, Rupert A.~C. and {Cruz-Gonzalez}, Irene and {da Costa}, Luiz N. and {Cousinou}, Marie-Claude and {Darling}, Jeremy and {de la Macorra}, Axel and {de la Torre}, Sylvain and {Delubac}, Timoth{\'e}e and {du Mas des Bourboux}, H{\'e}lion and {Dwelly}, Tom and {Ealet}, Anne and {Eisenstein}, Daniel J. and {Eracleous}, Michael and {Escoffier}, S. and {Fan}, Xiaohui and {Finoguenov}, Alexis and {Font-Ribera}, Andreu and {Frinchaboy}, Peter and {Gaulme}, Patrick and {Georgakakis}, Antonis and {Green}, Paul and {Guo}, Hong and {Guy}, Julien and {Ho}, Shirley and {Holder}, Diana and {Huehnerhoff}, Joe and {Hutchinson}, Timothy and {Jing}, Yipeng and {Jullo}, Eric and {Kamble}, Vikrant and {Kinemuchi}, Karen and {Kirkby}, David and {Kitaura}, Francisco-Shu and {Klaene}, Mark A. and {Laher}, Russ R. and {Lang}, Dustin and {Laurent}, Pierre and {Le Goff}, Jean-Marc and {Li}, Cheng and {Liang}, Yu and {Lima}, Marcos and {Lin}, Qiufan and {Lin}, Weipeng and {Lin}, Yen-Ting and {Long}, Daniel C. and {Lundgren}, Britt and {MacDonald}, Nicholas and {Geimba Maia}, Marcio Antonio and {Malanushenko}, Elena and {Malanushenko}, Viktor and {Mariappan}, Vivek and {McBride}, Cameron K. and {McGreer}, Ian D. and {M{\'e}nard}, Brice and {Merloni}, Andrea and {Meza}, Andres and {Montero-Dorta}, Antonio D. and {Muna}, Demitri and {Myers}, Adam D. and {Nandra}, Kirpal and {Naugle}, Tracy and {Newman}, Jeffrey A. and {Noterdaeme}, Pasquier and {Nugent}, Peter and {Ogando}, Ricardo and {Olmstead}, Matthew D. and {Oravetz}, Audrey and {Oravetz}, Daniel J. and {Padmanabhan}, Nikhil and {Palanque-Delabrouille}, Nathalie and {Pan}, Kaike and {Parejko}, John K. and {P{\^a}ris}, Isabelle and {Peacock}, John A. and {Petitjean}, Patrick and {Pieri}, Matthew M. and {Pisani}, Alice and {Prada}, Francisco and {Prakash}, Abhishek and {Raichoor}, Anand and {Reid}, Beth and {Rich}, James and {Ridl}, Jethro and {Rodriguez-Torres}, Sergio and {Carnero Rosell}, Aurelio and {Ross}, Ashley J. and {Rossi}, Graziano and {Ruan}, John and {Salvato}, Mara and {Sayres}, Conor and {Schneider}, Donald P. and {Schlegel}, David J. and {Seljak}, Uros and {Seo}, Hee-Jong and {Sesar}, Branimir and {Shandera}, Sarah and {Shu}, Yiping and {Slosar}, An{\v{z}}e and {Sobreira}, Flavia and {Streblyanska}, Alina and {Suzuki}, Nao and {Taylor}, Donna and {Tao}, Charling and {Tinker}, Jeremy L. and {Tojeiro}, Rita and {Vargas-Maga{\~n}a}, Mariana and {Wang}, Yuting and {Weaver}, Benjamin A. and {Weinberg}, David H. and {White}, Martin and {Wood-Vasey}, W.~M. and {Yeche}, Christophe and {Zhai}, Zhongxu and {Zhao}, Cheng and {Zhao}, Gong-bo and {Zheng}, Zheng and {Ben Zhu}, Guangtun and {Zou}, Hu},
        title = "{The SDSS-IV Extended Baryon Oscillation Spectroscopic Survey: Overview and Early Data}",
      journal = {\aj},
         year = 2016,
        month = feb,
       volume = {151},
       number = {2},
          eid = {44},
        pages = {44},
          doi = {10.3847/0004-6256/151/2/44},
archivePrefix = {arXiv},
       eprint = {1508.04473},
 primaryClass = {astro-ph.CO},
       adsurl = {https://ui.adsabs.harvard.edu/abs/2016AJ....151...44D}
}

@ARTICLE{deblok_2008,
       author = {{de Blok}, W.~J.~G. and {Walter}, F. and {Brinks}, E. and {Trachternach}, C. and {Oh}, S. -H. and {Kennicutt}, Jr., R.~C.},
        title = "{High-Resolution Rotation Curves and Galaxy Mass Models from THINGS}",
      journal = {\aj},
         year = 2008,
        month = dec,
       volume = {136},
       number = {6},
        pages = {2648-2719},
          doi = {10.1088/0004-6256/136/6/2648},
archivePrefix = {arXiv},
       eprint = {0810.2100},
 primaryClass = {astro-ph},
       adsurl = {https://ui.adsabs.harvard.edu/abs/2008AJ....136.2648D}
}

@ARTICLE{dekel_2006,
       author = {{Dekel}, Avishai and {Birnboim}, Yuval},
        title = "{Galaxy bimodality due to cold flows and shock heating}",
      journal = {\mnras},
         year = 2006,
        month = may,
       volume = {368},
       number = {1},
        pages = {2-20},
          doi = {10.1111/j.1365-2966.2006.10145.x},
archivePrefix = {arXiv},
       eprint = {astro-ph/0412300},
 primaryClass = {astro-ph},
       adsurl = {https://ui.adsabs.harvard.edu/abs/2006MNRAS.368....2D}
}

@ARTICLE{delucia_2007,
       author = {{De Lucia}, Gabriella and {Blaizot}, J{\'e}r{\'e}my},
        title = "{The hierarchical formation of the brightest cluster galaxies}",
      journal = {\mnras},
         year = 2007,
        month = feb,
       volume = {375},
       number = {1},
        pages = {2-14},
          doi = {10.1111/j.1365-2966.2006.11287.x},
archivePrefix = {arXiv},
       eprint = {astro-ph/0606519},
 primaryClass = {astro-ph},
       adsurl = {https://ui.adsabs.harvard.edu/abs/2007MNRAS.375....2D}
}

@ARTICLE{delucia_2014,
       author = {{De Lucia}, Gabriella and {Tornatore}, Luca and {Frenk}, Carlos S. and {Helmi}, Amina and {Navarro}, Julio F. and {White}, Simon D.~M.},
        title = "{Elemental abundances in Milky Way-like galaxies from a hierarchical galaxy formation model}",
      journal = {\mnras},
         year = 2014,
        month = nov,
       volume = {445},
       number = {1},
        pages = {970-987},
          doi = {10.1093/mnras/stu1752},
archivePrefix = {arXiv},
       eprint = {1407.7867},
 primaryClass = {astro-ph.GA},
       adsurl = {https://ui.adsabs.harvard.edu/abs/2014MNRAS.445..970D}
}

@ARTICLE{delucia_2024b,
       author = {{De Lucia}, Gabriella and {Fontanot}, Fabio and {Xie}, Lizhi and {Hirschmann}, Michaela},
        title = "{Tracing the quenching journey across cosmic time}",
      journal = {\aap},
         year = 2024,
        month = jul,
       volume = {687},
          eid = {A68},
        pages = {A68},
          doi = {10.1051/0004-6361/202349045},
archivePrefix = {arXiv},
       eprint = {2401.06211},
 primaryClass = {astro-ph.GA},
       adsurl = {https://ui.adsabs.harvard.edu/abs/2024A&A...687A..68D}
}

@ARTICLE{desi_2016,
       author = {{DESI Collaboration} and {Aghamousa}, Amir and {Aguilar}, Jessica and {Ahlen}, Steve and {Alam}, Shadab and {Allen}, Lori E. and {Allende Prieto}, Carlos and {Annis}, James and {Bailey}, Stephen and {Balland}, Christophe and {Ballester}, Otger and {Baltay}, Charles and {Beaufore}, Lucas and {Bebek}, Chris and {Beers}, Timothy C. and {Bell}, Eric F. and {Bernal}, Jos{\'e} Luis and {Besuner}, Robert and {Beutler}, Florian and {Blake}, Chris and {Bleuler}, Hannes and {Blomqvist}, Michael and {Blum}, Robert and {Bolton}, Adam S. and {Briceno}, Cesar and {Brooks}, David and {Brownstein}, Joel R. and {Buckley-Geer}, Elizabeth and {Burden}, Angela and {Burtin}, Etienne and {Busca}, Nicolas G. and {Cahn}, Robert N. and {Cai}, Yan-Chuan and {Cardiel-Sas}, Laia and {Carlberg}, Raymond G. and {Carton}, Pierre-Henri and {Casas}, Ricard and {Castander}, Francisco J. and {Cervantes-Cota}, Jorge L. and {Claybaugh}, Todd M. and {Close}, Madeline and {Coker}, Carl T. and {Cole}, Shaun and {Comparat}, Johan and {Cooper}, Andrew P. and {Cousinou}, M. -C. and {Crocce}, Martin and {Cuby}, Jean-Gabriel and {Cunningham}, Daniel P. and {Davis}, Tamara M. and {Dawson}, Kyle S. and {de la Macorra}, Axel and {De Vicente}, Juan and {Delubac}, Timoth{\'e}e and {Derwent}, Mark and {Dey}, Arjun and {Dhungana}, Govinda and {Ding}, Zhejie and {Doel}, Peter and {Duan}, Yutong T. and {Ealet}, Anne and {Edelstein}, Jerry and {Eftekharzadeh}, Sarah and {Eisenstein}, Daniel J. and {Elliott}, Ann and {Escoffier}, St{\'e}phanie and {Evatt}, Matthew and {Fagrelius}, Parker and {Fan}, Xiaohui and {Fanning}, Kevin and {Farahi}, Arya and {Farihi}, Jay and {Favole}, Ginevra and {Feng}, Yu and {Fernandez}, Enrique and {Findlay}, Joseph R. and {Finkbeiner}, Douglas P. and {Fitzpatrick}, Michael J. and {Flaugher}, Brenna and {Flender}, Samuel and {Font-Ribera}, Andreu and {Forero-Romero}, Jaime E. and {Fosalba}, Pablo and {Frenk}, Carlos S. and {Fumagalli}, Michele and {Gaensicke}, Boris T. and {Gallo}, Giuseppe and {Garcia-Bellido}, Juan and {Gaztanaga}, Enrique and {Pietro Gentile Fusillo}, Nicola and {Gerard}, Terry and {Gershkovich}, Irena and {Giannantonio}, Tommaso and {Gillet}, Denis and {Gonzalez-de-Rivera}, Guillermo and {Gonzalez-Perez}, Violeta and {Gott}, Shelby and {Graur}, Or and {Gutierrez}, Gaston and {Guy}, Julien and {Habib}, Salman and {Heetderks}, Henry and {Heetderks}, Ian and {Heitmann}, Katrin and {Hellwing}, Wojciech A. and {Herrera}, David A. and {Ho}, Shirley and {Holland}, Stephen and {Honscheid}, Klaus and {Huff}, Eric and {Hutchinson}, Timothy A. and {Huterer}, Dragan and {Hwang}, Ho Seong and {Illa Laguna}, Joseph Maria and {Ishikawa}, Yuzo and {Jacobs}, Dianna and {Jeffrey}, Niall and {Jelinsky}, Patrick and {Jennings}, Elise and {Jiang}, Linhua and {Jimenez}, Jorge and {Johnson}, Jennifer and {Joyce}, Richard and {Jullo}, Eric and {Juneau}, St{\'e}phanie and {Kama}, Sami and {Karcher}, Armin and {Karkar}, Sonia and {Kehoe}, Robert and {Kennamer}, Noble and {Kent}, Stephen and {Kilbinger}, Martin and {Kim}, Alex G. and {Kirkby}, David and {Kisner}, Theodore and {Kitanidis}, Ellie and {Kneib}, Jean-Paul and {Koposov}, Sergey and {Kovacs}, Eve and {Koyama}, Kazuya and {Kremin}, Anthony and {Kron}, Richard and {Kronig}, Luzius and {Kueter-Young}, Andrea and {Lacey}, Cedric G. and {Lafever}, Robin and {Lahav}, Ofer and {Lambert}, Andrew and {Lampton}, Michael and {Landriau}, Martin and {Lang}, Dustin and {Lauer}, Tod R. and {Le Goff}, Jean-Marc and {Le Guillou}, Laurent and {Le Van Suu}, Auguste and {Lee}, Jae Hyeon and {Lee}, Su-Jeong and {Leitner}, Daniela and {Lesser}, Michael and {Levi}, Michael E. and {L'Huillier}, Benjamin and {Li}, Baojiu and {Liang}, Ming and {Lin}, Huan and {Linder}, Eric and {Loebman}, Sarah R. and {Luki{\'c}}, Zarija and {Ma}, Jun and {MacCrann}, Niall and {Magneville}, Christophe and {Makarem}, Laleh and {Manera}, Marc and {Manser}, Christopher J. and {Marshall}, Robert and {Martini}, Paul and {Massey}, Richard and {Matheson}, Thomas and {McCauley}, Jeremy and {McDonald}, Patrick and {McGreer}, Ian D. and {Meisner}, Aaron and {Metcalfe}, Nigel and {Miller}, Timothy N. and {Miquel}, Ramon and {Moustakas}, John and {Myers}, Adam and {Naik}, Milind and {Newman}, Jeffrey A. and {Nichol}, Robert C. and {Nicola}, Andrina and {Nicolati da Costa}, Luiz and {Nie}, Jundan and {Niz}, Gustavo and {Norberg}, Peder and {Nord}, Brian and {Norman}, Dara and {Nugent}, Peter and {O'Brien}, Thomas and {Oh}, Minji and {Olsen}, Knut A.~G.},
        title = "{The DESI Experiment Part I: Science,Targeting, and Survey Design}",
      journal = {arXiv e-prints},
         year = 2016,
        month = oct,
          eid = {arXiv:1611.00036},
        pages = {arXiv:1611.00036},
          doi = {10.48550/arXiv.1611.00036},
archivePrefix = {arXiv},
       eprint = {1611.00036},
 primaryClass = {astro-ph.IM},
       adsurl = {https://ui.adsabs.harvard.edu/abs/2016arXiv161100036D}
}

@ARTICLE{desi_2025,
       author = {{DESI Collaboration} and {Abdul-Karim}, M. and {Adame}, A.~G. and {Aguado}, D. and {Aguilar}, J. and {Ahlen}, S. and {Alam}, S. and {Aldering}, G. and {Alexander}, D.~M. and {Alfarsy}, R. and {Allen}, L. and {Allende Prieto}, C. and {Alves}, O. and {Anand}, A. and {Andrade}, U. and {Armengaud}, E. and {Avila}, S. and {Aviles}, A. and {Awan}, H. and {Bailey}, S. and {Baleato Lizancos}, A. and {Ballester}, O. and {Bault}, A. and {Bautista}, J. and {BenZvi}, S. and {Beraldo e Silva}, L. and {Bermejo-Climent}, J.~R. and {Beutler}, F. and {Bianchi}, D. and {Blake}, C. and {Blum}, R. and {Bolton}, A.~S. and {Bonici}, M. and {Brieden}, S. and {Brodzeller}, A. and {Brooks}, D. and {Buckley-Geer}, E. and {Burtin}, E. and {Canning}, R. and {Carnero Rosell}, A. and {Carr}, A. and {Carrilho}, P. and {Casas}, L. and {Castander}, F.~J. and {Cereskaite}, R. and {Cervantes-Cota}, J.~L. and {Chaussidon}, E. and {Chaves-Montero}, J. and {Chen}, S. and {Chen}, X. and {Claybaugh}, T. and {Cole}, S. and {Cooper}, A.~P. and {Cousinou}, M. -C. and {Cuceu}, A. and {Davis}, T.~M. and {Dawson}, K.~S. and {de Belsunce}, R. and {de la Cruz}, R. and {de la Macorra}, A. and {de Mattia}, A. and {Deiosso}, N. and {Della Costa}, J. and {Demina}, R. and {Demirbozan}, U. and {DeRose}, J. and {Dey}, A. and {Dey}, B. and {Ding}, J. and {Ding}, Z. and {Doel}, P. and {Douglass}, K. and {Dowicz}, M. and {Ebina}, H. and {Edelstein}, J. and {Eisenstein}, D.~J. and {Elbers}, W. and {Emas}, N. and {Escoffier}, S. and {Fagrelius}, P. and {Fan}, X. and {Fanning}, K. and {Fawcett}, V.~A. and {Fern\textbackslash'andez-Garc\textbackslash'ia}, E. and {Ferraro}, S. and {Findlay}, N. and {Font-Ribera}, A. and {Forero-Romero}, J.~E. and {Forero-S\textbackslash'anchez}, D. and {Frenk}, C.~S. and {G\textbackslash''ansicke}, B.~T. and {Galbany}, L. and {Garc\textbackslash'ia-Bellido}, J. and {Garcia-Quintero}, C. and {Garrison}, L.~H. and {Gazta\textbackslash\raisebox{-0.5ex}\textasciitildenaga}, E. and {Gil-Mar\textbackslash'in}, H. and {Gnedin}, O.~Y. and {Gontcho}, S. Gontcho A and {Gonzalez-Morales}, A.~X. and {Gonzalez-Perez}, V. and {Gordon}, C. and {Graur}, O. and {Green}, D. and {Gruen}, D. and {Gsponer}, R. and {Guandalin}, C. and {Gutierrez}, G. and {Guy}, J. and {Hahn}, C. and {Han}, J.~J. and {Han}, J. and {He}, S. and {Herrera-Alcantar}, H.~K. and {Honscheid}, K. and {Hou}, J. and {Howlett}, C. and {Huterer}, D. and {Ir\textbackslashv\{s\}i\textbackslashv\{c\}}, V. and {Ishak}, M. and {Jacques}, A. and {Jimenez}, J. and {Jing}, Y.~P. and {Joachimi}, B. and {Joudaki}, S. and {Joyce}, R. and {Jullo}, E. and {Juneau}, S. and {Kara\textbackslashc\{c\}ayl\{\textbackslashi\}}, N.~G. and {Karim}, T. and {Kehoe}, R. and {Kent}, S. and {Khederlarian}, A. and {Kirkby}, D. and {Kisner}, T. and {Kitaura}, F. -S. and {Kizhuprakkat}, N. and {Kong}, H. and {Koposov}, S.~E. and {Kremin}, A. and {Krolewski}, A. and {Lahav}, O. and {Lai}, Y. and {Lamman}, C. and {Lan}, T. -W. and {Landriau}, M. and {Lang}, D. and {Lange}, J.~U. and {Lasker}, J. and {Le Goff}, J.~M. and {Le Guillou}, L. and {Leauthaud}, A. and {Levi}, M.~E. and {Li}, S. and {Li}, T.~S. and {Lodha}, K. and {Lokken}, M. and {Luo}, Y. and {Magneville}, C. and {Manera}, M. and {Manser}, C.~J. and {Margala}, D. and {Martini}, P. and {Maus}, M. and {McCullough}, J. and {McDonald}, P. and {Medina}, G.~E. and {Medina-Varela}, L. and {Meisner}, A. and {Mena-Fern\textbackslash'andez}, J. and {Menegas}, A. and {Mezcua}, M. and {Miquel}, R. and {Montero-Camacho}, P. and {Moon}, J. and {Moustakas}, J. and {Mu\textbackslash\raisebox{-0.5ex}\textasciitildenoz-Guti\textbackslash'errez}, A. and {Mu\textbackslash\raisebox{-0.5ex}\textasciitildenoz-Santos}, D. and {Myers}, A.~D. and {Myles}, J. and {Nadathur}, S. and {Najita}, J. and {Napolitano}, L. and {Newman}, J.~A. and {Nikakhtar}, F. and {Nikutta}, R. and {Niz}, G. and {Noriega}, H.~E. and {Padmanabhan}, N. and {Paillas}, E. and {Palanque-Delabrouille}, N. and {Palmese}, A. and {Pan}, J. and {Pan}, Z. and {Parkinson}, D. and {Peacock}, J. and {Percival}, W.~J. and {P\textbackslash'erez-Fern\textbackslash'andez}, A. and {P\textbackslash'erez-R\textbackslash`afols}, I. and {Peterson}, P.},
        title = "{Data Release 1 of the Dark Energy Spectroscopic Instrument}",
      journal = {arXiv e-prints},
         year = 2025,
        month = mar,
          eid = {arXiv:2503.14745},
        pages = {arXiv:2503.14745},
          doi = {10.48550/arXiv.2503.14745},
archivePrefix = {arXiv},
       eprint = {2503.14745},
 primaryClass = {astro-ph.CO},
       adsurl = {https://ui.adsabs.harvard.edu/abs/2025arXiv250314745D}
}

@ARTICLE{desmond_2012,
       author = {{Desmond}, Harry},
        title = "{The baryonic Tully-Fisher Relation predicted by cold dark matter cosmogony}",
      journal = {arXiv e-prints},
         year = 2012,
        month = apr,
          eid = {arXiv:1204.1497},
        pages = {arXiv:1204.1497},
          doi = {10.48550/arXiv.1204.1497},
archivePrefix = {arXiv},
       eprint = {1204.1497},
 primaryClass = {astro-ph.CO},
       adsurl = {https://ui.adsabs.harvard.edu/abs/2012arXiv1204.1497D}
}

@ARTICLE{diemer_2018,
       author = {{Diemer}, Benedikt and {Stevens}, Adam R.~H. and {Forbes}, John C. and {Marinacci}, Federico and {Hernquist}, Lars and {Lagos}, Claudia del P. and {Sternberg}, Amiel and {Pillepich}, Annalisa and {Nelson}, Dylan and {Popping}, Gerg{\"o} and {Villaescusa-Navarro}, Francisco and {Torrey}, Paul and {Vogelsberger}, Mark},
        title = "{Modeling the Atomic-to-molecular Transition in Cosmological Simulations of Galaxy Formation}",
      journal = {\apjs},
         year = 2018,
        month = oct,
       volume = {238},
       number = {2},
          eid = {33},
        pages = {33},
          doi = {10.3847/1538-4365/aae387},
archivePrefix = {arXiv},
       eprint = {1806.02341},
 primaryClass = {astro-ph.GA},
       adsurl = {https://ui.adsabs.harvard.edu/abs/2018ApJS..238...33D}
}

@ARTICLE{dubois_2021,
       author = {{Dubois}, Yohan and {Beckmann}, Ricarda and {Bournaud}, Fr{\'e}d{\'e}ric and {Choi}, Hoseung and {Devriendt}, Julien and {Jackson}, Ryan and {Kaviraj}, Sugata and {Kimm}, Taysun and {Kraljic}, Katarina and {Laigle}, Clotilde and {Martin}, Garreth and {Park}, Min-Jung and {Peirani}, S{\'e}bastien and {Pichon}, Christophe and {Volonteri}, Marta and {Yi}, Sukyoung K.},
        title = "{Introducing the NEWHORIZON simulation: Galaxy properties with resolved internal dynamics across cosmic time}",
      journal = {\aap},
         year = 2021,
        month = jul,
       volume = {651},
          eid = {A109},
        pages = {A109},
          doi = {10.1051/0004-6361/202039429},
archivePrefix = {arXiv},
       eprint = {2009.10578},
 primaryClass = {astro-ph.GA},
       adsurl = {https://ui.adsabs.harvard.edu/abs/2021A&A...651A.109D}
}

@ARTICLE{duffy_2012a,
       author = {{Duffy}, A.~R. and {Moss}, A. and {Staveley-Smith}, L.},
        title = "{Cosmological Surveys with the Australian Square Kilometre Array Pathfinder}",
      journal = {\pasa},
         year = 2012,
        month = may,
       volume = {29},
       number = {2},
        pages = {202-211},
          doi = {10.1071/AS11013},
archivePrefix = {arXiv},
       eprint = {1103.3944},
 primaryClass = {astro-ph.CO},
       adsurl = {https://ui.adsabs.harvard.edu/abs/2012PASA...29..202D}
}

@ARTICLE{duffy_2012b,
       author = {{Duffy}, Alan R. and {Meyer}, Martin J. and {Staveley-Smith}, Lister and {Bernyk}, Maksym and {Croton}, Darren J. and {Koribalski}, B{\"a}rbel S. and {Gerstmann}, Derek and {Westerlund}, Stefan},
        title = "{Predictions for ASKAP neutral hydrogen surveys}",
      journal = {\mnras},
         year = 2012,
        month = nov,
       volume = {426},
       number = {4},
        pages = {3385-3402},
          doi = {10.1111/j.1365-2966.2012.21987.x},
archivePrefix = {arXiv},
       eprint = {1208.5592},
 primaryClass = {astro-ph.CO},
       adsurl = {https://ui.adsabs.harvard.edu/abs/2012MNRAS.426.3385D}
}

@ARTICLE{elahi_2018,
       author = {{Elahi}, Pascal J. and {Welker}, Charlotte and {Power}, Chris and {Lagos}, Claudia del P. and {Robotham}, Aaron S.~G. and {Ca{\~n}as}, Rodrigo and {Poulton}, Rhys},
        title = "{SURFS: Riding the waves with Synthetic UniveRses For Surveys}",
      journal = {\mnras},
         year = 2018,
        month = apr,
       volume = {475},
       number = {4},
        pages = {5338-5359},
          doi = {10.1093/mnras/sty061},
archivePrefix = {arXiv},
       eprint = {1712.01988},
 primaryClass = {astro-ph.GA},
       adsurl = {https://ui.adsabs.harvard.edu/abs/2018MNRAS.475.5338E}
}

@ARTICLE{euclid_2025,
       author = {{Euclid Collaboration} and {Mellier}, Y. and {Abdurro'uf} and {Acevedo Barroso}, J.~A. and {Ach{\'u}carro}, A. and {Adamek}, J. and {Adam}, R. and {Addison}, G.~E. and {Aghanim}, N. and {Aguena}, M. and {Ajani}, V. and {Akrami}, Y. and {Al-Bahlawan}, A. and {Alavi}, A. and {Albuquerque}, I.~S. and {Alestas}, G. and {Alguero}, G. and {Allaoui}, A. and {Allen}, S.~W. and {Allevato}, V. and {Alonso-Tetilla}, A.~V. and {Altieri}, B. and {Alvarez-Candal}, A. and {Alvi}, S. and {Amara}, A. and {Amendola}, L. and {Amiaux}, J. and {Andika}, I.~T. and {Andreon}, S. and {Andrews}, A. and {Angora}, G. and {Angulo}, R.~E. and {Annibali}, F. and {Anselmi}, A. and {Anselmi}, S. and {Arcari}, S. and {Archidiacono}, M. and {Aric{\`o}}, G. and {Arnaud}, M. and {Arnouts}, S. and {Asgari}, M. and {Asorey}, J. and {Atayde}, L. and {Atek}, H. and {Atrio-Barandela}, F. and {Aubert}, M. and {Aubourg}, E. and {Auphan}, T. and {Auricchio}, N. and {Aussel}, B. and {Aussel}, H. and {Avelino}, P.~P. and {Avgoustidis}, A. and {Avila}, S. and {Awan}, S. and {Azzollini}, R. and {Baccigalupi}, C. and {Bachelet}, E. and {Bacon}, D. and {Baes}, M. and {Bagley}, M.~B. and {Bahr-Kalus}, B. and {Balaguera-Antolinez}, A. and {Balbinot}, E. and {Balcells}, M. and {Baldi}, M. and {Baldry}, I. and {Balestra}, A. and {Ballardini}, M. and {Ballester}, O. and {Balogh}, M. and {Ba{\~n}ados}, E. and {Barbier}, R. and {Bardelli}, S. and {Baron}, M. and {Barreiro}, T. and {Barrena}, R. and {Barriere}, J. -C. and {Barros}, B.~J. and {Barthelemy}, A. and {Bartolo}, N. and {Basset}, A. and {Battaglia}, P. and {Battisti}, A.~J. and {Baugh}, C.~M. and {Baumont}, L. and {Bazzanini}, L. and {Beaulieu}, J. -P. and {Beckmann}, V. and {Belikov}, A.~N. and {Bel}, J. and {Bellagamba}, F. and {Bella}, M. and {Bellini}, E. and {Benabed}, K. and {Bender}, R. and {Benevento}, G. and {Bennett}, C.~L. and {Benson}, K. and {Bergamini}, P. and {Bermejo-Climent}, J.~R. and {Bernardeau}, F. and {Bertacca}, D. and {Berthe}, M. and {Berthier}, J. and {Bethermin}, M. and {Beutler}, F. and {Bevillon}, C. and {Bhargava}, S. and {Bhatawdekar}, R. and {Bianchi}, D. and {Bisigello}, L. and {Biviano}, A. and {Blake}, R.~P. and {Blanchard}, A. and {Blazek}, J. and {Blot}, L. and {Bosco}, A. and {Bodendorf}, C. and {Boenke}, T. and {B{\"o}hringer}, H. and {Boldrini}, P. and {Bolzonella}, M. and {Bonchi}, A. and {Bonici}, M. and {Bonino}, D. and {Bonino}, L. and {Bonvin}, C. and {Bon}, W. and {Booth}, J.~T. and {Borgani}, S. and {Borlaff}, A.~S. and {Borsato}, E. and {Bose}, B. and {Botticella}, M.~T. and {Boucaud}, A. and {Bouche}, F. and {Boucher}, J.~S. and {Boutigny}, D. and {Bouvard}, T. and {Bouwens}, R. and {Bouy}, H. and {Bowler}, R.~A.~A. and {Bozza}, V. and {Bozzo}, E. and {Branchini}, E. and {Brando}, G. and {Brau-Nogue}, S. and {Brekke}, P. and {Bremer}, M.~N. and {Brescia}, M. and {Breton}, M. -A. and {Brinchmann}, J. and {Brinckmann}, T. and {Brockley-Blatt}, C. and {Brodwin}, M. and {Brouard}, L. and {Brown}, M.~L. and {Bruton}, S. and {Bucko}, J. and {Buddelmeijer}, H. and {Buenadicha}, G. and {Buitrago}, F. and {Burger}, P. and {Burigana}, C. and {Busillo}, V. and {Busonero}, D. and {Cabanac}, R. and {Cabayol-Garcia}, L. and {Cagliari}, M.~S. and {Caillat}, A. and {Caillat}, L. and {Calabrese}, M. and {Calabro}, A. and {Calderone}, G. and {Calura}, F. and {Camacho Quevedo}, B. and {Camera}, S. and {Campos}, L. and {Ca{\~n}as-Herrera}, G. and {Candini}, G.~P. and {Cantiello}, M. and {Capobianco}, V. and {Cappellaro}, E. and {Cappelluti}, N. and {Cappi}, A. and {Caputi}, K.~I. and {Cara}, C. and {Carbone}, C. and {Cardone}, V.~F. and {Carella}, E. and {Carlberg}, R.~G. and {Carle}, M. and {Carminati}, L. and {Caro}, F. and {Carrasco}, J.~M. and {Carretero}, J. and {Carrilho}, P. and {Carron Duque}, J. and {Carry}, B.},
        title = "{Euclid: I. Overview of the Euclid mission}",
      journal = {\aap},
         year = 2025,
        month = may,
       volume = {697},
          eid = {A1},
        pages = {A1},
          doi = {10.1051/0004-6361/202450810},
archivePrefix = {arXiv},
       eprint = {2405.13491},
 primaryClass = {astro-ph.CO},
       adsurl = {https://ui.adsabs.harvard.edu/abs/2025A&A...697A...1E}
}

@ARTICLE{fernandez_2016,
       author = {{Fern{\'a}ndez}, Ximena and {Gim}, Hansung B. and {van Gorkom}, J.~H. and {Yun}, Min S. and {Momjian}, Emmanuel and {Popping}, Attila and {Chomiuk}, Laura and {Hess}, Kelley M. and {Hunt}, Lucas and {Kreckel}, Kathryn and {Lucero}, Danielle and {Maddox}, Natasha and {Oosterloo}, Tom and {Pisano}, D.~J. and {Verheijen}, M.~A.~W. and {Hales}, Christopher A. and {Chung}, Aeree and {Dodson}, Richard and {Golap}, Kumar and {Gross}, Julia and {Henning}, Patricia and {Hibbard}, John and {Jaff{\'e}}, Yara L. and {Donovan Meyer}, Jennifer and {Meyer}, Martin and {Sanchez-Barrantes}, Monica and {Schiminovich}, David and {Wicenec}, Andreas and {Wilcots}, Eric and {Bershady}, Matthew and {Scoville}, Nick and {Strader}, Jay and {Tremou}, Evangelia and {Salinas}, Ricardo and {Ch{\'a}vez}, Ricardo},
        title = "{Highest Redshift Image of Neutral Hydrogen in Emission: A CHILES Detection of a Starbursting Galaxy at z = 0.376}",
      journal = {\apjl},
         year = 2016,
        month = jun,
       volume = {824},
       number = {1},
          eid = {L1},
        pages = {L1},
          doi = {10.3847/2041-8205/824/1/L1},
archivePrefix = {arXiv},
       eprint = {1606.00013},
 primaryClass = {astro-ph.GA},
       adsurl = {https://ui.adsabs.harvard.edu/abs/2016ApJ...824L...1F}
}

@ARTICLE{fernandez_2013,
       author = {{Fern{\'a}ndez}, Ximena and {van Gorkom}, J.~H. and {Hess}, Kelley M. and {Pisano}, D.~J. and {Kreckel}, Kathryn and {Momjian}, Emmanuel and {Popping}, Attila and {Oosterloo}, Tom and {Chomiuk}, Laura and {Verheijen}, M.~A.~W. and {Henning}, Patricia A. and {Schiminovich}, David and {Bershady}, Matthew A. and {Wilcots}, Eric M. and {Scoville}, Nick},
        title = "{A Pilot for a Very Large Array H I Deep Field}",
      journal = {\apjl},
         year = 2013,
        month = jun,
       volume = {770},
       number = {2},
          eid = {L29},
        pages = {L29},
          doi = {10.1088/2041-8205/770/2/L29},
archivePrefix = {arXiv},
       eprint = {1303.2659},
 primaryClass = {astro-ph.GA},
       adsurl = {https://ui.adsabs.harvard.edu/abs/2013ApJ...770L..29F}
}

@ARTICLE{fontanot_2017b,
       author = {{Fontanot}, Fabio and {Hirschmann}, Michaela and {De Lucia}, Gabriella},
        title = "{Strong Stellar-driven Outflows Shape the Evolution of Galaxies at Cosmic Dawn}",
      journal = {\apjl},
         year = 2017,
        month = jun,
       volume = {842},
       number = {2},
          eid = {L14},
        pages = {L14},
          doi = {10.3847/2041-8213/aa74bd},
archivePrefix = {arXiv},
       eprint = {1703.02983},
 primaryClass = {astro-ph.GA},
       adsurl = {https://ui.adsabs.harvard.edu/abs/2017ApJ...842L..14F}
}

@ARTICLE{fontanot_2020,
       author = {{Fontanot}, Fabio and {De Lucia}, Gabriella and {Hirschmann}, Michaela and {Xie}, Lizhi and {Monaco}, Pierluigi and {Menci}, Nicola and {Fiore}, Fabrizio and {Feruglio}, Chiara and {Cristiani}, Stefano and {Shankar}, Francesco},
        title = "{The rise of active galactic nuclei in the galaxy evolution and assembly semi-analytic model}",
      journal = {\mnras},
         year = 2020,
        month = aug,
       volume = {496},
       number = {3},
        pages = {3943-3960},
          doi = {10.1093/mnras/staa1716},
archivePrefix = {arXiv},
       eprint = {2002.10576},
 primaryClass = {astro-ph.CO},
       adsurl = {https://ui.adsabs.harvard.edu/abs/2020MNRAS.496.3943F}
}

@ARTICLE{fontanot_2021,
       author = {{Fontanot}, Fabio and {Calabr{\`o}}, Antonello and {Talia}, Margherita and {Mannucci}, Filippo and {Castellano}, Marco and {Cresci}, Giovanni and {De Lucia}, Gabriella and {Gallazzi}, Anna and {Hirschmann}, Michaela and {Pentericci}, Laura and {Xie}, Lizhi and {Amorin}, Ricardo and {Bolzonella}, Micol and {Bongiorno}, Angela and {Cucciati}, Olga and {Cullen}, Fergus and {Fynbo}, Johan P.~U. and {Hathi}, Nimish and {Hibon}, Pascale and {McLure}, Ross J. and {Pozzetti}, Lucia},
        title = "{The evolution of the mass-metallicity relations from the VANDELS survey and the GAEA semi-analytic model}",
      journal = {\mnras},
         year = 2021,
        month = jul,
       volume = {504},
       number = {3},
        pages = {4481-4492},
          doi = {10.1093/mnras/stab1213},
archivePrefix = {arXiv},
       eprint = {2104.08295},
 primaryClass = {astro-ph.GA},
       adsurl = {https://ui.adsabs.harvard.edu/abs/2021MNRAS.504.4481F}
}

@ARTICLE{fontanot_2025,
       author = {{Fontanot}, Fabio and {De Lucia}, Gabriella and {Xie}, Lizhi and {Hirschmann}, Michaela and {Baugh}, Carlton and {Helly}, John C.},
        title = "{Galaxy assembly and evolution in the P-Millennium simulation: Galaxy clustering}",
      journal = {\aap},
         year = 2025,
        month = jul,
       volume = {699},
          eid = {A108},
        pages = {A108},
          doi = {10.1051/0004-6361/202452029},
archivePrefix = {arXiv},
       eprint = {2409.02194},
 primaryClass = {astro-ph.GA},
       adsurl = {https://ui.adsabs.harvard.edu/abs/2025A&A...699A.108F}
}

@ARTICLE{fraternali_2008,
       author = {{Fraternali}, F. and {Binney}, J.~J.},
        title = "{Accretion of gas on to nearby spiral galaxies}",
      journal = {\mnras},
         year = 2008,
        month = may,
       volume = {386},
       number = {2},
        pages = {935-944},
          doi = {10.1111/j.1365-2966.2008.13071.x},
archivePrefix = {arXiv},
       eprint = {0802.0496},
 primaryClass = {astro-ph},
       adsurl = {https://ui.adsabs.harvard.edu/abs/2008MNRAS.386..935F}
}

@ARTICLE{freeman_1970,
       author = {{Freeman}, K.~C.},
        title = "{On the Disks of Spiral and S0 Galaxies}",
      journal = {\apj},
         year = 1970,
        month = jun,
       volume = {160},
        pages = {811},
          doi = {10.1086/150474},
       adsurl = {https://ui.adsabs.harvard.edu/abs/1970ApJ...160..811F}
}

@ARTICLE{fu_2013,
       author = {{Fu}, Jian and {Kauffmann}, Guinevere and {Huang}, Mei-ling and {Yates}, Robert M. and {Moran}, Sean and {Heckman}, Timothy M. and {Dav{\'e}}, Romeel and {Guo}, Qi and {Henriques}, Bruno M.~B.},
        title = "{Star formation and metallicity gradients in semi-analytic models of disc galaxy formation}",
      journal = {\mnras},
         year = 2013,
        month = sep,
       volume = {434},
       number = {2},
        pages = {1531-1548},
          doi = {10.1093/mnras/stt1117},
archivePrefix = {arXiv},
       eprint = {1303.5586},
 primaryClass = {astro-ph.CO},
       adsurl = {https://ui.adsabs.harvard.edu/abs/2013MNRAS.434.1531F}
}

@ARTICLE{fu_2010,
       author = {{Fu}, Jian and {Guo}, Qi and {Kauffmann}, Guinevere and {Krumholz}, Mark R.},
        title = "{The atomic-to-molecular transition and its relation to the scaling properties of galaxy discs in the local Universe}",
      journal = {\mnras},
         year = 2010,
        month = dec,
       volume = {409},
       number = {2},
        pages = {515-530},
          doi = {10.1111/j.1365-2966.2010.17342.x},
archivePrefix = {arXiv},
       eprint = {1004.2325},
 primaryClass = {astro-ph.CO},
       adsurl = {https://ui.adsabs.harvard.edu/abs/2010MNRAS.409..515F}
}

@ARTICLE{giovanelli_2005,
       author = {{Giovanelli}, Riccardo and {Haynes}, Martha P. and {Kent}, Brian R. and {Perillat}, Philip and {Saintonge}, Amelie and {Brosch}, Noah and {Catinella}, Barbara and {Hoffman}, G. Lyle and {Stierwalt}, Sabrina and {Spekkens}, Kristine and {Lerner}, Mikael S. and {Masters}, Karen L. and {Momjian}, Emmanuel and {Rosenberg}, Jessica L. and {Springob}, Christopher M. and {Boselli}, Alessandro and {Charmandaris}, Vassilis and {Darling}, Jeremy K. and {Davies}, Jonathan and {Garcia Lambas}, Diego and {Gavazzi}, Giuseppe and {Giovanardi}, Carlo and {Hardy}, Eduardo and {Hunt}, Leslie K. and {Iovino}, Angela and {Karachentsev}, Igor D. and {Karachentseva}, Valentina E. and {Koopmann}, Rebecca A. and {Marinoni}, Christian and {Minchin}, Robert and {Muller}, Erik and {Putman}, Mary and {Pantoja}, Carmen and {Salzer}, John J. and {Scodeggio}, Marco and {Skillman}, Evan and {Solanes}, Jose M. and {Valotto}, Carlos and {van Driel}, Wim and {van Zee}, Liese},
        title = "{The Arecibo Legacy Fast ALFA Survey. I. Science Goals, Survey Design, and Strategy}",
      journal = {\aj},
         year = 2005,
        month = dec,
       volume = {130},
       number = {6},
        pages = {2598-2612},
          doi = {10.1086/497431},
archivePrefix = {arXiv},
       eprint = {astro-ph/0508301},
 primaryClass = {astro-ph},
       adsurl = {https://ui.adsabs.harvard.edu/abs/2005AJ....130.2598G}
}

@INCOLLECTION{giovanelli_1988,
       author = {{Giovanelli}, Riccardo and {Haynes}, Martha P.},
        title = "{Extragalactic neutral hydrogen.}",
    booktitle = {Galactic and Extragalactic Radio Astronomy},
         year = 1988,
       editor = {{Kellermann}, K.~I. and {Verschuur}, G.~L.},
        pages = {522-562},
       adsurl = {https://ui.adsabs.harvard.edu/abs/1988gera.book..522G},
    publisher = {Springer-Verlag}
}

@ARTICLE{gnedin_2011,
       author = {{Gnedin}, Nickolay Y. and {Kravtsov}, Andrey V.},
        title = "{Environmental Dependence of the Kennicutt-Schmidt Relation in Galaxies}",
      journal = {\apj},
         year = 2011,
        month = feb,
       volume = {728},
       number = {2},
          eid = {88},
        pages = {88},
          doi = {10.1088/0004-637X/728/2/88},
archivePrefix = {arXiv},
       eprint = {1004.0003},
 primaryClass = {astro-ph.CO},
       adsurl = {https://ui.adsabs.harvard.edu/abs/2011ApJ...728...88G}
}

@ARTICLE{gunn_1972,
       author = {{Gunn}, James E. and {Gott}, III, J. Richard},
        title = "{On the Infall of Matter Into Clusters of Galaxies and Some Effects on Their Evolution}",
      journal = {\apj},
         year = 1972,
        month = aug,
       volume = {176},
        pages = {1},
          doi = {10.1086/151605},
       adsurl = {https://ui.adsabs.harvard.edu/abs/1972ApJ...176....1G}
}

@ARTICLE{haynes_2018,
       author = {{Haynes}, Martha P. and {Giovanelli}, Riccardo and {Kent}, Brian R. and {Adams}, Elizabeth A.~K. and {Balonek}, Thomas J. and {Craig}, David W. and {Fertig}, Derek and {Finn}, Rose and {Giovanardi}, Carlo and {Hallenbeck}, Gregory and {Hess}, Kelley M. and {Hoffman}, G. Lyle and {Huang}, Shan and {Jones}, Michael G. and {Koopmann}, Rebecca A. and {Kornreich}, David A. and {Leisman}, Lukas and {Miller}, Jeffrey and {Moorman}, Crystal and {O'Connor}, Jessica and {O'Donoghue}, Aileen and {Papastergis}, Emmanouil and {Troischt}, Parker and {Stark}, David and {Xiao}, Li},
        title = "{The Arecibo Legacy Fast ALFA Survey: The ALFALFA Extragalactic H I Source Catalog}",
      journal = {\apj},
         year = 2018,
        month = jul,
       volume = {861},
       number = {1},
          eid = {49},
        pages = {49},
          doi = {10.3847/1538-4357/aac956},
archivePrefix = {arXiv},
       eprint = {1805.11499},
 primaryClass = {astro-ph.GA},
       adsurl = {https://ui.adsabs.harvard.edu/abs/2018ApJ...861...49H}
}

@ARTICLE{haynes_2011,
       author = {{Haynes}, Martha P. and {Giovanelli}, Riccardo and {Martin}, Ann M. and {Hess}, Kelley M. and {Saintonge}, Am{\'e}lie and {Adams}, Elizabeth A.~K. and {Hallenbeck}, Gregory and {Hoffman}, G. Lyle and {Huang}, Shan and {Kent}, Brian R. and {Koopmann}, Rebecca A. and {Papastergis}, Emmanouil and {Stierwalt}, Sabrina and {Balonek}, Thomas J. and {Craig}, David W. and {Higdon}, Sarah J.~U. and {Kornreich}, David A. and {Miller}, Jeffrey R. and {O'Donoghue}, Aileen A. and {Olowin}, Ronald P. and {Rosenberg}, Jessica L. and {Spekkens}, Kristine and {Troischt}, Parker and {Wilcots}, Eric M.},
        title = "{The Arecibo Legacy Fast ALFA Survey: The {\ensuremath{\alpha}}.40 H I Source Catalog, Its Characteristics and Their Impact on the Derivation of the H I Mass Function}",
      journal = {\aj},
         year = 2011,
        month = nov,
       volume = {142},
       number = {5},
          eid = {170},
        pages = {170},
          doi = {10.1088/0004-6256/142/5/170},
archivePrefix = {arXiv},
       eprint = {1109.0027},
 primaryClass = {astro-ph.CO},
       adsurl = {https://ui.adsabs.harvard.edu/abs/2011AJ....142..170H}
}

@ARTICLE{henriques_2015,
       author = {{Henriques}, Bruno M.~B. and {White}, Simon D.~M. and {Thomas}, Peter A. and {Angulo}, Raul and {Guo}, Qi and {Lemson}, Gerard and {Springel}, Volker and {Overzier}, Roderik},
        title = "{Galaxy formation in the Planck cosmology - I. Matching the observed evolution of star formation rates, colours and stellar masses}",
      journal = {\mnras},
         year = 2015,
        month = aug,
       volume = {451},
       number = {3},
        pages = {2663-2680},
          doi = {10.1093/mnras/stv705},
archivePrefix = {arXiv},
       eprint = {1410.0365},
 primaryClass = {astro-ph.GA},
       adsurl = {https://ui.adsabs.harvard.edu/abs/2015MNRAS.451.2663H}
}

@ARTICLE{hernquist_1990,
       author = {{Hernquist}, Lars},
        title = "{An Analytical Model for Spherical Galaxies and Bulges}",
      journal = {\apj},
         year = 1990,
        month = jun,
       volume = {356},
        pages = {359},
          doi = {10.1086/168845},
       adsurl = {https://ui.adsabs.harvard.edu/abs/1990ApJ...356..359H}
}

@ARTICLE{hirschmann_2016,
       author = {{Hirschmann}, Michaela and {De Lucia}, Gabriella and {Fontanot}, Fabio},
        title = "{Galaxy assembly, stellar feedback and metal enrichment: the view from the GAEA model}",
      journal = {\mnras},
         year = 2016,
        month = sep,
       volume = {461},
       number = {2},
        pages = {1760-1785},
          doi = {10.1093/mnras/stw1318},
archivePrefix = {arXiv},
       eprint = {1512.04531},
 primaryClass = {astro-ph.GA},
       adsurl = {https://ui.adsabs.harvard.edu/abs/2016MNRAS.461.1760H}
}

@ARTICLE{hollenbach_1971,
       author = {{Hollenbach}, David and {Salpeter}, E.~E.},
        title = "{Surface Recombination of Hydrogen Molecules}",
      journal = {\apj},
         year = 1971,
        month = jan,
       volume = {163},
        pages = {155},
          doi = {10.1086/150754},
       adsurl = {https://ui.adsabs.harvard.edu/abs/1971ApJ...163..155H}
}

@ARTICLE{howlett_2019,
       author = {{Howlett}, Cullan},
        title = "{The redshift-space momentum power spectrum - I. Optimal estimation from peculiar velocity surveys}",
      journal = {\mnras},
         year = 2019,
        month = aug,
       volume = {487},
       number = {4},
        pages = {5209-5234},
          doi = {10.1093/mnras/stz1403},
archivePrefix = {arXiv},
       eprint = {1906.02875},
 primaryClass = {astro-ph.CO},
       adsurl = {https://ui.adsabs.harvard.edu/abs/2019MNRAS.487.5209H}
}

@ARTICLE{howlett_2017b,
       author = {{Howlett}, Cullan and {Staveley-Smith}, Lister and {Blake}, Chris},
        title = "{Cosmological forecasts for combined and next-generation peculiar velocity surveys}",
      journal = {\mnras},
         year = 2017,
        month = jan,
       volume = {464},
       number = {3},
        pages = {2517-2544},
          doi = {10.1093/mnras/stw2466},
archivePrefix = {arXiv},
       eprint = {1609.08247},
 primaryClass = {astro-ph.CO},
       adsurl = {https://ui.adsabs.harvard.edu/abs/2017MNRAS.464.2517H}
}

@INPROCEEDINGS{jarvis_2016,
       author = {{Jarvis}, M. and {Taylor}, R. and {Agudo}, I. and {Allison}, J.~R. and {Deane}, R.~P. and {Frank}, B. and {Gupta}, N. and {Heywood}, I. and {Maddox}, N. and {McAlpine}, K. and {Santos}, M. and {Scaife}, A.~M.~M. and {Vaccari}, M. and {Zwart}, J.~T.~L. and {Adams}, E. and {Bacon}, D.~J. and {Baker}, A.~J. and {Bassett}, B.~A. and {Best}, P.~N. and {Beswick}, R. and {Blyth}, S. and {Brown}, M.~L. and {Bruggen}, M. and {Cluver}, M. and {Colafrancesco}, S. and {Cotter}, G. and {Cress}, C. and {Dav{\'e}}, R. and {Ferrari}, C. and {Hardcastle}, M.~J. and {Hale}, C.~L. and {Harrison}, I. and {Hatfield}, P.~W. and {Klockner}, H.~R. and {Kolwa}, S. and {Malefahlo}, E. and {Marubini}, T. and {Mauch}, T. and {Moodley}, K. and {Morganti}, R. and {Norris}, R.~P. and {Peters}, J.~A. and {Prandoni}, I. and {Prescott}, M. and {Oliver}, S. and {Oozeer}, N. and {Rottgering}, H.~J.~A. and {Seymour}, N. and {Simpson}, C. and {Smirnov}, O. and {Smith}, D.~J.~B.},
        title = "{The MeerKAT International GHz Tiered Extragalactic Exploration (MIGHTEE) Survey}",
    booktitle = {MeerKAT Science: On the Pathway to the SKA},
         year = 2016,
        month = jan,
          eid = {6},
        pages = {6},
          doi = {10.22323/1.277.0006},
archivePrefix = {arXiv},
       eprint = {1709.01901},
 primaryClass = {astro-ph.GA},
       adsurl = {https://ui.adsabs.harvard.edu/abs/2016mks..confE...6J}
}

@ARTICLE{jiang_2019,
       author = {{Jiang}, Zhen and {Wang}, Jie and {Gao}, Liang and {Zhang}, Feng-Hui and {Guo}, Qi and {Wang}, Lan and {Pan}, Jun},
        title = "{GABE: Galaxy Assembly with Binary Evolution}",
      journal = {Research in Astronomy and Astrophysics},
         year = 2019,
        month = oct,
       volume = {19},
       number = {10},
          eid = {151},
        pages = {151},
          doi = {10.1088/1674-4527/19/10/151},
archivePrefix = {arXiv},
       eprint = {1904.11224},
 primaryClass = {astro-ph.GA},
       adsurl = {https://ui.adsabs.harvard.edu/abs/2019RAA....19..151J}
}

@ARTICLE{johnston_2008,
       author = {{Johnston}, S. and {Taylor}, R. and {Bailes}, M. and {Bartel}, N. and {Baugh}, C. and {Bietenholz}, M. and {Blake}, C. and {Braun}, R. and {Brown}, J. and {Chatterjee}, S. and {Darling}, J. and {Deller}, A. and {Dodson}, R. and {Edwards}, P. and {Ekers}, R. and {Ellingsen}, S. and {Feain}, I. and {Gaensler}, B. and {Haverkorn}, M. and {Hobbs}, G. and {Hopkins}, A. and {Jackson}, C. and {James}, C. and {Joncas}, G. and {Kaspi}, V. and {Kilborn}, V. and {Koribalski}, B. and {Kothes}, R. and {Landecker}, T. and {Lenc}, E. and {Lovell}, J. and {Macquart}, J.-P. and {Manchester}, R. and {Matthews}, D. and {McClure-Griffiths}, N. and {Norris}, R. and {Pen}, U.-L. and {Phillips}, C. and {Power}, C. and {Protheroe}, R. and {Sadler}, E. and {Schmidt}, B. and {Stairs}, I. and {Staveley-Smith}, L. and {Stil}, J. and {Tingay}, S. and {Tzioumis}, A. and {Walker}, M. and {Wall}, J. and {Wolleben}, M.},
        title = "{Science with ASKAP. The Australian square-kilometre-array pathfinder}",
      journal = {Experimental Astronomy},
         year = 2008,
        month = dec,
       volume = {22},
       number = {3},
        pages = {151-273},
          doi = {10.1007/s10686-008-9124-7},
archivePrefix = {arXiv},
       eprint = {0810.5187},
 primaryClass = {astro-ph},
       adsurl = {https://ui.adsabs.harvard.edu/abs/2008ExA....22..151J}
}

@ARTICLE{jones_2018,
       author = {{Jones}, Michael G. and {Haynes}, Martha P. and {Giovanelli}, Riccardo and {Moorman}, Crystal},
        title = "{The ALFALFA H I mass function: a dichotomy in the low-mass slope and a locally suppressed `knee' mass}",
      journal = {\mnras},
         year = 2018,
        month = jun,
       volume = {477},
       number = {1},
        pages = {2-17},
          doi = {10.1093/mnras/sty521},
archivePrefix = {arXiv},
       eprint = {1802.00053},
 primaryClass = {astro-ph.GA},
       adsurl = {https://ui.adsabs.harvard.edu/abs/2018MNRAS.477....2J}
}

@ARTICLE{kennicutt_2012,
       author = {{Kennicutt}, Robert C. and {Evans}, Neal J.},
        title = "{Star Formation in the Milky Way and Nearby Galaxies}",
      journal = {\araa},
         year = 2012,
        month = sep,
       volume = {50},
        pages = {531-608},
          doi = {10.1146/annurev-astro-081811-125610},
archivePrefix = {arXiv},
       eprint = {1204.3552},
 primaryClass = {astro-ph.GA},
       adsurl = {https://ui.adsabs.harvard.edu/abs/2012ARA&A..50..531K}
}

@ARTICLE{keres_2005,
       author = {{Kere{\v{s}}}, Du{\v{s}}an and {Katz}, Neal and {Weinberg}, David H. and {Dav{\'e}}, Romeel},
        title = "{How do galaxies get their gas?}",
      journal = {\mnras},
         year = 2005,
        month = oct,
       volume = {363},
       number = {1},
        pages = {2-28},
          doi = {10.1111/j.1365-2966.2005.09451.x},
archivePrefix = {arXiv},
       eprint = {astro-ph/0407095},
 primaryClass = {astro-ph},
       adsurl = {https://ui.adsabs.harvard.edu/abs/2005MNRAS.363....2K}
}

@ARTICLE{kerr_1954,
       author = {{Kerr}, F.~J. and {Hindman}, J.~F. and {Robinson}, B.~J.},
        title = "{Observations of the 21 cm Line from the Magellanic Clouds}",
      journal = {Australian Journal of Physics},
         year = 1954,
        month = jan,
       volume = {7},
        pages = {297},
          doi = {10.1071/PH540297},
       adsurl = {https://ui.adsabs.harvard.edu/abs/1954AuJPh...7..297K}
}

@ARTICLE{koda_2014,
       author = {{Koda}, Jun and {Blake}, Chris and {Davis}, Tamara and {Magoulas}, Christina and {Springob}, Christopher M. and {Scrimgeour}, Morag and {Johnson}, Andrew and {Poole}, Gregory B. and {Staveley-Smith}, Lister},
        title = "{Are peculiar velocity surveys competitive as a cosmological probe?}",
      journal = {\mnras},
         year = 2014,
        month = dec,
       volume = {445},
       number = {4},
        pages = {4267-4286},
          doi = {10.1093/mnras/stu1610},
archivePrefix = {arXiv},
       eprint = {1312.1022},
 primaryClass = {astro-ph.CO},
       adsurl = {https://ui.adsabs.harvard.edu/abs/2014MNRAS.445.4267K}
}

@ARTICLE{koribalski_2020,
       author = {{Koribalski}, B{\"a}rbel S. and {Staveley-Smith}, L. and {Westmeier}, T. and {Serra}, P. and {Spekkens}, K. and {Wong}, O.~I. and {Lee-Waddell}, K. and {Lagos}, C.~D.~P. and {Obreschkow}, D. and {Ryan-Weber}, E.~V. and {Zwaan}, M. and {Kilborn}, V. and {Bekiaris}, G. and {Bekki}, K. and {Bigiel}, F. and {Boselli}, A. and {Bosma}, A. and {Catinella}, B. and {Chauhan}, G. and {Cluver}, M.~E. and {Colless}, M. and {Courtois}, H.~M. and {Crain}, R.~A. and {de Blok}, W.~J.~G. and {D{\'e}nes}, H. and {Duffy}, A.~R. and {Elagali}, A. and {Fluke}, C.~J. and {For}, B.-Q. and {Heald}, G. and {Henning}, P.~A. and {Hess}, K.~M. and {Holwerda}, B.~W. and {Howlett}, C. and {Jarrett}, T. and {Jones}, D.~H. and {Jones}, M.~G. and {J{\'o}zsa}, G.~I.~G. and {Jurek}, R. and {J{\"u}tte}, E. and {Kamphuis}, P. and {Karachentsev}, I. and {Kerp}, J. and {Kleiner}, D. and {Kraan-Korteweg}, R.~C. and {L{\'o}pez-S{\'a}nchez}, {\'A}. R. and {Madrid}, J. and {Meyer}, M. and {Mould}, J. and {Murugeshan}, C. and {Norris}, R.~P. and {Oh}, S.-H. and {Oosterloo}, T.~A. and {Popping}, A. and {Putman}, M. and {Reynolds}, T.~N. and {Rhee}, J. and {Robotham}, A.~S.~G. and {Ryder}, S. and {Schr{\"o}der}, A.~C. and {Shao}, Li and {Stevens}, A.~R.~H. and {Taylor}, E.~N. and {van{\^A} der Hulst}, J.~M. and {Verdes-Montenegro}, L. and {Wakker}, B.~P. and {Wang}, J. and {Whiting}, M. and {Winkel}, B. and {Wolf}, C.},
        title = "{WALLABY {\textendash} an SKA Pathfinder H I survey}",
      journal = {\apss},
         year = 2020,
        month = jul,
       volume = {365},
       number = {7},
          eid = {118},
        pages = {118},
          doi = {10.1007/s10509-020-03831-4},
archivePrefix = {arXiv},
       eprint = {2002.07311},
 primaryClass = {astro-ph.GA},
       adsurl = {https://ui.adsabs.harvard.edu/abs/2020Ap&SS.365..118K}
}

@ARTICLE{kourkchi_2020,
       author = {{Kourkchi}, Ehsan and {Tully}, R. Brent and {Eftekharzadeh}, Sarah and {Llop}, Jordan and {Courtois}, H{\'e}l{\`e}ne M. and {Guinet}, Daniel and {Dupuy}, Alexandra and {Neill}, James D. and {Seibert}, Mark and {Andrews}, Michael and {Chuang}, Juana and {Danesh}, Arash and {Gonzalez}, Randy and {Holthaus}, Alexandria and {Mokelke}, Amber and {Schoen}, Devin and {Urasaki}, Chase},
        title = "{Cosmicflows-4: The Catalog of {\ensuremath{\sim}}10,000 Tully-Fisher Distances}",
      journal = {\apj},
         year = 2020,
        month = oct,
       volume = {902},
       number = {2},
          eid = {145},
        pages = {145},
          doi = {10.3847/1538-4357/abb66b},
archivePrefix = {arXiv},
       eprint = {2009.00733},
 primaryClass = {astro-ph.GA},
       adsurl = {https://ui.adsabs.harvard.edu/abs/2020ApJ...902..145K}
}

@ARTICLE{kourkchi_2022,
       author = {{Kourkchi}, Ehsan and {Tully}, R. Brent and {Courtois}, H{\'e}l{\`e}ne M. and {Dupuy}, Alexandra and {Guinet}, Daniel},
        title = "{Cosmicflows-4: the baryonic Tully-Fisher relation providing  10 000 distances}",
      journal = {\mnras},
         year = 2022,
        month = apr,
       volume = {511},
       number = {4},
        pages = {6160-6178},
          doi = {10.1093/mnras/stac303},
archivePrefix = {arXiv},
       eprint = {2201.13023},
 primaryClass = {astro-ph.GA},
       adsurl = {https://ui.adsabs.harvard.edu/abs/2022MNRAS.511.6160K}
}

@ARTICLE{krumholz_2013,
       author = {{Krumholz}, Mark R.},
        title = "{The star formation law in molecule-poor galaxies}",
      journal = {\mnras},
         year = 2013,
        month = dec,
       volume = {436},
       number = {3},
        pages = {2747-2762},
          doi = {10.1093/mnras/stt1780},
archivePrefix = {arXiv},
       eprint = {1309.5100},
 primaryClass = {astro-ph.CO},
       adsurl = {https://ui.adsabs.harvard.edu/abs/2013MNRAS.436.2747K}
}

@ARTICLE{krumholz_2009,
       author = {{Krumholz}, Mark R. and {McKee}, Christopher F. and {Tumlinson}, Jason},
        title = "{The Star Formation Law in Atomic and Molecular Gas}",
      journal = {\apj},
         year = 2009,
        month = jul,
       volume = {699},
       number = {1},
        pages = {850-856},
          doi = {10.1088/0004-637X/699/1/850},
archivePrefix = {arXiv},
       eprint = {0904.0009},
 primaryClass = {astro-ph.GA},
       adsurl = {https://ui.adsabs.harvard.edu/abs/2009ApJ...699..850K}
}

@ARTICLE{lacey_2016,
       author = {{Lacey}, Cedric G. and {Baugh}, Carlton M. and {Frenk}, Carlos S. and {Benson}, Andrew J. and {Bower}, Richard G. and {Cole}, Shaun and {Gonzalez-Perez}, Violeta and {Helly}, John C. and {Lagos}, Claudia D.~P. and {Mitchell}, Peter D.},
        title = "{A unified multiwavelength model of galaxy formation}",
      journal = {\mnras},
         year = 2016,
        month = nov,
       volume = {462},
       number = {4},
        pages = {3854-3911},
          doi = {10.1093/mnras/stw1888},
archivePrefix = {arXiv},
       eprint = {1509.08473},
 primaryClass = {astro-ph.GA},
       adsurl = {https://ui.adsabs.harvard.edu/abs/2016MNRAS.462.3854L}
}

@ARTICLE{lagos_2018,
       author = {{Lagos}, Claudia del P. and {Tobar}, Rodrigo J. and {Robotham}, Aaron S.~G. and {Obreschkow}, Danail and {Mitchell}, Peter D. and {Power}, Chris and {Elahi}, Pascal J.},
        title = "{Shark: introducing an open source, free, and flexible semi-analytic model of galaxy formation}",
      journal = {\mnras},
         year = 2018,
        month = dec,
       volume = {481},
       number = {3},
        pages = {3573-3603},
          doi = {10.1093/mnras/sty2440},
archivePrefix = {arXiv},
       eprint = {1807.11180},
 primaryClass = {astro-ph.GA},
       adsurl = {https://ui.adsabs.harvard.edu/abs/2018MNRAS.481.3573L}
}

@ARTICLE{lagos_2011b,
       author = {{Lagos}, Claudia del P. and {Baugh}, Carlton M. and {Lacey}, Cedric G. and {Benson}, Andrew J. and {Kim}, Han-Seek and {Power}, Chris},
        title = "{Cosmic evolution of the atomic and molecular gas contents of galaxies}",
      journal = {\mnras},
         year = 2011,
        month = dec,
       volume = {418},
       number = {3},
        pages = {1649-1667},
          doi = {10.1111/j.1365-2966.2011.19583.x},
archivePrefix = {arXiv},
       eprint = {1105.2294},
 primaryClass = {astro-ph.CO},
       adsurl = {https://ui.adsabs.harvard.edu/abs/2011MNRAS.418.1649L}
}

@ARTICLE{laureijs_2011,
       author = {{Laureijs}, R. and {Amiaux}, J. and {Arduini}, S. and {Augu{\`e}res}, J. -L. and {Brinchmann}, J. and {Cole}, R. and {Cropper}, M. and {Dabin}, C. and {Duvet}, L. and {Ealet}, A. and {Garilli}, B. and {Gondoin}, P. and {Guzzo}, L. and {Hoar}, J. and {Hoekstra}, H. and {Holmes}, R. and {Kitching}, T. and {Maciaszek}, T. and {Mellier}, Y. and {Pasian}, F. and {Percival}, W. and {Rhodes}, J. and {Saavedra Criado}, G. and {Sauvage}, M. and {Scaramella}, R. and {Valenziano}, L. and {Warren}, S. and {Bender}, R. and {Castander}, F. and {Cimatti}, A. and {Le F{\`e}vre}, O. and {Kurki-Suonio}, H. and {Levi}, M. and {Lilje}, P. and {Meylan}, G. and {Nichol}, R. and {Pedersen}, K. and {Popa}, V. and {Rebolo Lopez}, R. and {Rix}, H. -W. and {Rottgering}, H. and {Zeilinger}, W. and {Grupp}, F. and {Hudelot}, P. and {Massey}, R. and {Meneghetti}, M. and {Miller}, L. and {Paltani}, S. and {Paulin-Henriksson}, S. and {Pires}, S. and {Saxton}, C. and {Schrabback}, T. and {Seidel}, G. and {Walsh}, J. and {Aghanim}, N. and {Amendola}, L. and {Bartlett}, J. and {Baccigalupi}, C. and {Beaulieu}, J. -P. and {Benabed}, K. and {Cuby}, J. -G. and {Elbaz}, D. and {Fosalba}, P. and {Gavazzi}, G. and {Helmi}, A. and {Hook}, I. and {Irwin}, M. and {Kneib}, J. -P. and {Kunz}, M. and {Mannucci}, F. and {Moscardini}, L. and {Tao}, C. and {Teyssier}, R. and {Weller}, J. and {Zamorani}, G. and {Zapatero Osorio}, M.~R. and {Boulade}, O. and {Foumond}, J.~J. and {Di Giorgio}, A. and {Guttridge}, P. and {James}, A. and {Kemp}, M. and {Martignac}, J. and {Spencer}, A. and {Walton}, D. and {Bl{\"u}mchen}, T. and {Bonoli}, C. and {Bortoletto}, F. and {Cerna}, C. and {Corcione}, L. and {Fabron}, C. and {Jahnke}, K. and {Ligori}, S. and {Madrid}, F. and {Martin}, L. and {Morgante}, G. and {Pamplona}, T. and {Prieto}, E. and {Riva}, M. and {Toledo}, R. and {Trifoglio}, M. and {Zerbi}, F. and {Abdalla}, F. and {Douspis}, M. and {Grenet}, C. and {Borgani}, S. and {Bouwens}, R. and {Courbin}, F. and {Delouis}, J. -M. and {Dubath}, P. and {Fontana}, A. and {Frailis}, M. and {Grazian}, A. and {Koppenh{\"o}fer}, J. and {Mansutti}, O. and {Melchior}, M. and {Mignoli}, M. and {Mohr}, J. and {Neissner}, C. and {Noddle}, K. and {Poncet}, M. and {Scodeggio}, M. and {Serrano}, S. and {Shane}, N. and {Starck}, J. -L. and {Surace}, C. and {Taylor}, A. and {Verdoes-Kleijn}, G. and {Vuerli}, C. and {Williams}, O.~R. and {Zacchei}, A. and {Altieri}, B. and {Escudero Sanz}, I. and {Kohley}, R. and {Oosterbroek}, T. and {Astier}, P. and {Bacon}, D. and {Bardelli}, S. and {Baugh}, C. and {Bellagamba}, F. and {Benoist}, C. and {Bianchi}, D. and {Biviano}, A. and {Branchini}, E. and {Carbone}, C. and {Cardone}, V. and {Clements}, D. and {Colombi}, S. and {Conselice}, C. and {Cresci}, G. and {Deacon}, N. and {Dunlop}, J. and {Fedeli}, C. and {Fontanot}, F. and {Franzetti}, P. and {Giocoli}, C. and {Garcia-Bellido}, J. and {Gow}, J. and {Heavens}, A. and {Hewett}, P. and {Heymans}, C. and {Holland}, A. and {Huang}, Z. and {Ilbert}, O. and {Joachimi}, B. and {Jennins}, E. and {Kerins}, E. and {Kiessling}, A. and {Kirk}, D. and {Kotak}, R. and {Krause}, O. and {Lahav}, O. and {van Leeuwen}, F. and {Lesgourgues}, J. and {Lombardi}, M. and {Magliocchetti}, M. and {Maguire}, K. and {Majerotto}, E. and {Maoli}, R. and {Marulli}, F. and {Maurogordato}, S. and {McCracken}, H. and {McLure}, R. and {Melchiorri}, A. and {Merson}, A. and {Moresco}, M. and {Nonino}, M. and {Norberg}, P. and {Peacock}, J. and {Pello}, R. and {Penny}, M. and {Pettorino}, V. and {Di Porto}, C. and {Pozzetti}, L. and {Quercellini}, C. and {Radovich}, M. and {Rassat}, A. and {Roche}, N. and {Ronayette}, S. and {Rossetti}, E.},
        title = "{Euclid Definition Study Report}",
      journal = {arXiv e-prints},
         year = 2011,
        month = oct,
          eid = {arXiv:1110.3193},
        pages = {arXiv:1110.3193},
          doi = {10.48550/arXiv.1110.3193},
archivePrefix = {arXiv},
       eprint = {1110.3193},
 primaryClass = {astro-ph.CO},
       adsurl = {https://ui.adsabs.harvard.edu/abs/2011arXiv1110.3193L}
}

@ARTICLE{lelli_2019,
       author = {{Lelli}, Federico and {McGaugh}, Stacy S. and {Schombert}, James M. and {Desmond}, Harry and {Katz}, Harley},
        title = "{The baryonic Tully-Fisher relation for different velocity definitions and implications for galaxy angular momentum}",
      journal = {\mnras},
         year = 2019,
        month = apr,
       volume = {484},
       number = {3},
        pages = {3267-3278},
          doi = {10.1093/mnras/stz205},
archivePrefix = {arXiv},
       eprint = {1901.05966},
 primaryClass = {astro-ph.GA},
       adsurl = {https://ui.adsabs.harvard.edu/abs/2019MNRAS.484.3267L}
}

@ARTICLE{lelli_2016a,
       author = {{Lelli}, Federico and {McGaugh}, Stacy S. and {Schombert}, James M.},
        title = "{SPARC: Mass Models for 175 Disk Galaxies with Spitzer Photometry and Accurate Rotation Curves}",
      journal = {\aj},
         year = 2016,
        month = dec,
       volume = {152},
       number = {6},
          eid = {157},
        pages = {157},
          doi = {10.3847/0004-6256/152/6/157},
archivePrefix = {arXiv},
       eprint = {1606.09251},
 primaryClass = {astro-ph.GA},
       adsurl = {https://ui.adsabs.harvard.edu/abs/2016AJ....152..157L}
}

@ARTICLE{leroy_2008,
       author = {{Leroy}, Adam K. and {Walter}, Fabian and {Brinks}, Elias and {Bigiel}, Frank and {de Blok}, W.~J.~G. and {Madore}, Barry and {Thornley}, M.~D.},
        title = "{The Star Formation Efficiency in Nearby Galaxies: Measuring Where Gas Forms Stars Effectively}",
      journal = {\aj},
         year = 2008,
        month = dec,
       volume = {136},
       number = {6},
        pages = {2782-2845},
          doi = {10.1088/0004-6256/136/6/2782},
archivePrefix = {arXiv},
       eprint = {0810.2556},
 primaryClass = {astro-ph},
       adsurl = {https://ui.adsabs.harvard.edu/abs/2008AJ....136.2782L}
}

@ARTICLE{maddox_2021,
       author = {{Maddox}, N. and {Frank}, B.~S. and {Ponomareva}, A.~A. and {Jarvis}, M.~J. and {Adams}, E.~A.~K. and {Dav{\'e}}, R. and {Oosterloo}, T.~A. and {Santos}, M.~G. and {Blyth}, S.~L. and {Glowacki}, M. and {Kraan-Korteweg}, R.~C. and {Mulaudzi}, W. and {Namumba}, B. and {Prandoni}, I. and {Rajohnson}, S.~H.~A. and {Spekkens}, K. and {Adams}, N.~J. and {Bowler}, R.~A.~A. and {Collier}, J.~D. and {Heywood}, I. and {Sekhar}, S. and {Taylor}, A.~R.},
        title = "{MIGHTEE-HI: The H I emission project of the MeerKAT MIGHTEE survey}",
      journal = {\aap},
         year = 2021,
        month = feb,
       volume = {646},
          eid = {A35},
        pages = {A35},
          doi = {10.1051/0004-6361/202039655},
archivePrefix = {arXiv},
       eprint = {2011.09470},
 primaryClass = {astro-ph.GA},
       adsurl = {https://ui.adsabs.harvard.edu/abs/2021A&A...646A..35M}
}

@ARTICLE{mcalpine_2022,
       author = {{McAlpine}, Stuart and {Helly}, John C. and {Schaller}, Matthieu and {Sawala}, Till and {Lavaux}, Guilhem and {Jasche}, Jens and {Frenk}, Carlos S. and {Jenkins}, Adrian and {Lucey}, John R. and {Johansson}, Peter H.},
        title = "{SIBELIUS-DARK: a galaxy catalogue of the local volume from a constrained realization simulation}",
      journal = {\mnras},
         year = 2022,
        month = jun,
       volume = {512},
       number = {4},
        pages = {5823-5847},
          doi = {10.1093/mnras/stac295},
archivePrefix = {arXiv},
       eprint = {2202.04099},
 primaryClass = {astro-ph.GA},
       adsurl = {https://ui.adsabs.harvard.edu/abs/2022MNRAS.512.5823M}
}

@ARTICLE{mccarthy_2017,
       author = {{McCarthy}, Ian G. and {Schaye}, Joop and {Bird}, Simeon and {Le Brun}, Amandine M.~C.},
        title = "{The BAHAMAS project: calibrated hydrodynamical simulations for large-scale structure cosmology}",
      journal = {\mnras},
         year = 2017,
        month = mar,
       volume = {465},
       number = {3},
        pages = {2936-2965},
          doi = {10.1093/mnras/stw2792},
archivePrefix = {arXiv},
       eprint = {1603.02702},
 primaryClass = {astro-ph.CO},
       adsurl = {https://ui.adsabs.harvard.edu/abs/2017MNRAS.465.2936M}
}

@ARTICLE{mcgaugh_2005,
       author = {{McGaugh}, Stacy S.},
        title = "{The Baryonic Tully-Fisher Relation of Galaxies with Extended Rotation Curves and the Stellar Mass of Rotating Galaxies}",
      journal = {\apj},
         year = 2005,
        month = oct,
       volume = {632},
       number = {2},
        pages = {859-871},
          doi = {10.1086/432968},
archivePrefix = {arXiv},
       eprint = {astro-ph/0506750},
 primaryClass = {astro-ph},
       adsurl = {https://ui.adsabs.harvard.edu/abs/2005ApJ...632..859M}
}

@ARTICLE{mcgaugh_2000,
       author = {{McGaugh}, S.~S. and {Schombert}, J.~M. and {Bothun}, G.~D. and {de Blok}, W.~J.~G.},
        title = "{The Baryonic Tully-Fisher Relation}",
      journal = {\apjl},
         year = 2000,
        month = apr,
       volume = {533},
       number = {2},
        pages = {L99-L102},
          doi = {10.1086/312628},
archivePrefix = {arXiv},
       eprint = {astro-ph/0003001},
 primaryClass = {astro-ph},
       adsurl = {https://ui.adsabs.harvard.edu/abs/2000ApJ...533L..99M}
}

@ARTICLE{mckee_2010,
       author = {{McKee}, Christopher F. and {Krumholz}, Mark R.},
        title = "{The Atomic-to-Molecular Transition in Galaxies. III. A New Method for Determining the Molecular Content of Primordial and Dusty Clouds}",
      journal = {\apj},
         year = 2010,
        month = jan,
       volume = {709},
       number = {1},
        pages = {308-320},
          doi = {10.1088/0004-637X/709/1/308},
archivePrefix = {arXiv},
       eprint = {0908.0330},
 primaryClass = {astro-ph.GA},
       adsurl = {https://ui.adsabs.harvard.edu/abs/2010ApJ...709..308M}
}

@ARTICLE{meyer_2004,
       author = {{Meyer}, M.~J. and {Zwaan}, M.~A. and {Webster}, R.~L. and {Staveley-Smith}, L. and {Ryan-Weber}, E. and {Drinkwater}, M.~J. and {Barnes}, D.~G. and {Howlett}, M. and {Kilborn}, V.~A. and {Stevens}, J. and {Waugh}, M. and {Pierce}, M.~J. and {Bhathal}, R. and {de Blok}, W.~J.~G. and {Disney}, M.~J. and {Ekers}, R.~D. and {Freeman}, K.~C. and {Garcia}, D.~A. and {Gibson}, B.~K. and {Harnett}, J. and {Henning}, P.~A. and {Jerjen}, H. and {Kesteven}, M.~J. and {Knezek}, P.~M. and {Koribalski}, B.~S. and {Mader}, S. and {Marquarding}, M. and {Minchin}, R.~F. and {O'Brien}, J. and {Oosterloo}, T. and {Price}, R.~M. and {Putman}, M.~E. and {Ryder}, S.~D. and {Sadler}, E.~M. and {Stewart}, I.~M. and {Stootman}, F. and {Wright}, A.~E.},
        title = "{The HIPASS catalogue - I. Data presentation}",
      journal = {\mnras},
         year = 2004,
        month = jun,
       volume = {350},
       number = {4},
        pages = {1195-1209},
          doi = {10.1111/j.1365-2966.2004.07710.x},
archivePrefix = {arXiv},
       eprint = {astro-ph/0406384},
 primaryClass = {astro-ph},
       adsurl = {https://ui.adsabs.harvard.edu/abs/2004MNRAS.350.1195M}
}

@ARTICLE{mo_1998,
       author = {{Mo}, H.~J. and {Mao}, Shude and {White}, Simon D.~M.},
        title = "{The formation of galactic discs}",
      journal = {\mnras},
         year = 1998,
        month = apr,
       volume = {295},
       number = {2},
        pages = {319-336},
          doi = {10.1046/j.1365-8711.1998.01227.x},
archivePrefix = {arXiv},
       eprint = {astro-ph/9707093},
 primaryClass = {astro-ph},
       adsurl = {https://ui.adsabs.harvard.edu/abs/1998MNRAS.295..319M}
}

@INPROCEEDINGS{murphy_2018,
       author = {{Murphy}, E.~J. and {Bolatto}, A. and {Chatterjee}, S. and {Casey}, C.~M. and {Chomiuk}, L. and {Dale}, D. and {de Pater}, I. and {Dickinson}, M. and {Francesco}, J.~D. and {Hallinan}, G. and {Isella}, A. and {Kohno}, K. and {Kulkarni}, S.~R. and {Lang}, C. and {Lazio}, T.~J.~W. and {Leroy}, A.~K. and {Loinard}, L. and {Maccarone}, T.~J. and {Matthews}, B.~C. and {Osten}, R.~A. and {Reid}, M.~J. and {Riechers}, D. and {Sakai}, N. and {Walter}, F. and {Wilner}, D.},
        title = "{The ngVLA Science Case and Associated Science Requirements}",
    booktitle = {Science with a Next Generation Very Large Array},
         year = 2018,
       editor = {{Murphy}, Eric},
       series = {Astronomical Society of the Pacific Conference Series},
       volume = {517},
        month = dec,
        pages = {3},
          doi = {10.48550/arXiv.1810.07524},
archivePrefix = {arXiv},
       eprint = {1810.07524},
 primaryClass = {astro-ph.IM},
       adsurl = {https://ui.adsabs.harvard.edu/abs/2018ASPC..517....3M}
}

@ARTICLE{navarro_1996,
       author = {{Navarro}, Julio F. and {Frenk}, Carlos S. and {White}, Simon D.~M.},
        title = "{The Structure of Cold Dark Matter Halos}",
      journal = {\apj},
         year = 1996,
        month = may,
       volume = {462},
        pages = {563},
          doi = {10.1086/177173},
archivePrefix = {arXiv},
       eprint = {astro-ph/9508025},
 primaryClass = {astro-ph},
       adsurl = {https://ui.adsabs.harvard.edu/abs/1996ApJ...462..563N}
}

@ARTICLE{neto_2007,
       author = {{Neto}, Angelo F. and {Gao}, Liang and {Bett}, Philip and {Cole}, Shaun and {Navarro}, Julio F. and {Frenk}, Carlos S. and {White}, Simon D.~M. and {Springel}, Volker and {Jenkins}, Adrian},
        title = "{The statistics of {\ensuremath{\Lambda}} CDM halo concentrations}",
      journal = {\mnras},
         year = 2007,
        month = nov,
       volume = {381},
       number = {4},
        pages = {1450-1462},
          doi = {10.1111/j.1365-2966.2007.12381.x},
archivePrefix = {arXiv},
       eprint = {0706.2919},
 primaryClass = {astro-ph},
       adsurl = {https://ui.adsabs.harvard.edu/abs/2007MNRAS.381.1450N}
}

@ARTICLE{obreschkow_2009a,
       author = {{Obreschkow}, D. and {Croton}, D. and {De Lucia}, G. and {Khochfar}, S. and {Rawlings}, S.},
        title = "{Simulation of the Cosmic Evolution of Atomic and Molecular Hydrogen in Galaxies}",
      journal = {\apj},
         year = 2009,
        month = jun,
       volume = {698},
       number = {2},
        pages = {1467-1484},
          doi = {10.1088/0004-637X/698/2/1467},
archivePrefix = {arXiv},
       eprint = {0904.2221},
 primaryClass = {astro-ph.CO},
       adsurl = {https://ui.adsabs.harvard.edu/abs/2009ApJ...698.1467O}
}

@ARTICLE{obreschkow_2009b,
       author = {{Obreschkow}, D. and {Kl{\"o}ckner}, H.-R. and {Heywood}, I. and {Levrier}, F. and {Rawlings}, S.},
        title = "{A Virtual Sky with Extragalactic H I and CO Lines for the Square Kilometre Array and the Atacama Large Millimeter/Submillimeter Array}",
      journal = {\apj},
         year = 2009,
        month = oct,
       volume = {703},
       number = {2},
        pages = {1890-1903},
          doi = {10.1088/0004-637X/703/2/1890},
archivePrefix = {arXiv},
       eprint = {0908.0983},
 primaryClass = {astro-ph.CO},
       adsurl = {https://ui.adsabs.harvard.edu/abs/2009ApJ...703.1890O}
}

@ARTICLE{oku_2024,
       author = {{Oku}, Yuri and {Nagamine}, Kentaro},
        title = "{Osaka Feedback Model. III. Cosmological Simulation CROCODILE}",
      journal = {\apj},
         year = 2024,
        month = nov,
       volume = {975},
       number = {2},
          eid = {183},
        pages = {183},
          doi = {10.3847/1538-4357/ad77d3},
archivePrefix = {arXiv},
       eprint = {2401.06324},
 primaryClass = {astro-ph.GA},
       adsurl = {https://ui.adsabs.harvard.edu/abs/2024ApJ...975..183O}
}

@ARTICLE{oman_2022,
       author = {{Oman}, Kyle A.},
        title = "{The ALFALFA H I velocity width function}",
      journal = {\mnras},
         year = 2022,
        month = jan,
       volume = {509},
       number = {3},
        pages = {3268-3284},
          doi = {10.1093/mnras/stab3164},
archivePrefix = {arXiv},
       eprint = {2108.08856},
 primaryClass = {astro-ph.GA},
       adsurl = {https://ui.adsabs.harvard.edu/abs/2022MNRAS.509.3268O}
}

@ARTICLE{oman_2024,
       author = {{Oman}, Kyle A. and {Frenk}, Carlos S. and {Crain}, Robert A. and {Lovell}, Mark R. and {Pfeffer}, Joel},
        title = "{A warm dark matter cosmogony may yield more low-mass galaxy detections in 21-cm surveys than a cold dark matter one}",
      journal = {\mnras},
         year = 2024,
        month = sep,
       volume = {533},
       number = {1},
        pages = {67-78},
          doi = {10.1093/mnras/stae1772},
archivePrefix = {arXiv},
       eprint = {2401.11878},
 primaryClass = {astro-ph.CO},
       adsurl = {https://ui.adsabs.harvard.edu/abs/2024MNRAS.533...67O}
}

@ARTICLE{papastergis_2011,
       author = {{Papastergis}, Emmanouil and {Martin}, Ann M. and {Giovanelli}, Riccardo and {Haynes}, Martha P.},
        title = "{The Velocity Width Function of Galaxies from the 40\% ALFALFA Survey: Shedding Light on the Cold Dark Matter Overabundance Problem}",
      journal = {\apj},
         year = 2011,
        month = sep,
       volume = {739},
       number = {1},
          eid = {38},
        pages = {38},
          doi = {10.1088/0004-637X/739/1/38},
archivePrefix = {arXiv},
       eprint = {1106.0710},
 primaryClass = {astro-ph.CO},
       adsurl = {https://ui.adsabs.harvard.edu/abs/2011ApJ...739...38P}
}

@ARTICLE{prada_2012,
       author = {{Prada}, Francisco and {Klypin}, Anatoly A. and {Cuesta}, Antonio J. and {Betancort-Rijo}, Juan E. and {Primack}, Joel},
        title = "{Halo concentrations in the standard {\ensuremath{\Lambda}} cold dark matter cosmology}",
      journal = {\mnras},
         year = 2012,
        month = jul,
       volume = {423},
       number = {4},
        pages = {3018-3030},
          doi = {10.1111/j.1365-2966.2012.21007.x},
archivePrefix = {arXiv},
       eprint = {1104.5130},
 primaryClass = {astro-ph.CO},
       adsurl = {https://ui.adsabs.harvard.edu/abs/2012MNRAS.423.3018P}
}

@ARTICLE{qin_2019,
       author = {{Qin}, Fei and {Howlett}, Cullan and {Staveley-Smith}, Lister},
        title = "{The redshift-space momentum power spectrum - II. Measuring the growth rate from the combined 2MTF and 6dFGSv surveys}",
      journal = {\mnras},
         year = 2019,
        month = aug,
       volume = {487},
       number = {4},
        pages = {5235-5247},
          doi = {10.1093/mnras/stz1576},
archivePrefix = {arXiv},
       eprint = {1906.02874},
 primaryClass = {astro-ph.CO},
       adsurl = {https://ui.adsabs.harvard.edu/abs/2019MNRAS.487.5235Q}
}

@ARTICLE{qin_2025,
       author = {{Qin}, Fei and {Howlett}, Cullan and {Parkinson}, David},
        title = "{The Redshift-space Momentum Power Spectrum. III. Measuring the Growth Rate from the SDSSv Survey Using the Auto- and Cross-power Spectrum of the Galaxy Density and Momentum Fields}",
      journal = {\apj},
         year = 2025,
        month = jan,
       volume = {978},
       number = {1},
          eid = {7},
        pages = {7},
          doi = {10.3847/1538-4357/ad9391},
archivePrefix = {arXiv},
       eprint = {2411.09571},
 primaryClass = {astro-ph.CO},
       adsurl = {https://ui.adsabs.harvard.edu/abs/2025ApJ...978....7Q}
}

@ARTICLE{sancisi_2008,
       author = {{Sancisi}, Renzo and {Fraternali}, Filippo and {Oosterloo}, Tom and {van der Hulst}, Thijs},
        title = "{Cold gas accretion in galaxies}",
      journal = {\aapr},
         year = 2008,
        month = jun,
       volume = {15},
       number = {3},
        pages = {189-223},
          doi = {10.1007/s00159-008-0010-0},
archivePrefix = {arXiv},
       eprint = {0803.0109},
 primaryClass = {astro-ph},
       adsurl = {https://ui.adsabs.harvard.edu/abs/2008A&ARv..15..189S}
}

@ARTICLE{sofue_2001,
       author = {{Sofue}, Yoshiaki and {Rubin}, Vera},
        title = "{Rotation Curves of Spiral Galaxies}",
      journal = {\araa},
         year = 2001,
        month = jan,
       volume = {39},
        pages = {137-174},
          doi = {10.1146/annurev.astro.39.1.137},
archivePrefix = {arXiv},
       eprint = {astro-ph/0010594},
 primaryClass = {astro-ph},
       adsurl = {https://ui.adsabs.harvard.edu/abs/2001ARA&A..39..137S}
}





\bsp
\label{lastpage}
\end{document}